\documentclass[a4paper,12pt,times,print,index,custommargin,custombib]{Classes/PhDThesisPSnPDF}
\usepackage[normalem]{ulem}

\input{Preamble/preamble}

\title{Stellar Disruption of Axion Minihalos and Consequences for Direct Axion Detection}

\author{Ian DSouza}

\dept{School of Physical and Chemical Sciences}

\university{University of Canterbury}
\crest{\includegraphics[width=0.3\textwidth]{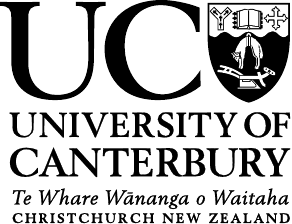}}

\degreetitle{Doctor of Philosophy}

\degreedate{August 2025} 

\subject{LaTeX} \keywords{{LaTeX} {PhD Thesis} {Physics} {University of
Canterbury}}

\ifdefineAbstract
\fi

\ifdefineChapter
\fi

\begin{document}

\frontmatter

\maketitle


\begin{acknowledgements}      

I would like to thank Dr.~Chris Gordon for supervising my PhD research. I would also like to thank Dr.~John Forbes for collaborating with me on the orbital evoluation of minihalos in the galactic environment.

\end{acknowledgements}

\begin{abstract}
    Scenarios such as the QCD axion with the Peccei-Quinn symmetry broken after inflation predict an enhanced matter power spectrum on sub-parsec scales. These theories lead to the formation of dense dark matter structures known as minihalos, which provide insights into early Universe dynamics and have implications for direct detection experiments. We examine the mass loss of minihalos during stellar encounters, building on previous studies that derived formulas for mass loss and performed N-body simulations. We propose a new formula for the mass loss that accounts for changes in the minihalo profile after disruption by a passing star. We also investigate the mass loss for multiple stellar encounters.
		We demonstrate that accurately assessing the mass loss in minihalos due to multiple stellar encounters necessitates considering the alterations in the minihalo's binding energy after each encounter, as overlooking this aspect results in a substantial underestimation of the mass loss.

        We further extend our analysis to the Galactic environment by more accurately incorporating multiple stellar encounters and dynamical relaxation timescales, simulating minihalo orbits in the Galactic potential.
Our results show stellar interactions are more destructive than previously estimated, reducing minihalo mass retention at the solar system to ~30\%, compared to earlier estimates of ~60\%. This enhanced loss arises from cumulative energy injections when relaxation periods between stellar encounters are accounted for.

        The altered minihalo mass function implies a larger fraction of axion dark matter occupies inter-minihalo space, potentially increasing the local axion density and improving haloscope detection prospects. This thesis highlights the significance of detailed modeling of stellar disruptions in shaping the axion dark matter distribution.
\end{abstract}


\tableofcontents

\listoffigures

\listoftables


\printnomenclature

\mainmatter


\chapter{Introduction} 
\label{ch:introduction}

\graphicspath{{Introduction/Figs/}}

Sections~\ref{sec:Background Cosmology}, \ref{sec:Misalignment Mechanism for a Classical Scalar Field}, \ref{sec:Misalignment Mechanism for the Axion}, \ref{sec:post-inflationary scenario}, \ref{sec:axion miniclusters} closely follow the treatment in Ref.~\cite{O'Hare24}. The reader is directed to this reference for detailed discussions.

\section{Background Cosmology}\label{sec:Background Cosmology}

On large enough length scales of $\mathcal{O}(10^8 \mathrm{pc})$, the Universe is expected to be isotropic and homogeneous. In this thesis, we will assume the Friedmann–Lemaître–Robertson–Walker (FLRW) Universe is spatially flat. The invariant spacetime distance is 
\begin{equation}
    \mathrm{d}s^2 = -\mathrm{d}t^2 + a^2(t)\left[\mathrm{d}x^2 + \mathrm{d}y^2 + \mathrm{d}z^2\right] \ ,
\end{equation}
where $a$ is called the scale factor of the Universe. \rev{Barring peculiar velocities,} $a$ encodes how the physical distance between two faraway galaxies changes with time. We assume $a(\mathrm{today}) = 1$. We also assume natural units where we set the speed of light $c_{\rm light}=1$. Here, $x,y,z$ are called the comoving distances along the three mutually perpendicular length axes. In general, physical distances are equal to the comoving distances multiplied by a factor of $a$. The metric of the FLRW universe is 
\begin{equation}
    g_{\rm FLRW} = \mathrm{diag}(-1, a^2, a^2, a^2)
\end{equation}

Starting from Einstein's equations and enforcing the FLRW metric tensor, we can derive some laws governing the evolution of the FLRW universe. We define the time-dependent Hubble parameter as $H = \dot{a}/a$. Then, the Friedmann equation is
\begin{equation}\label{eq:first Friedmann equation}
    H^2 \equiv \left(\frac{\dot{a}}{a}\right)^2 = \frac{8\pi G}{3}\sum\limits_i \rho_i \ ,
\end{equation}
where $G$ is Newton's gravitational constant, and $\rho_i$ is the large-scale average energy density of component $i$ that contributes to the overall energy density of the Universe. We assume that the components can be matter (baryonic or cold dark matter), radiation, and dark energy (represented by the cosmological constant). The Hubble constant is defined as $H_0 \equiv H(\mathrm{today})$. The dimensionless parameter $h$ is defined as
\begin{equation}\label{eq:Hubble constant definition}
    H_0 \equiv 100 h\ \mathrm{km\ s^{-1}\ Mpc^{-1}} \ . 
\end{equation}

In this thesis, we will use $h = 0.697$ \cite{Planck2018}. Furthermore, the redshift $z$ is defined as
\begin{equation}\label{eq:redshift denition}
    z \equiv \frac{a(\mathrm{today})}{a} - 1 = \frac{1}{a} - 1 \ .
\end{equation}

Let the density ratio parameter of the $i^{\rm th}$ component of the Universe's contents contributing to the total energy density of the universe be 
\begin{equation}\label{eq:density parameter for jth species}
    \Omega_i = \frac{\rho_i(\mathrm{today})} {\sum\limits_i  \rho_i(\mathrm{today})} \ .
\end{equation}

Let's expand eqn.~(\ref{eq:first Friedmann equation}) out where $i$ can correspond to matter (m), radiation (r), and dark energy ($\Lambda$).
\begin{equation}\label{eq:first Friedmann equation 2}
    H^2(a) = \frac{8\pi G}{3} \left[\rho_{\rm m}(a) + \rho_{\rm r}(a) + \rho_\Lambda(a)\right] \ .
\end{equation}

We assume that matter, radiation, and dark energy behave like perfect fluids. The scale factor encodes the length scale of the universe at a given time. If the length scale of the Universe increases by a factor of 2, and matter comoves with the Hubble flow, the physical volume encapsulating a system of $N$ matter particles increases by a factor of 8. By conservation of energy, we can say that the physical energy density of the same $N$ matter particles drops by a factor of $8 = 2^{3}$. Thus, the energy density of matter scales as $\rho_{\rm m} \propto a^{-3}$. A similar argument will hold for how the energy density of radiation dilutes. However, in the matter case, when the universe expanded, the energy associated with a single particle was constant. In the radiation case, the energy associated with a photon itself dilutes when the universe expands. We can think of this as the wavelength $\lambda_{\rm photon}$ of the photon getting stretched with the expanding universe. Since the energy of a single photon, $E_{\rm photon} \propto \lambda_{\rm photon}^{-1} \propto a^{-1}$, there is an additional factor of $a$ (compared to the matter case) in the dilution law for radiation: $\rho_{\rm r} \propto a^{-4}$. Finally, dark energy is assumed to have a constant energy density $\rho_\Lambda$ which is independent of the physical size of the universe. Thus, eqn. (\ref{eq:first Friedmann equation 2}) can be written as
\begin{equation}\label{eq:first Friedmann equation 3}
    H^2(a) = \frac{8\pi G}{3} \left[\rho_{\rm m}(\mathrm{today})a^{-3} + \rho_{\rm r}(\mathrm{today})a^{-4} + \rho_\Lambda(\mathrm{today})\right] \ .
\end{equation}

Setting $a = 1$ (today) in eqn.~(\ref{eq:first Friedmann equation 3}), we get
\begin{align}\label{eq:first Friedmann equation 4}
    H_0^2 &= \frac{8\pi G}{3} \left[\rho_{\rm m}(\mathrm{today}) + \rho_{\rm r}(\mathrm{today}) + \rho_\Lambda(\mathrm{today})\right] \nonumber \\
    &= \frac{8\pi G}{3} \sum\limits_{i \in \{\mathrm{m,r},\Lambda\} } \rho_i(\mathrm{today}) \ .
\end{align}

Dividing eqn.~(\ref{eq:first Friedmann equation 3}) by eqn.~(\ref{eq:first Friedmann equation 4}), and applying the formula in eqn.(\ref{eq:density parameter for jth species}), we get
\begin{equation}\label{eq:first Friedmann equation 5}
    \frac{H^2(a)}{H_0^2} = \Omega_{\rm m} a^{-3} + \Omega_{\rm r} a^{-4} + \Omega_\Lambda \ .
\end{equation}

Rearranging eqn.~(\ref{eq:first Friedmann equation 5}) and converting from scale factor $a$ to redshift $z$ using the formula in eqn.~(\ref{eq:redshift denition}), we get
\begin{equation}\label{eq:derived from Friedmann's first equation}
    H^2(z) = H_0^2 \left[\Omega_{\rm m} (1+z)^{3} + \Omega_{\rm r} (1+z)^{4} + \Omega_\Lambda\right] \ .
\end{equation}

\section{Dark Matter}
Dark matter is inferred as a non-luminous, pressureless component required to explain gravity-driven phenomena from galactic to cosmological scales. In spiral galaxies, rotation curves remain approximately flat, implying a mass distribution that cannot be supplied by stars and visible gas alone \cite{sofue2001rotation}. On cluster scales, gravitational lensing in mergers such as the Bullet Cluster shows the gravitational potential following the collisionless galaxies rather than the X-ray–bright plasma that contains most of the baryonic mass—directly revealing a dominant, unseen mass component displaced from the gas \cite{clowe2006direct}.

Cosmological probes independently require cold, non-baryonic dark matter. The pattern and amplitudes of acoustic peaks in the Cosmic Microwave Background require dark cold matter to be explained \cite{Planck2018}. Large-scale structure measurements, including the detection of the baryon-acoustic peak in galaxy clustering, trace the growth of structure seeded in the early universe and are consistent with the same dark matter-dominated model \cite{eisenstein2005detection}.

Beyond the well-known weakly interacting massive particles (WIMPs) paradigm, there are alternative theories that predict an enhanced matter power spectrum on sub-parsec scales. These theories, which include 
	the quantum chromodynamics  axion  with the Peccei-Quinn symmetry \citep{Peccei:1977hh} broken after inflation \citep[{\em e.g.},][]{Hogan:1988mp,Kolb1993,Kolb1994nonlinear,ZurekHoganQuinn07},
	early matter domination 
	\citep[{\em e.g.},][]{Erickcek:2011us,Fan:2014zua,visinelli2020axion}
	and vector dark matter models \citep[{\em e.g.},][]{Nelson:2011sf, Graham:2015rva}, lead to the formation of dense dark matter structures, which are known as {\em minihalos\/}.
	Dark matter minihalos, in these theories, originate earlier and are denser, making them much less susceptible to disruption compared to models such as those based on WIMPs, which do not have an enhanced matter power spectrum on sub-parsec scales \citep[{\em e.g.},][]{Ostriker1972, Gnedin1999, Goerdt2007, Zhao2007, Schneider2010}.
	Minihalos are potentially observable in local studies \citep[{\em e.g.},][]{Dror:2019twh, Ramani:2020hdo, Lee:2020wfn} and their presence would also have important implications for direct detection experiments \citep[{\em e.g.},][]{eggemeierAxionMinivoidsImplications2023, OHare2023}. 

Since the matter power spectrum has an enhancement at a comoving length scale of $r_{\rm mh} \lesssim 1 \rm pc$, we can find out the corresponding mass of the collapsed minihalos that form from regions in a homogeneous universe at this length scale using a back-of-the-envelope calculation. The comoving energy density $\rho_{\rm today}$ of such a universe is the physical energy density today. Rearranging eqn.~(\ref{eq:first Friedmann equation 4}), we get
\begin{equation}
    \rho_{\rm today} \equiv \sum\limits_{i \in \{\mathrm{m,r},\Lambda\} } \rho_i(\mathrm{today}) = \frac{3H_0^2}{8\pi G} \approx 1.35\times10^{-7} M_\odot \rm pc^{-3} \ ,
\end{equation}
where we have evaluated $H_0$ from eqn.~(\ref{eq:Hubble constant definition}) and we have assumed that $G = 6.674\times10^{-11} \rm m^3 kg^{-1} s^{-2}$. We can calculate the mass of a minihalo $M_{\rm mh}$ collapsed from matter enclosed by a sphere of radius $r_{\rm mh} \sim 1\rm pc$ as follows
\begin{equation}
    M_{\rm mh} = \rho_{\rm today} \frac{4\pi}{3} r_{\rm mh}^3 \approx 5.6\times10^{-7} M_\odot \ .
\end{equation}

Thus, we can say that an enhanced matter power spectrum at the sub-parsec comoving length scale corresponds to minihalos of mass $\lesssim 10^{-6}M_\odot$.

\section{Misalignment Mechanism for a Classical Scalar Field}
\label{sec:Misalignment Mechanism for a Classical Scalar Field}

The number of dark matter particles in its corresponding de Broglie volume is inversely proportional to the fourth power of the assumed mass of the particle. For a particle of mass $\sim 100 \mu\mathrm{eV}$, to predict the observed local dark matter density, we would need $~\sim 10^{22}$ particles in the de Broglie volume. Such a high occupation number compels us to treat the macroscopic behaviour of light dark matter classically. 

The field $\phi (\vec{x}, t)$ associated with the axion is assumed to be a \rev{real valued} scalar field. We assume the field lives in FLRW spacetime. The potential $V$ associated with the field is assumed to be quadratic with a local minimum at $\phi=0$.
\begin{equation}
V(\phi) = \frac{1}{2} m_\phi^2 \phi^2 \ ,
\end{equation}
\rev{where $m_\phi$ is the mass associated with $\phi$.} The Lagrangian density for this system can be written as 
\begin{equation}\label{eq:Lagrangian for phi}
    \mathcal{L} = -\frac{1}{2}\partial_\mu \phi \partial^\mu \phi-\frac{1}{2} m_\phi^2 \phi^2 \ .
\end{equation}

Using the Euler-Lagrange equation, one can derive the equation of motion of the field in FLRW spacetime. This equation of motion is called the Klein-Gordon (KG) equation. We can expand the field $\phi (\vec{x}, t)$ using the Fourier transform into its momentum modes $\phi (\vec{k}, t)$. Considering only the zero-momentum  mode $\phi (\vec{k} = 0, t)$ for now, the KG equation looks like:

\begin{equation}\label{eq:Klein-Gordon equation}
    \ddot{\phi}+3 H(t) \dot{\phi}+m_\phi^2 \phi=0 \ .
\end{equation}

In the radiation-dominated phase in the FLRW universe, we have $H(t) = 1/(2t)$. For the initial conditions, we assume that the initial state of the field is not at the minimum of its potential, i.e., $\phi_i \neq 0$. Furthermore, we assume $\dot{\phi}_i = 0$. Solving for eqn.~(\ref{eq:Klein-Gordon equation}) with these initial conditions shows an overdamped behaviour at early times when $H(t) \gtrsim m_\phi$ where we see an exponential decay trying to drive the field to zero without oscillations. However, as time passes, $H(t)$ decreases, eventually reaching an underdamped behaviour when $H(t) \lesssim  m_\phi$, where the field oscillates about zero and the amplitude of oscillations continues to decrease over time. During this late time oscillations, we can model the field as 

\begin{equation}
    \phi(t) \approx A_{\mathrm{env}}(t) \cos m_\phi t \ ,
\end{equation}
where $A_{\mathrm{env}}(t)$ is the envelope of oscillations and is a monotonically decreasing function of $t$.

If one computes the energy density of the field during late times,  averaged over time scales much longer than a single oscillation whose time period is fixed by $m_\phi$, and assumes relatively fast oscillations with a relatively slowly decaying envelope, it can be shown that the average energy density $\rho_\phi$ of the field behaves like

\begin{equation}\label{eq:scalar field behaves like matter}
    \rho_\phi (a) \propto a^{-3} \ .
\end{equation}

Since this is how the energy density of pressure-less matter decays with the cosmological scale factor $a$, the field can potentially be associated with dark matter behaviour.

CMB measurements by the \textit{Planck} satellite \cite{Planck2018} fix the density ratio parameter of dark matter to be
\begin{equation}\label{eq:Omega_DM}
    \Omega_{\rm DM} h^2 = 0.1200 \pm 0.0012 \ .
\end{equation}

If we want the scalar field $\phi$ to fully contribute to all of the inferred dark matter, then we mandate $\Omega_\phi = \Omega_{\rm DM}$. It can be shown that by combining the first Friedmann equation in eqn.~(\ref{eq:first Friedmann equation}) with eqns.~(\ref{eq:density parameter for jth species}) and (\ref{eq:Omega_DM}) that

\begin{equation}\label{eq:DM constraint equation}
    \Omega_\phi h^2 = \frac{8\pi G}{3H_0^2} \rho_\phi(\mathrm{today}) h^2 = 0.12 \ .
\end{equation}

To determine the value of $\rho_\phi(\mathrm{today})$, we have two free parameters in the system: the axion mass $m_\phi$ and the initial value of the field $\phi_i \neq 0$. Eqn.~(\ref{eq:DM constraint equation}) then constrains these two parameters relative to each other. Relating the average energy density today to that at $t_{\rm osc}$, it can be shown that 

\begin{equation}\label{eq:Omega_phi final constraint}
    \Omega_\phi h^2=0.12\left(\frac{\phi_i}{4.7 \times 10^{16} \mathrm{GeV}}\right)^2\left(\frac{m_\phi}{10^{-21} \mathrm{eV}}\right)^{\frac{1}{2}} \ .
\end{equation}

Eqn.~(\ref{eq:Omega_phi final constraint}) assumes that the field started oscillating at relatively late times (but strictly before matter-radiation equality), which can be achieved for small enough scalar mass $m_\phi$.

\section{Misalignment Mechanism for the Axion}
\label{sec:Misalignment Mechanism for the Axion}

The axion was introduced to solve the strong CP problem in Quantum Chromodynamics (QCD). The QCD axion is a pseudo-Nambu Goldstone boson that arises from the spontaneous breaking of the Peccei–Quinn (PQ) $U(1)_{\rm PQ}$ symmetry \cite{Peccei:1977hh,peccei1977constraints,LucaLandscape}  at some energy scale $f_a$. We consider a complex scalar field $\Phi_a$ and decompose it into a radial mode and a phase as: $\Phi_a \equiv |\Phi_a|e^{i\theta}$. The axion is associated with this phase $\theta$. The canonical field $\phi$ that is coupled to QCD can be written as: $\phi = \theta f_a$.

Assuming the QCD axion makes up all of the inferred dark matter \cite{preskill1983cosmology,abbott1983cosmological,dine1983not}, i.e., $\Omega_a h^2 = 0.12$, where $\Omega_a$ is the density ratio parameter for axions, we get
\begin{equation}
\label{eq:Omega_a final equation}
    \Omega_a h^2\approx 0.12 \left(\frac{\theta_i}{2.155}\right)^2 \left(\frac{9 \mu \mathrm{eV}}{m_a}\right)^{7/6} \ .
\end{equation}

Looking at eqn.(\ref{eq:Omega_a final equation}), we see that a QCD axion mass of $\mathcal{O}(10\mu\mathrm{eV})$ could explain the inferred dark matter. Such axion masses are potentially detectable by existing haloscope experiments. Hence, in this thesis, we will consider the QCD axion and eventually consider predictions for such experiments.

The PQ symmetry could have been broken before or after inflation. The two cases have different implications for \rev{$\theta_i$ which in turn impact } $\Omega_a$ and have to be treated separately. In what follows, we will only consider the case where the PQ symmetry was broken after inflation.

\section{PQ symmetry broken after inflation}
\label{sec:post-inflationary scenario}

During the inflationary period, the Hubble parameter $H$ is approximately constant, and so is the length scale of the Hubble horizon $H^{-1}$. After the end of inflation, we can consider many causally disconnected patches of the Universe. If the PQ symmetry breaks during this post-inflationary phase, the initial misalignment angles $\theta_i$ for each patch will be uncorrelated with the other patches. During this phase, $H$ decreases and the Hubble horizon $H^{-1}$ grows in size. Thus, these previously causally disconnected patches will ``enter'' the horizon and contribute to the dark matter abundance that we see in the observable universe today. Thus, to choose the value of $\theta_i^2$ in eqn.~(\ref{eq:Omega_a final equation}), we will have to average over all possible $\theta_i^2$ such that $\theta_i$ assumes values uniformly in the interval $[-\pi, \pi]$. This is under the approximation that the potential governing the dynamics of $\theta$ is a quadratic potential. However, since the true potential is cosinusoidal, applying anharmonic corrections, it can be shown that $\langle\theta_i^2\rangle = (2.155)^2$ \cite{di2016qcd}, which is a value slightly greater than the case where we assume the potential to be quadratic.

In the Klein-Gordon equation for $\theta$ (see eqn. (44) of Ref.~\cite{O'Hare24}), we have only considered the zero-momentum mode for which we could ignore the term containing the spatial derivative of $\theta$. However, in the post-inflationary case, although inflation tends to homogenize spatial inhomogeneities, the $\theta$ values arise after inflation. So, the inhomogeneities in \rev{$\theta$} are not wiped away. Thus, we will have to reconsider the spatial derivative term in the KG equation which \rev{is} $-\left(\nabla^2\theta\right)/a^2$. Considering this term gives rise to new objects like cosmic strings, domain walls, axion miniclusters, and axion stars \rev{(although axion stars can form in the pre-inflationary scenario)} \cite{Fleury2016,Klaer2017, Gorghetto2018, Buschmann2020, Gorghetto2021, Buschmann2022, OHare2022,LucaLandscape}. We will only look at the axion miniclusters in this thesis.  

\section{Axion Miniclusters}
\label{sec:axion miniclusters}

Around the QCD cross-over, the effective mass of the axion becomes non-negligible and plateaus. At this point, the spatial inhomogeneities in $\theta_i$ manifest as physical fluctuations in the distribution of axions that are assumed to make up all of dark matter. Defining the amplitude of these fluctuations as $\delta =  \left(\rho_a - \bar{\rho}_a \right) / \bar{\rho}_a$, it can be shown that amplitudes are high, $|\delta| \sim \mathcal{O}(1)$. {\rev{The length scales over which these fluctuations occur are comparable to the size of the Hubble horizon at the time the oscillations in $\theta$ began, but certainly smaller than the Hubble horizons at matter-radiation equality and today}. Note that in this section, we will explicitly write the local axion density in a region of space as $\rho_a$ while the average energy density of axions is $\bar{\rho}_a$. We note that the fluctuations can be underdense or overdense relative to the average axion density, depending on where in space \rev{one is} looking at. It is the overdense regions that collapse to form objects we call ``axion miniclusters'', before matter-radiation equality. Let us quantify these overdensities by their amplitude $\delta$. If the initial value of the amplitude of an overdensity is $\delta_i$, that overdensity will collapse at a scale factor of around $a_{\rm eq} / \delta_i$, where $a_{\rm eq}$ is the scale factor at matter-radiation equality. Considering $\delta_i \gtrsim 1$, this means that these overdensities collapse before matter-radiation equality. 

It can be shown that if we assume that the axion makes up all of the inferred dark matter today and that the axion field started its oscillations in the radiation-dominated epoch
\begin{equation}
    M = 2.4 \times10^{-12} M_\odot \left(\frac{100 \mu\mathrm{eV}}{m_a}\right)^{1/2} \ ,
\end{equation}
thus resulting in asteroid-mass miniclusters for the QCD axion. 

We assume that the miniclusters are formed from a spherically symmetric collapse of the axion overdensity \cite{kolb1994large,ellis2021axion}. In general, we have assumed an expanding FLRW Universe with $\dot{a} > 0$. The overdensity starts the collapse process when the gravity due to the axions inside the overdensity starts decoupling the overdensity from the Hubble flow which is the uniform expansion of space characteristic of the FLRW Universe. In the spherically symmetric overdensity, let's consider some shell of physical radius $r$. We can describe the evolution of $r$ by solving for an ordinary differential equation in $r$ that is dependent on the Hubble parameter during the radiation-dominated epoch (which in turn is dependent on the average energy density in radiation) and the mass enclosed by the spherical shell. It can be shown that at early times, significantly earlier than the matter-radiation equality, the shell starts decoupling from the Hubble flow but still expands. At some later time, the self-gravitational effect of the mass inside the shell becomes comparable to the outward dragging effect of the Hubble flow, and the shell temporarily halts in its expansion with $\dot{r} = 0$. We call this the ``turn-around''. After this, $\dot{r} < 0$ and the shell physically collapses. It can be shown \cite{ellis2021axion} that the average density of the minicluster is a factor of $10^7$ greater than the local dark matter density in our galaxy and that the radius of the assumed spherically symmetric minicluster is of the order of 1 AU, the astronomical unit. 

So far, we have defined the overdensity $\delta$ as a function of the position vector $\vec{x}$. We can use the Fourier transform to get the overdensity in comoving momentum space, $\delta(\vec{k})$. The comoving momentum $\vec{k}$ is related to the physical momentum $\vec{k}_{\rm phys}$ through $\vec{k} = a \vec{k}_{\rm phys}$. The correlations between density fluctuations in comoving momentum space can be expressed in terms of a power spectrum $P$ as
\begin{equation}
    \langle \delta(\vec{k})\ \delta(\vec{k^\prime}) \rangle = (2\pi)^3 \delta_D(\vec{k} - \vec{k^\prime}) P(k) \ ,
\end{equation}
where $\delta_D$ is the Dirac-Delta function and $k = |\vec{k}|$. For very small $k$, which corresponds to large length scales, we will have miniclusters that collapsed with uncorrelated $\theta_i$ values. Hence, the power spectrum for small $k$ will be approximately uncorrelated white noise, meaning that the $P$ is independent of $k$ in this regime. As we increase $k$, at some point, the power spectrum $P$ deviates from the uncorrelated white noise. The maximum value of $k$ that is relevant for the axion's power spectrum is called the Jeans wavenumber $k_J\big|_{z=0} \sim 10^{10} \rm Mpc^{-1}$, which corresponds to the length scale below which we do not expect axion overdensities to grow and eventually form miniclusters. 

We can define the dimensionless variance $\Delta^2$ associated with the power spectrum as
\begin{equation}
    \Delta^2(k) \equiv \frac{k^3}{2\pi^2} P(k) \ .
\end{equation}
As time progresses, $\Delta^2(k)$ exceeds unity and the miniclusters start growing non-linearly. Semi-analytic methods like the Press-Schechter, Peak-Patch, and Sheth-Tormen \cite{Blinov:2019jqc,ellis2021axion,fairbairnStructureFormationMicrolensing2018b,Enander2017,DavidsonSchwetz16,Ellis2022,PressSchecter74,ShethTormen99}, and numerical techniques like N-body simulations \cite{eggemeierAxionMinivoidsImplications2023,Eggemeier20,X2021,S2023,eggemeier2024evidence,pierobonMiniclustersAxionString2023a} have been developed for evolving this phase of minicluster growth.

N-body simulations model a set of $N$ particles in a given region of space, model the gravitational force of attraction between any two particles, and numerically evolve the system in time. These N-body simulations predict miniclusters of varying masses. The internal density structure of the miniclusters has been predicted to be a decaying power law of the radius, i.e., $\rho \propto r^{-\alpha}$ \cite{pierobonMiniclustersAxionString2023a,bertschinger1985self}. However, as time progresses, individual miniclusters merge hierarchically with each other to form larger objects called minihalos. Previous studies have explored the formation and evolution of axion minihalos, including their mass function and spatial distribution \cite{Hogan:1988mp, Kolb1993, KolbTkachev1994, KolbTkachev1996, ZurekHoganQuinn07, Hardy16, DavidsonSchwetz16, Enander2017, FairbairnMarshQuevillon2017, Fairbairn2018, Eggemeier2019, BlinovDolanDraper2020, Eggemeier20, Croon20, X2021, Edwards21, Ellis2022, Dandoy2024}. These minihalos have been predicted to have the Navarro-Frenk-White (NFW) density profile
\begin{equation}\label{eq:NFW profile}
    \rho_{\rm NFW}(r) = \frac{\rho_s}{\frac{r}{r_s}\left(1 + \frac{r}{r_s}\right)^2} \ ,
\end{equation}
where $\rho_s$ is the scale density and $r_s$ is the scale radius. Below the scale radius, the NFW density falls off as $r^{-1}$ while above the scale radius, the density falls off as $r^{-3}$. In addition, the virial radius ($r_{\rm vir}$) of the NFW density profile minihalo at a given redshift ($z$) is defined as that radius inside which the mean density of the minihalo is given by
\begin{equation}\label{eq:virial density}
	\bar{\rho}_{\mathrm{vir}}(z) = 200 \,\rho_{\rm crit}(z)\ ,
\end{equation}
and  $\rho_{\rm crit}$ is the cosmological critical density at redshift $z$. The concentration parameter of the minihalo is defined as
\begin{equation}\label{eq:concentration defintion}
    c \equiv \frac{r_{\rm vir}}{r_{\rm s}}\, .
\end{equation}

By the time galaxies form, most of the axions will be contained in these minihalos, and in this thesis, we will consider these minihalos. Some of these minihalos are expected to fall into galaxies where they will experience tidal forces from the stellar population in the galaxy. These tidal forces unbind axions from the minihalos, and much of this thesis will quantify this mass stripping effect.

We will now detail two semi-analytic techniques used to evolve the miniclusters in time: Press-Schechter and Sheth-Tormen methods.

\section{Press--Schechter and Sheth--Tormen Formalisms}\label{sec:Press-Schechter and Sheth-Tormen techniques}

This section summarises the Press-Schechter and Sheth-Tormen formalisms used to predict the abundance of collapsed miniclusters from the dark matter density field.

\subsection{Basic definitions}
Let $\delta(\vec{x},z)\equiv [\rho(\vec{x},z)-\bar\rho(z)]/\bar\rho(z)$ be the dark matter overdensity. \rev{Let us say one wants} to look at miniclusters of mass $M$. At this mass scale, we can smooth $\delta(\vec{x},z)$ using a window function which we assume to be a spherical top hat filter $W_R$ of length scale $R = (3M/4\pi\bar{\rho})^{1/3}$, where $\bar{\rho}$ is the comoving density of dark matter. When we say ``smooth'', we mean that we place a sphere of radius $R$ at some point in space. Then, we find the average overdensity inside that sphere and assign this average overdensity to the point in space where the centre of the sphere is located. We then translate the sphere to a new location in space and repeat the same procedure. In position space, the spherical top hat filter is mathematically defined as
\begin{equation}
    W_R(|\vec{x}|) = \frac{3}{4\pi R^3} \Theta(R - |\vec{x}|) \ ,
\end{equation}
where $\Theta$ is the Heaviside step function. We mathematically express the smoothed overdensity field $\delta_R$ in position space as the convolution of the window function with the (non-smoothed) overdensity field $\delta$ in position space 
\begin{equation}
\delta_R(\vec{x},z) \equiv \int \mathrm{d}^3x' W_R(|\vec{x}-\vec{x^\prime}|) \delta(\vec{x^\prime},z) = \int \frac{\mathrm{d}^3k}{(2\pi)^3} W(kR)\delta(\vec{k},z)e^{i\vec{k}\cdot\vec{x}} \ ,
\end{equation}
where in the second equality we have transitioned from position space to momentum space using the Fourier transform. $W$ is the Fourier transform in momentum space of the position space spherical top hat filter $W_R$. It can be shown that
\begin{equation}
W(kR) = \frac{3}{(kR)^3}\left[\sin (kR) - kR\cos (kR)\right] \ .
\end{equation}

The variance of the smoothed overdensity field is
\begin{equation}
\sigma^2(M,z) = \int\limits_0^{\infty} \frac{dk}{k}\frac{k^3P(k)}{2\pi^2}D^2(z)|W[kR(M)]|^2 \ ,
\end{equation}
where $P$ is the power spectrum, and $D$ is the growth function . The critical overdensity for spherical collapse is $\delta_c\simeq1.686$ \rev{\cite{dodelson2020modern} in a matter-dominated} Universe.

We define a dimensionless parameter that quantifies the rareness of a minicluster.
\begin{equation}\label{eq:nu definition}
\nu(M,z) \equiv \frac{\delta_c^2}{\sigma^2(M,z)} \, .
\end{equation}

\rev{
By definition, the mean of the unsmoothed overdensity field is $\langle \delta(\vec{x}, z) \rangle = 0$. Thus, the mean of the smoothed overdensity field is 
\begin{equation}
    \langle \delta_R(\vec{x}, z) \rangle = \int \mathrm{d}^3x^\prime W_R(|\vec{x} - \vec{x^\prime}|)\langle \delta(\vec{x}^\prime, z) \rangle = 0
\end{equation}
}

\rev{
We assume the smoothed density field $\delta_R(\vec{x}, z)$ to be a Gaussian random variable with mean 0 and variance $\sigma^2$. If a region of space has $\delta_R\left(\vec{x}, z\right) > \delta_c$, that region undergoes spherical collapse into a minicluster. Looking at eqn.~(\ref{eq:nu definition}), $\sqrt{\nu}$ is a measure of how many standard deviations away from the mean (0) the critical overdensity parameter $\delta_c$ is. This means it is statistically rare for collapse to occur when $\sqrt{\nu}(M,z)$ is large. Hence, $\nu(M,z)$ quantifies the rareness of a minicluster of mass $M$ collapsing at redshift $z$.
}

\subsection{Press--Schechter Formalism}

We can find the (cumulative) fraction of total mass of axions collapsed into miniclusters of mass greater than $M$ at a given redshift $z$ by integrating the Gaussian's tail above $\delta_c$
\begin{equation}\label{eq:cumulative mass fraction miniclusters above M}
F_{\rm PS}(>M,z)= 2 \int\limits_{\delta_c}^{\infty} \frac{\mathrm{d}\delta_R}{\sqrt{2\pi}\sigma(M,z)}\exp\left(-\frac{\delta_R^2}{2\sigma^2(M,z)}\right)
 = \operatorname{erfc}\left(\frac{\delta_c}{\sqrt{2}\,\sigma\rev{(M,z)}}\right) \ ,
\end{equation}
where the factor of 2 before the $\delta_R$ integral comes from the conservation of mass in the formal excursion-set derivation \cite{bond1991excursion}.

The mass fraction of axions collapsed into miniclusters in a mass interval  $[M,M+\mathrm{d}M]$ is $(\mathrm{d}F_{\rm PS}/\mathrm{d}M) \mathrm{d}M$. The corresponding number density of miniclusters in the same mass interval is then
\begin{align}\label{eq:mass function first equation}
    \mathrm{d}n &= \frac{\bar{\rho}}{M} \frac{\mathrm{d}F_{\rm PS}}{\mathrm{d}M} \mathrm{d}M  \nonumber\\
    \implies \frac{\mathrm{d}n}{\mathrm{d}M} &=\frac{\bar{\rho}}{M} \frac{\mathrm{d}F_{\rm PS}}{\mathrm{d}M} \ ,
\end{align}
where we call the term $\mathrm{d}n/\mathrm{d}M$ the mass function. We can define the multiplicity function $f_{\rm PS}$ as
\begin{equation}\label{eq:multiplicity definition}
    f_{\rm PS} \equiv \frac{\mathrm{d}F_{\rm PS}}{\mathrm{d}\nu} \ .
\end{equation}

Thus, eqn.~(\ref{eq:mass function first equation}) becomes
\begin{equation}
    \frac{\mathrm{d}n}{\mathrm{d}M}(M,z) =\frac{\bar{\rho}}{M} f_{\rm PS}(\nu) \frac{\mathrm{d}\nu}{\mathrm{d}M} \ .
\end{equation}

This gives us the mass function of miniclusters in the mass interval $[M,M+\mathrm{d}M]$ evaluated at redshift $z$. Plugging in eqn.~(\ref{eq:cumulative mass fraction miniclusters above M}) into eqn.~(\ref{eq:multiplicity definition}), it can be shown that
\begin{equation}
    \nu f_{\rm PS}(\nu) = \sqrt{\frac{\nu}{2\pi}} \exp\left(-\frac{\nu}{2}\right)
\end{equation}

\subsection{Sheth-Tormen formalism}

The Press-Schechter formalism assumes spherical collapse of overdensities and a constant critical density $\delta_c$. The iso-density contours around peaks in the over-density field are not exactly spherical due to tidal shear. Moreover, extensive $N$-body simulations show that the Press-Schechter formalism predicts a larger number density of miniclusters at the low mass end \cite{sheth2001ellipsoidal,jenkins2001mass}. Also, at the high mass end, there is a disagreement between Press-Schechter and numerical simulations. This motivated Sheth and Tormen to present a modification to the traditional Press-Schechter theory. The Sheth-Tormen formalism assumes ellipsoidal collapse and a mass-dependent critical density, wherein the small mass patches in the overdensity field require a higher critical density to collapse, leading to a smaller number density of miniclusters at the low mass end. Towards this, they proposed a new multiplicity function $f_{\rm ST}$ defined by 
\begin{equation}\label{eq: nu times f(nu) definition}
	\nu f_{\rm ST}(\nu)=A(p)\left[1+(q \nu)^{-p }\right]\left(\frac{q \nu}{2 \pi}\right)^{1 / 2} \exp \left(-\frac{q \nu}{2}\right) \ ,
\end{equation} 
where $q \simeq 0.75$, $p \simeq 0.3$, and $A(p) = \left[1+ \frac{2^{-p}}{\sqrt{\pi}}\Gamma\left(\frac{1}{2}-p\right)\right]^{-1} \simeq 0.3222$. Here, $q<1$ rescales the exponential term and boosts the high mass end of the mass function relative to the Press-Schechter formalism. Moreover, the $\left[1+(q \nu)^{-p }\right]$ term suppresses the low mass end of the mass function, making its prediction in closer agreement with the numerical simulations relative to the Press-Schechter formalism.

\section{Assumptions for stellar interaction of minihalos}

One key aspect affecting the survival and distribution of axion minihalos is their interaction with stars in the Milky Way galaxy. In this thesis, we will be primarily interested in how minihalos lose mass through tidal forces experienced when the minihalo encounters the gravitational presence of a star \cite{Berezinsky2013,Tinyakov2016,DokuchaevEroshenkoTkachev17,K2020,S2023,OHare2023}. Since the star has significantly more mass than the minihalo, this tidal stripping of mass from the minihalo becomes non-negligible. In section~\ref{sec:axion miniclusters}, we stated the physical length scale of the axion minihalo is $\sim$1 AU when they first form. The \rev{radius} of a star like our Sun is $\approx 0.00465 \rm AU$ \cite{rozelot2018big}. This motivates us to model the minihalo as an extended object and model the star as a point-like object. We then assume the distant-tide approximation and impulse approximation. The ideas developed below follow the references \cite{binneyTremaine,spitzer1958disruption,spitzer2014dynamical,ostriker1972evolution,aguilar1985tidal}.

\subsection{Geometry and Assumptions}
We consider an axion minihalo (target) of centre–of–mass (COM) position vector $\vec{0}$ and an internal tracer coordinate vector $\vec{r}$ (measured from the COM). This minihalo is perturbed by a passing star (perturber). Encounters are treated in the target COM frame. The star is modelled as a point-like object of mass $M_\star$. We define the length scale of the minihalo by its virial radius $r_{\rm vir}$. Let $\vec{r}_p$ correspond to the pericentre (distance of closest approach of the star to the minihalo) of the star's trajectory. Geometric non‑penetration is enforced by requiring  $r_p>r_{\rm vir}$.

Let $\vec{R}(t)$ be the relative position vector pointing from the minihalo’s centre of mass to the star, with $R \equiv |\vec{R}(t)|$. Let $V_\infty$ be the asymptotic relative speed. This is the relative speed when the star and minihalo are infinitely far away from each other, where the mutual gravitational force of attraction is negligible. Let $b$ be the impact parameter, which is the distance of closest approach of the star to the minihalo if the gravitational fields of the star and minihalo were `turned off'. Gravitational focusing produces the following relation \cite{binneyTremaine}
\begin{equation}
\label{eq:focusing b}
b^2 = r_p^2\left(1+\frac{2G M_\star}{r_p V_\infty^2}\right) \ ,
\end{equation}
with encounter duration $\tau_{\rm enc}\sim b/V_\infty$ \cite{MurrayDermott1999}. The relative speed $v_p$ of the star at the pericentre is 
\begin{equation}\label{eq:focusing v_p}
    v_p^2 = V_\infty^2 + \frac{2GM_\star}{r_p} \ .
\end{equation}
From eqn.~(\ref{eq:focusing b}), it can be inferred that $b$ is positively correlated with $r_p$. Thus, the minimum impact parameter $b_{\min}$ for non-penetrative encounter is
\begin{equation}
b_{\min} = r_{\rm vir}\sqrt{1+\frac{2GM_\star}{r_{\rm vir} V_\infty^2}}.
\end{equation}

\noindent The potential $\Phi_\star$ due to the point-mass star evaluated at a position defined by some placeholder vector $\vec{x}$ is
\begin{equation}\label{eq:potential for point-mass star}
    \Phi_\star(\vec{x})=-\frac{G M_\star}{\left\vert\vec{R} - \vec{x}\right\vert} \ .
\end{equation}
The tidal acceleration $\ddot{\vec{r}}\big|_{\rm tidal}$ of a tracer particle (in the minihalo) is the tracer’s acceleration relative to the minihalo's COM frame caused by spatial variations of the external potential generated by the star; the tracer’s total acceleration in the COM-frame is the sum of this tidal term and the minihalo’s internal gravitational force term. We estimate this tidal acceleration by the difference between the negative of the gradient of the star's potential evaluated at the tracer particle, and the negative of the gradient of the star's potential evaluated at the COM of the minihalo.
\begin{equation}
\label{eq:exact_tidal}
\ddot{\vec{r}}\big|_{\rm tidal} = -\nabla\Phi_\star\left(\vec{r}\right) + \nabla\Phi_\star\left(\vec{0}\right) \ .
\end{equation}

\subsection{Distant–tide approximation (spatial)}

Let's write the term $\nabla\Phi_\star\left(\vec{r}\right) \equiv \partial_i \Phi_\star\left(\vec{r}\right)$ from eqn.~(\ref{eq:exact_tidal}) as
\begin{equation}
    F_i \equiv \partial_i \Phi_\star \ .
\end{equation}
Taylor expanding out $F_i(\vec{r})$ about $\vec{0}$ (the COM), we get
\begin{align}\label{eq:taylor expansion of Phi_star}
    F_i\left(\vec{r}\right) &= F_i\left(\vec{0}\right) + r^j\partial_j F_i\left(\vec{0}\right) + \mathcal{O}\left(r^2\right) \nonumber  \\
   \implies \partial_i \Phi_\star\left(\vec{r}\right)& = \partial_i \Phi_\star\left(\vec{0}\right) + r^j \partial_j\partial_i \Phi_\star\left(\vec{0}\right) + \mathcal{O}\left(r^2\right) \ .
\end{align}
We define the symmetric tidal tensor $T_{ij} \equiv \partial_i\partial_j \Phi_\star$. Thus, eqn.~(\ref{eq:taylor expansion of Phi_star}) becomes
\begin{equation}\label{eq:taylor expansion of Phi_star part 2}
    \nabla \Phi_\star\left(\vec{r}\right) = \nabla \Phi_\star\left(\vec{0}\right) +  \mathbf{T}\left(\vec{0}\right)\vec{r} + \mathcal{O}\left(r^2\right) \ ,
\end{equation}
where $\mathbf{T}\left(\vec{0}\right)\vec{r} \equiv r^j T_{ij}\left(\vec{0}\right)$.  Substituting eqn.~(\ref{eq:taylor expansion of Phi_star part 2}) in eqn.~(\ref{eq:exact_tidal}), we get
\begin{equation}\label{eq:tidal acceleration part 1}
    \ddot{\vec{r}}\big|_{\rm tidal} = -\mathbf{T}(\vec{0})\ \vec{r} + \mathcal{O}\left(r^2\right) \ .
\end{equation}

We would like to evaluate how the magnitude of $\mathbf{T}\left(\vec{0}\right)\vec{r}$ scales with $r\equiv \left\vert\vec{r}\right\vert$ and $R\equiv \left\vert\vec{R}\right\vert$, relative to the $\mathcal{O}\left(r^2\right)$ terms. Towards this, we start by deriving some intermediate results. In index notation, we define $R \equiv \sqrt{R^k R_k}$. We also use the symmetric spatial Euclidean metric $\delta^{ij}$ or $\delta_{ij}$ to raise or lower indices. We can then write
\begin{align}\label{eq:partial_i R}
    \partial_i R &= \frac{1}{2\sqrt{R^k R_k}} \partial_i\left[R^k R_k\right] \nonumber\\
    &= \frac{1}{2R} \left[R_k \partial_i R^k + R^k\partial_i R_k\right]  \ .
\end{align}
But $R^k$ are just the spatial components of $\vec{R}$, evaluated in the same frame of reference as $\partial_i$. Thus,
\begin{equation}\label{eq:partial_i R^k}
    \partial_i R^k = \delta_i^k \ .
\end{equation}
Also, 
\begin{align}\label{eq:partial_i R_k}
    R_k &= \delta_{kj} R^j \nonumber\\
    \implies \partial_i R_k &= \partial_i\left[\delta_{kj} R^j\right] = \delta_{kj} \delta_i^j = \delta_{ik} \ ,
\end{align}
where in the second equality, we have used the result in eqn.~(\ref{eq:partial_i R^k}). Moreover,
\begin{equation}\label{eq:R^k}
    R^k = \delta^{km}R_m \ .
\end{equation}
Substituting eqns.~(\ref{eq:partial_i R^k}), (\ref{eq:R^k}), and (\ref{eq:partial_i R_k}) into eqn.~(\ref{eq:partial_i R}), we get
\begin{align}\label{eq:partial_i R - final result}
    \partial_i R &= \frac{1}{2R} \left[R_k \delta_i^k + \delta^{km} R_m \delta_{ik}\right] \nonumber\\
    &= \frac{1}{2R} \left[R_i + \delta_i^m R_m\right] \nonumber\\
    &= \frac{R_i}{R} \ .
\end{align}
Next, we derive the short result for some real number $n$:
\begin{align}\label{eq:partial_i R^{-n}}
    \partial_i \left(R^{-n}\right)  &= \frac{\mathrm{d}}{\mathrm{d}R}\left(R^{-n}\right)\ \partial_i R = -n R^{-n-1} \frac{R_i}{R} \nonumber\\
    &= -\frac{n R_i}{R^{n+2}} \ ,
\end{align}
where we have used the result from eqn.~(\ref{eq:partial_i R - final result}). We are now in a position to derive an expression for the tidal tensor evaluated at the COM of the minihalo: $T_{ij}\left(\vec{0}\right) \equiv \partial_i \partial_j \Phi_\star\left(\vec{0}\right)$. Using eqn.~(\ref{eq:potential for point-mass star}), we have
\begin{align}\label{eq:partial_j Phi_star}
    \partial_j \Phi_\star\left(\vec{0}\right) &= \partial_j \left[-G M_\star R^{-1}\right] = -G M_\star \left[-\frac{R_j}{R^3}\right] \nonumber\\
    &= G M_\star \frac{R_j}{R^3} \ ,
\end{align}
where in the second equality of the first line, we have used the result from eqn.~(\ref{eq:partial_i R^{-n}}) with $n=1$. Next, we have
\begin{align}\label{eq:T_ij expression as a function of R}
    T_{ij}\left(\vec{0}\right) &\equiv \partial_i \partial_j \Phi_\star\left(\vec{0}\right) = \partial_i\left[G M_\star \frac{R_j}{R^3}\right] \nonumber\\
    &= G M_\star \left[\frac{1}{R^3}\partial_i R_j + R_j \partial_i\left(R^{-3}\right)\right] \nonumber\\
    &= G M_\star \left[\frac{\delta_{ij}}{R^3} + R_j \left(-\frac{3R_i}{R^5}\right)\right] \nonumber\\
    &= \frac{G M_\star}{R^3} \left[\delta_{ij} - \frac{3R_i R_j}{R^2}\right] \ ,
\end{align}
where we have used the results from eqns.~(\ref{eq:partial_i R_k}) and (\ref{eq:partial_j Phi_star}). We have also used the result from eqn.~(\ref{eq:partial_i R^{-n}}) with $n=3$. We now evaluate the linear order term $\mathbf{T}(\vec{0})\ \vec{r}$ as
\begin{align}\label{eq:linear order term}
    r^j T_{ij}\left(\vec{0}\right) &= \frac{G M_\star}{R^3}  \left[r^j \delta_{ij} - \frac{3R_i r^j  R_j}{R^2}\right] \nonumber\\
    &= \frac{G M_\star}{R^3} \left[r_i - \frac{3 \left(r^j  R_j\right)}{R^2}R_i\right] \ .
\end{align}
We now write the scalar dot product of $\vec{r}$ and $\vec{R}$ as
\begin{equation}\label{eq:dot product of r and R}
    \vec{r} \cdot \vec{R} \equiv r^j  R_j = r R \cos\left(\theta_{rR}\right) \ ,
\end{equation}
where $\theta_{rR}$ is the angle between the vectors $\vec{r}$ and $\vec{R}$. Therefore, eqn.~(\ref{eq:linear order term}) becomes
\begin{align}
    r^j T_{ij}\left(\vec{0}\right) &= \frac{G M_\star}{R^3} \left[r_i - \frac{3 r R \cos\left(\theta_{rR}\right) }{R^2}R_i\right] \nonumber\\
    &= \frac{G M_\star}{R^3} \left[r_i - \frac{3 r  \cos\left(\theta_{rR}\right) }{R}R_i\right]
\end{align}
Defining $f_i \equiv r^j T_{ij}\left(\vec{0}\right)$, we can find the magnitude of $\mathbf{T}\left(\vec{0}\right)\vec{r}$ as
\begin{align}\label{eq:magnitude of linear order term, part 1}
    \left\vert\mathbf{T}\left(\vec{0}\right)\vec{r}\right\vert &= \left(f^i f_i\right)^{1/2} \nonumber\\
    &= \frac{G M_\star}{R^3}\left(\left[r^i - \frac{3 r  \cos\left(\theta_{rR}\right) }{R}R^i\right]\left[r_i - \frac{3 r  \cos\left(\theta_{rR}\right) }{R}R_i\right]\right)^{1/2} \nonumber\\
    &= \frac{G M_\star}{R^3}\left( r^2 + \frac{9r^2\cos^2\left(\theta_{rR}\right)}{R^2}R^2 -\frac{3 r  \cos\left(\theta_{rR}\right)}{R} \left[r^i R_i + r_i R^i\right]\right)^{1/2} \ ,
\end{align}
where we have used the identities $r^i r_i \equiv r^2$ and $R^i R_i \equiv R^2$. Making use of the result in eqn.~(\ref{eq:dot product of r and R}), we get
%
\begin{align}
    \left\vert\mathbf{T}\left(\vec{0}\right)\vec{r}\right\vert & =\frac{G M_\star}{R^3} \left(r^2 + 9r^2\cos^2\left(\theta_{rR}\right) - \frac{3r\cos\left(\theta_{rR}\right)}{R}\left[2rR \cos\left(\theta_{rR}\right)\right]\right)^{1/2} \nonumber\\
    &= \frac{G M_\star}{R^3} \left(r^2 + 3r^2\cos^2\left(\theta_{rR}\right)\right)^{1/2} \nonumber\\
    &= \frac{G M_\star}{R^3} r \left(1 + 3\cos^2\left(\theta_{rR}\right)\right)^{1/2} \ .
\end{align}
Now, we know that
\begin{align}
    &0 \leq \cos^2\left(\theta_{rR}\right) \leq 1 \nonumber\\
    \implies &1 \leq \left(1 + 3\cos^2\left(\theta_{rR}\right)\right)^{1/2} \leq 2 \nonumber\\
    \implies & \frac{G M_\star}{R^3} r \leq \left\vert\mathbf{T}\left(\vec{0}\right)\vec{r}\right\vert \leq 2 \frac{G M_\star}{R^3} r \ .
\end{align}
Hence, as an order of magnitude estimate, we can write
\begin{equation}\label{eq:magnitude of linear order term, final part}
    \left\vert\mathbf{T}\left(\vec{0}\right)\vec{r}\right\vert \sim \frac{G M_\star}{R^2} \frac{r}{R} \ .
\end{equation}
We would now like to also estimate the magnitude of the quadratic and higher order terms in eqn.~(\ref{eq:tidal acceleration part 1}). Doing this rigorously is exhaustive. Instead, we will look again at the magnitude of the linear order term in eqn.~(\ref{eq:magnitude of linear order term, final part}), develop some intuition, and extrapolate that intuition to estimate the magnitude of the higher order terms. All terms in eqn.~(\ref{eq:tidal acceleration part 1}) should have the dimensions of acceleration, i.e., $\mathrm{LT^{-2}} = \mathrm{dim}\left(G M_\star\right) \mathrm{L^{-2}}$. We know that the linear order term in eqn.~(\ref{eq:magnitude of linear order term, final part}) is linear in $r$. Apart from the terms $G M_\star$, the remaining terms must have the dimensions of $L^{-2}$. But we are only allowed one factor of $r$. Thus, we must divide $r$ by the third power of some other length scale. The only other characteristic length scale variable in our system is $R$. Hence, we must divide by $R^3$  to have consistent dimensions. Hence the factors of $r/R^3$ appear in eqn.~(\ref{eq:magnitude of linear order term, final part}).

We now extend this intuition to the quadratic term in eqn.~(\ref{eq:tidal acceleration part 1}) which must also have the dimensions of $\mathrm{dim}\left(G M_\star\right) \mathrm{L^{-2}}$. Here, we are only allowed to use a factor of $r^2$. Thus, we must divide $r^2$ by the fourth power of $R$ for dimensional consistency. Hence,
\begin{equation}\label{eq:magnitude of the quadrtic term}
    \left\vert \text{Quadratic Term} \right\vert \sim \frac{G M_\star}{R^2} \left(\frac{r}{R}\right)^2 \ .
\end{equation}
Going to even higher order terms, we can write the magnitude of the $n^{\rm th}$-order term as
\begin{equation}\label{eq:magnitude of the n^th order term}
    \left\vert \text{$n^{\rm th}$-order Term} \right\vert \sim \frac{G M_\star}{R^2} \left(\frac{r}{R}\right)^n \ .
\end{equation}

\noindent We assume that $r<r_{\rm vir}$ and $R \gtrsim b$ for weak focusing. Thus we can write
\begin{equation}\label{eq:tidal acceleration part 3}
    \frac{r}{R} < \frac{r_{\rm vir}}{b} \ .
\end{equation}
We now make the critical approximation that the physical size of the minihalo is significantly smaller than the impact parameter, i.e., $r_{\rm vir} \ll b$. This is called the distant-tide approximation. This implies
\begin{equation}
    \frac{r}{R} < \frac{r_{\rm vir}}{b} \ll 1 \ .
\end{equation}
With this constraint, it is clear from eqn.~(\ref{eq:magnitude of linear order term, final part}), (\ref{eq:magnitude of the quadrtic term}) and (\ref{eq:magnitude of the n^th order term}) that only the linear order term in eqn.~(\ref{eq:tidal acceleration part 1}) survives, and all higher order terms become sub-dominant. Hence, we can write eqn.~(\ref{eq:tidal acceleration part 1}) as
\begin{equation}\label{eq:tidal acceleration in the distant tide approximation}
    \ddot{\vec{r}}\big|_{\rm tidal} \approx -\mathbf{T}(\vec{0})\ \vec{r} \ .
\end{equation}

\subsection{Impulse approximation (temporal)}

To evaluate the cumulative velocity change of a tracer particle in general, we approximate the minihalo’s self-potential $\Psi_{\rm int}$ near its COM by a quadratic form \cite{binneyTremaine}
\begin{equation}\label{eq:internal potential}
    \Psi_{\rm int}(\vec{r}) \approx \Psi_{\rm int}(\vec{r} = \vec{0}) + \tfrac12\, \vec{r}^{\top}\mathbf K\,\vec{r} \ ,
\qquad K_{ij}\equiv \left.\frac{\partial^2\Psi_{\rm int}}{\partial x_i\partial x_j}\right|_{\vec{r} = \vec{0}} \ .
\end{equation}
Note that the linear order term containing $\partial_i \Psi_{\rm int}\left(\vec{r}=\vec{0}\right)$ is not present in the equation above. This term is an acceleration term. However, assuming a spherically symmetric density profile for the minihalo, this term becomes zero since one does not experience a `force' at the centre of the minihalo due to its self-potential.  The COM-frame equation of motion (ignoring the external tide due to the star for the moment) of the tracer particle is
\begin{equation}\label{eq:internal equation of motion part 1}
    \ddot{\vec{r}} = -\nabla\Psi_{\rm int}(\vec{r})\ . 
\end{equation}
From eqn.~(\ref{eq:internal potential}), we can write
\begin{equation}\label{eq:derivative of internal potential}
    \nabla\Psi_{\rm int}(\vec{r}) \equiv \partial_k\Psi_{\rm int} \approx \frac{1}{2} \partial_k \left(r^i K_{ij} r^j\right) = K_{kj} r^j \equiv \mathbf{K}\vec{r} \ .
\end{equation}
Thus, plugging eqn.~(\ref{eq:derivative of internal potential}) into eqn.~(\ref{eq:internal equation of motion part 1}), we get
\begin{equation}\label{eq:ODE for internal motion}
    \ddot{\vec{r}} \approx -\mathbf{K}\vec{r} \ .
\end{equation}
This is the equation for small oscillations in a quadratic potential \cite{binneyTremaine,spitzer2014dynamical}. If $\mathbf K$ is diagonalizable with eigenvalues $\omega_i^2$, then in each principal direction, the motion is a harmonic oscillator with frequency $\omega_i$. In matrix notation, the solution to eqn.~(\ref{eq:ODE for internal motion}) is
\begin{equation}
    \vec{r}(t) = \cos(\mathbf\Omega_{\rm int}\,\Delta t)\vec{r}_0+ \mathbf\Omega_{\rm int}^{-1}\sin(\mathbf\Omega_{\rm int}\Delta t)\,\vec{v}_0 \ ,
\end{equation}
where $\Delta t = t-t_0$, where $t_0$ is a reference time, $\vec{r}_0\equiv\vec{r}(t_0)$, $\vec{v}_0\equiv\dot{\vec{r}}(t_0)$, and $\mathbf K=\mathbf\Omega_{\rm int}^2$. Here $\cos(\mathbf\Omega_{\rm int}\Delta t)$ and $\sin(\mathbf\Omega_{\rm int}\Delta t)$ are understood as matrix functions.

For a spherically symmetric minihalo, the restoring tensor $\mathbf K$ reduces to a diagonal matrix $\mathbf K=\omega_{\rm int}^2\mathbf I$, with a single characteristic internal frequency $\omega_{\rm int}$ \cite{binneyTremaine}
\begin{equation}\label{eq:omega_int}
    \omega_{\rm int} \sim \sqrt{\frac{G M_{\rm vir}}{r_{\rm vir}^3}} = \sqrt{\frac{4\pi G}{3}\,\bar\rho_{\rm vir}} \ ,    
\end{equation}
where $M_{\rm vir}$ is called the virial mass and is the mass enclosed inside the virial radius $r_{\rm vir}$. $\bar{\rho}_{\rm vir} \equiv M_{\rm vir} / (4\pi r_{\rm vir}^3/3)$ is called the virial density and it is the average density inside the virial radius. \rev{Note that the first equality in eqn.~(\ref{eq:omega_int}) is an order of magnitude estimate because the exact equality only holds true for a sphere with uniform density. For real minihalos, the density profile is not uniform, and different particle orbits will have different orbital frequencies. However, the expression presented in eqn.~(\ref{eq:omega_int}) is still a good order of magnitude approximation, and does not invalidate the impulse approximation that follows.} In this spherically symmetric case, the solution $\vec{r}(t)$ simplifies to 
\begin{equation}\label{eq:SHO solution for r(t)}
    \vec{r}(t) = \vec{r}_0\cos\big[\omega_{\rm int}\Delta t\big]
+ \frac{\vec{v}_0}{\omega_{\rm int}}\sin\big[\omega_{\rm int}\Delta t\big] \ .
\end{equation}
Define the dimensionless parameter $\beta \equiv \omega_{\rm int}\tau_{\rm enc}$. Then, $\omega_{\rm int} = \beta / \tau_{\rm enc}$. Plugging this into eqn.~(\ref{eq:SHO solution for r(t)}), we get
\begin{equation}\label{eq:SHO solution for r(t) part 2}
    \vec{r}(t) = \vec{r}_0\cos\left[\beta \frac{\Delta t}{\tau_{\rm enc}}\right]
+ \frac{\vec{v}_0}{\omega_{\rm int}}\sin\left[\beta \frac{\Delta t}{\tau_{\rm enc}}\right] \ .
\end{equation}

We can now evaluate the cumulative velocity kick given to the tracer particle by the tidal field of the star as 
\begin{equation}\label{eq:cumulative velocity kick for tracer}
    \Delta\vec{v}  = \int\limits_{-\infty}^{+\infty} \mathrm{d}t\ \ddot{\vec{r}}\big|_{\rm tidal} (t) \approx -\int\limits_{-\infty}^{+\infty} \mathrm{d}t\ \mathbf T(\vec{0})\,\vec{r}\ ,    
\end{equation}
where in the second equality, we have used the distant-tide result in eqn.~(\ref{eq:tidal acceleration in the distant tide approximation}) and we note that $\mathbf T(\vec{0})$ is a function of $\vec{R}$. At least for the moment, we also note that $\vec{R}$ and $\vec{r}$ are functions of time. Let $t_0$ be the time of the instant of closest approach of the star to the minihalo. Although we have integrated from $t = -\infty$ to $t = +\infty$, only times corresponding to $|t - t_0| = |\Delta t| \lesssim \tau_{\rm enc}$ contribute appreciably towards the integral. The intuition for this is that according to eqn.~(\ref{eq:magnitude of linear order term, final part}), when $t \to \pm\infty$, $R \to \infty$ but $0 < r < r_{\rm vir}$, making the magnitude of the integrand in eqn.~(\ref{eq:cumulative velocity kick for tracer}) go to zero. This magnitude scales as $R^{-3}$ and only becomes appreciable when the star is near its pericentre. 
This implies that $|\Delta t / \tau_{\rm enc}| \lesssim 1$ is a good approximation when we substitute the expression for $\vec{r}(t)$ from eqn.~(\ref{eq:SHO solution for r(t) part 2}) into eqn.~(\ref{eq:cumulative velocity kick for tracer}). Then, we can write out the cosine and sine terms as
\begin{equation}
    \cos\left[\beta \frac{\Delta t}{\tau_{\rm enc}}\right] = 1 + \mathcal{O}(\beta^2) \ .
\end{equation}
\begin{equation}
    \sin\left[\beta \frac{\Delta t}{\tau_{\rm enc}}\right] = \mathcal{O}(\beta)\ .
\end{equation}

We now make the critical assumption that $\beta \ll 1$. This is called the impulse approximation. This implies that upto leading order in $\beta$
\begin{equation}
    \cos\left[\beta \frac{\Delta t}{\tau_{\rm enc}}\right] \approx 1  \ , \qquad \sin\left[\beta \frac{\Delta t}{\tau_{\rm enc}}\right] \approx 0 \ .
\end{equation}
Thus, in eqn.~(\ref{eq:cumulative velocity kick for tracer}), inside the integral, $\vec{r}(t) \approx \vec{r}_0$ which can be \rev{obtained} from eqn.~(\ref{eq:SHO solution for r(t) part 2}). This refers to the approximation that during the encounter (whose time scale is defined by $\tau_{\rm enc}$), the position of the tracer particle in the COM frame doesn't change appreciably. Since $\beta \ll 1$, $\tau_{\rm enc} \ll 1/\omega_{\rm int}$. Here, $1/\omega_{\rm int}$ quantifies the dynamical timescale of the minihalo. Thus, in words, we say that in the impulse approximation, the timescale of the encounter is significantly smaller than the dynamical timescale of the minihalo. Thus, eqn.~(\ref{eq:cumulative velocity kick for tracer}) reduces to 
\begin{equation}\label{eq:impulsive velocity kick to tracer}
    \Delta\vec{v}_{\rm imp}(\vec{r}_0) \approx - \int\limits_{-\infty}^{+\infty} \mathrm{d}t \mathbf T(\vec{0})\vec{r}_0 \ .
\end{equation}

\subsection{Energy input in a distant, impulsive flyby}
\label{sec:energy formula in distant-tide and impulsive approximations}

We have been working with a Euclidean metric. We now use the familiar Cartesian coordinates where the Euclidean metric has the simple form of an identity matrix. This means we are able to raise and lower indices on a general tensor easily, i.e., $R_i = R^i$.  We assume the star and the COM of the minihalo to be confined to the $\hat{y}-\hat{z}$ plane, still working in the COM frame of the minihalo. We assume the star moves along a line parallel to the positive $\hat{z}$ direction but offset from the $\hat{z}$ axis by impact parameter $b$ which is measured along the positive $\hat{y}$ direction. We assume the star has a constant speed $V\simeq V_\infty$. Let's now assume that our reference time $t_0$ equals zero. Thus $\Delta t  = t - t_0  = t$. We can describe the star's trajectory as
\begin{equation}
\vec R(t)=(0,b,V t)\ ,\qquad R(t)=\sqrt{b^2+(V t)^2} \ .
\end{equation}
The position of the tracer particle when the star is at its pericentre is $\vec{r}_0 \equiv r_0^j = \left(x_0, y_0, z_0\right)$. Then, the net impulsive velocity kick to the tracer particle is given by eqn.~(\ref{eq:impulsive velocity kick to tracer}). Dropping the subscript \textit{imp}, we have
\begin{equation}
\Delta v_i(\vec r_0)\approx - \int\limits_{-\infty}^{+\infty} \mathrm{d}t\ r_0^j\ T_{ij}(\vec{0})  \ .
\end{equation}
Using the result from eqn.~(\ref{eq:linear order term}), we have
\begin{equation}\label{eq:impulsive kick at r_0, part 1}
    \Delta v_i(\vec r_0)\approx - \int\limits_{-\infty}^{+\infty} \mathrm{d}t \frac{G M_\star}{R^3} \left[r_{0_i} - \frac{3 \left(r_0^j  R_j\right)}{R^2}R_i\right] \ .
\end{equation}
We define
\begin{equation}\label{eq:definition of u}
    u \equiv \frac{Vt}{b} \ , \qquad \implies \mathrm{d}t = \frac{b}{V} \mathrm{d}u \ .
\end{equation}
\begin{align}\label{eq:R(u)}
    \therefore \vec{R}(t) &= b\left(0,1,\frac{Vt}{b}\right) \nonumber\\
    \implies \vec{R}(u) &= b\left(0,1,u\right) \nonumber\\
    \implies R(u) &= b \sqrt{1+u^2}
\end{align}
We note that
\begin{align}\label{eq:dot product of r_0 an R}
    r_0^j  R_j &= b y_0 + Vtz_0 = b \left(y_0 + \frac{Vt}{b} z_0\right)\nonumber\\
    &= b\left(y_0 + z_0 u\right)
\end{align}
Substituting eqns.~(\ref{eq:definition of u}), (\ref{eq:R(u)}), and (\ref{eq:dot product of r_0 an R}) into eqn.~(\ref{eq:impulsive kick at r_0, part 1}),
\begin{align}\label{eq:impulsive kick at r_0, part 2}
    \Delta v_i(\vec r_0) &\approx - G  M_\star \int\limits_{-\infty}^{+\infty} \mathrm{d}u \frac{b}{V} \frac{1}{b^3\left(1+u^2\right)^{3/2}} 
    \left[ r_{0_i} - \frac{3b\left(y_0 + z_0 u\right)}{b^2\left(1+u^2\right)}R_i \right] \nonumber\\
    &= - \frac{GM_\star}{b^3 V} \int\limits_{-\infty}^{+\infty} \frac{\mathrm{d}u}{\left(1+u^2\right)^{5/2}} \left[ b\left(1+u^2\right)r_{0_i} - 3\left(y_0 + z_0 u\right)R_i \right] \ .
\end{align}
The $\hat{x}$-component of eqn.~(\ref{eq:impulsive kick at r_0, part 2}) is 
\begin{align}
    \Delta v_{\hat{x}}(\vec r_0) &\approx - \frac{GM_\star}{b^3 V} \int\limits_{-\infty}^{+\infty} \frac{\mathrm{d}u}{\left(1+u^2\right)^{5/2}} \left[b\left(1+u^2\right)x_0\right] \nonumber\\
    &= - \frac{GM_\star}{b^2 V} x_0 \int\limits_{-\infty}^{+\infty} \frac{\mathrm{d}u}{\left(1+u^2\right)^{3/2}} \nonumber\\
    &= - \frac{2GM_\star}{b^2 V} x_0 \ .
\end{align}
The $\hat{y}$-component of eqn.~(\ref{eq:impulsive kick at r_0, part 2}) is 
\begin{align}
    \Delta v_{\hat{y}}(\vec r_0) &\approx - \frac{GM_\star}{b^3 V} \int\limits_{-\infty}^{+\infty} \frac{\mathrm{d}u}{\left(1+u^2\right)^{5/2}} 
    \left[  b\left(1+u^2\right) y_0 - 3\left(y_0 + z_0 u\right)b\right] \nonumber\\
    &= - \frac{GM_\star}{b^2 V} \left[ y_0 \int\limits_{-\infty}^{+\infty} \mathrm{d}u \frac{u^2 - 2}{\left(1+u^2\right)^{5/2}} - 3z_0 \int\limits_{-\infty}^{+\infty} \mathrm{d}u \frac{u}{\left(1+u^2\right)^{5/2}}\right] \nonumber\\
    &= - \frac{GM_\star}{b^2 V} \left[ y_0 (-2) - 3z_0(0) \right] \nonumber\\
    &= \frac{2GM_\star}{b^2 V} y_0 \ .
\end{align}
The $\hat{z}$-component of eqn.~(\ref{eq:impulsive kick at r_0, part 2}) is 
\begin{align}
    \Delta v_{\hat{z}}(\vec r_0) &\approx - \frac{GM_\star}{b^3 V} \int\limits_{-\infty}^{+\infty} \frac{\mathrm{d}u}{\left(1+u^2\right)^{5/2}} 
    \left[ b\left(1+u^2\right) z_0 - 3\left(y_0 + z_0 u\right)bu  \right] \nonumber\\
    &= - \frac{GM_\star}{b^2 V} \left[ -3y_0 \int\limits_{-\infty}^{+\infty} \mathrm{d}u \frac{u}{\left(1+u^2\right)^{5/2}}  + z_0  \int\limits_{-\infty}^{+\infty} \mathrm{d}u \frac{1-2u^2} {\left(1+u^2\right)^{5/2}}\right] \nonumber\\
    &= - \frac{GM_\star}{b^2 V} \left[ -3y_0(0) + z_0 (0) \right] \nonumber\\
    &= 0 \ .
\end{align}
Thus, the cumulative velocity kick vector to the tracer particle is
\begin{equation}\label{eq:velocity kick for one tracer particle, final part}
    \Delta \vec{v}(\vec r_0) \approx 2\frac{GM_\star}{b^2 V}  \left(-x_0, y_0,0\right) \ .
\end{equation}
The minihalo comprises of a large number of tracer particles. Let $\chi$ label each tracer particle. We note that during the encounter, the tracer particles remain fixed in position. So, the potential energy associated with the minihalo doesn't change. Thus, the change in total energy $\Delta E$ of the minihalo during the encounter equals the change in the kinetic energies $\Delta\tilde{K}$ of all tracer particles. Assuming that $\vec{v}_\chi$ is the velocity of the $\chi^{\rm th}$ tracer particle just before adding the velocity kick, and $m_\chi$ is its mass,
\begin{align}\label{eq:input energy, part 1}
    \Delta E & = \Delta\tilde{K} = \sum\limits_\chi \frac{1}{2} m_\chi \left\vert \vec{v}_\chi + \Delta \vec{v}_\chi \right\vert^2
    - \sum\limits_\chi  \frac{1}{2} m_\chi \left\vert \vec{v}_\chi \right\vert^2 \nonumber\\
    &= \frac{1}{2}\sum\limits_\chi m_\chi \left(\left(\vec{v}_\chi + \Delta \vec{v}_\chi\right) \cdot \left(\vec{v}_\chi + \Delta \vec{v}_\chi\right) - \left\vert \vec{v}_\chi \right\vert^2\right) \nonumber\\
    &= \frac{1}{2} \sum\limits_\chi m_\chi  \left(\left\vert \Delta\vec{v}_\chi \right\vert^2 + 2 \vec{v}_\chi \cdot \Delta \vec{v}_\chi  \right) \ .
\end{align}
Due to the impulse approximation, the tracer particles are `static'. Our minihalo is spherically symmetric, and hence is axisymmetric about any axis. In a static axisymmetric system, by symmetry, we have \cite{binneyTremaine}
\begin{equation}
    \sum\limits_\chi m_\chi \left( \vec{v}_\chi \cdot \Delta \vec{v}_\chi \right) = 0 \ .
\end{equation}
Thus, eqn.~(\ref{eq:input energy, part 1}) becomes,
\begin{equation}\label{eq:input energy, part 2}
    \Delta E = \frac{1}{2} \sum\limits_\chi m_\chi \left\vert \Delta\vec{v}_\chi \left(\vec{r}_{0_\chi}\right) \right\vert^2 \ .
\end{equation}
Substituting eqn.~(\ref{eq:velocity kick for one tracer particle, final part}) into eq.~(\ref{eq:input energy, part 2}), we have
\begin{align}\label{eq:input energy, part 3}
    \Delta E &= \frac{1}{2} \sum\limits_\chi m_\chi \left(2\frac{GM_\star}{b^2 V}\right)^2 \left(x_{0_\chi}^2 + y_{0_\chi}^2\right) \nonumber\\
    &= \frac{2G^2M_\star^2}{b^4 V^2} \left[\sum\limits_\chi m_\chi x_{0_\chi}^2 + \sum\limits_\chi m_\chi y_{0_\chi}^2\right] \ .
\end{align}
In continuous form, we can write
\begin{equation}\label{eq:Simga m x^2}
    \sum\limits_\chi m_\chi x_{0_\chi}^2 = \int \mathrm{d}m\ x^2 = \int \rho(\vec{x}) x^2 \mathrm{d}^3x \ .
\end{equation}
For a spherically symmetric minihalo, the density profile $\rho(\vec{x}) = \rho(r)$. Also, in spherical coordinates $(r,\theta,\phi)$, 
\begin{equation}
    x = r \sin \theta \cos \phi , 
\end{equation}
and
\begin{equation}
    \mathrm{d}^3x = r^2 \sin \theta \mathrm{d}r \mathrm{d}\theta\mathrm{d}\phi \ .
\end{equation}
Thus, eqn.~(\ref{eq:Simga m x^2}) becomes
\begin{align}
    \sum\limits_\chi m_\chi x_{0_\chi}^2 &= \iiint \rho(r) \left(r^2 \sin^2 \theta \cos^2 \phi\right) \left( r^2 \sin \theta \mathrm{d}r \mathrm{d}\theta\mathrm{d}\phi \right) \nonumber\\
    &= \int\limits_0^{r_{\rm vir}} \mathrm{d}r \ \rho(r)r^4\ \int\limits_0^\pi \mathrm{d}\theta\ \sin^3\theta \ \int\limits_0^{2\pi} \mathrm{d}\phi \ \cos^2\phi \nonumber\\
    &= \int\limits_0^{r_{\rm vir}} \mathrm{d}r \ \rho(r)r^4 \left(\frac{4}{3}\right) (\pi) \nonumber\\
    &= \frac{1}{3} \int\limits_0^{r_{\rm vir}} \left(4\pi r^2\mathrm{d}r\right) \rho(r)r^2 \ .
\end{align}
But with spherical symmetry, $\int 4\pi r^2\mathrm{d}r = \int \mathrm{d}^3x$. Thus,
\begin{equation}
    \sum\limits_\chi m_\chi x_{0_\chi}^2  = \frac{1}{3} \int \mathrm{d}^3x \ \rho(r)r^2 \ .
\end{equation}
The virial mass $M_{\rm vir}$ can be written as
\begin{equation}
    M_{\rm vir} = \int \mathrm{d}^3x\ \rho(r) \ .
\end{equation}
\begin{align}\label{eq:eq:Simga m x^2, final}
    \implies\sum\limits_\chi m_\chi x_{0_\chi}^2  &= \frac{M_{\rm vir}}{3} \left( \frac{\int \mathrm{d}^3x \ \rho(r)r^2}{\int \mathrm{d}^3x\ \rho(r)}\right) \nonumber\\
    &= \frac{M_{\rm vir}}{3} \left\langle r^2 \right\rangle \ ,
\end{align}
where 
\begin{equation}
    \left\langle r^2 \right\rangle  \equiv \frac{\int \mathrm{d}^3x \ \rho(r)r^2}{\int \mathrm{d}^3x\ \rho(r)} \ ,
\end{equation}
is the mass-weighted average of $r^2$. It can be similarly shown that
\begin{equation}\label{eq:eq:Simga m y^2, final}
    \sum\limits_\chi m_\chi y_{0_\chi}^2 = \frac{M_{\rm vir}}{3} \left\langle r^2 \right\rangle \ .
\end{equation}
Substituting eqns.~(\ref{eq:eq:Simga m x^2, final}) and (\ref{eq:eq:Simga m y^2, final}) into eqn.~(\ref{eq:input energy, part 3}), the energy input during a stellar encounter in the distant-tide and impulse approximations is 
\begin{equation}\label{eq:input energy, final part}
    \Delta E = \frac{4G^2 M_\star^2 M_{\rm vir}}{3b^4 V^2} \left\langle r^2 \right\rangle \ .
\end{equation}
Although this expression's derivation can be found in refs.~\cite{binneyTremaine,spitzer1958disruption}, we have supplied some extra details not found in those references.


\chapter{The Disruption of Dark Matter Minihalos by Successive Stellar Encounters}
\label{ch:journal_paper_1}

\graphicspath{{Journal_Paper_1/Figs/}}

\section{Introduction}
\label{sec:Introduction for journal paper 1}

As discussed in chapter~\ref{ch:introduction}, when a minihalo encounters a star, energy is injected into the minihalo through tidal interactions~\citep[{\em e.g.},][]{binneyTremaine}.
	Ref.~\cite{K2020} (hereafter referred to as K2021) derived a general formula for the mass loss of a minihalo during a stellar encounter using the phase space distribution function of the dark matter particles. A similar formula, using a wave description, for the mass loss, was derived by ref.~\cite{Dandoy:2022prp}. 
	Ref.~\cite{S2023} (hereafter referred to as S2023) performed N-body simulations of dark matter minihalos undergoing a stellar interaction. They varied the normalized injected energy of the minihalo-star interaction, the concentration parameter, and the virial mass of the minihalo and computed the survival fraction of the minihalo. They also used an empirical response function to fit the numerically simulated data. They found that the formula developed by K2021 provided a reasonable fit for a halo concentration of $c=100$. In this chapter, we show that K2021's formula does not work so well for other concentrations. We derive a formula that performs better on all the concentrations that S2023 showed detailed results for in their paper. Our new formula uses a sequential stripping approach and accounts for the minihalo profile change after being disrupted by a passing star.
	
	S2023 also investigated the mass loss for multiple stellar encounters. This is important as the minihalos in our galaxy will transverse the Galactic disk many times during the history of the Universe. 
	The energy injection parameter is the ratio of the energy injected into the minihalo divided by the minihalo binding energy. S2023
	assumed that the mass loss from multiple stellar encounters depends on
	the sum of the energy injection parameters from each encounter.
	In this thesis, we check this result and show that it does not account for the change in the halo profile after each encounter. Using our formula, we find that the mass loss is significantly more severe than the method employed by S2023.

	S2023 estimated that 
	about 60\% of mass in minihalos with an initial mass greater than $10^{-12}M_\odot$ will  
	be retained by minihalos observed at the redshift zero at the solar system location.
	Our results indicate that this is an overestimate of the amount of retained mass.

	This chapter follows our published article \cite{dsouza2024}. The chapter is structured as follows: Section \ref{sec:Mass_loss_in_a_minhalo_during_a_stellar_encounter} details the K2021 method and its application across various concentration parameter values. Our proposed sequential stripping model is elaborated in Section \ref{sec:The_sequential_stripping_model}. The dynamics of multiple stellar encounters are discussed in Section \ref{sec:Multiple_stellar_encounters_of_an_NFW_minihalo}, leading to our concluding thoughts in Section \ref{sec:Discussion_and_Conclusions}. For more in-depth technical explanations, readers are directed to the appendices.

\section{{Mass loss in a minihalo during a stellar encounter}}
\label{sec:Mass_loss_in_a_minhalo_during_a_stellar_encounter}

N-body simulations~\citep[{\em e.g.},][]{X2021} indicate that the undisrupted minihalos can be fit by the well-known
spherically symmetric NFW density profile (Ref.~\cite{NFW1997universal}) and presented in eqn.~(\ref{eq:NFW profile}).

We will use the following normalized distance from the center of the minihalo
\begin{equation}\label{eq:x definition}
    x \equiv \frac{r}{r_{\rm vir}}
\end{equation}
This is sometimes also called the normalized radius. Using this variable and the concentration $c$ of the minihalo, we can rewrite the NFW profile as 
\begin{equation}\label{eqn:NFW as a function of x}
    \rho(x) = \frac{\rho_{\rm s}}{cx\left(1+cx\right)^2}\,.
\end{equation}

Consider a minihalo with an NFW density profile extending to infinity. We are interested in the mass loss during a stellar encounter within the virial radius of the minihalo.

As in S2023, the terms $\langle r^2 \rangle$ and $\langle r^{-2} \rangle$ both averaged within the virial radius are parametrized by $\alpha$ and $\beta$ as follows:
\begin{align}\label{eq:langle r^2 rangle}
    \langle r^2 \rangle & \equiv \alpha^2 r_{\rm vir}^2\\
    \langle r^{-2} \rangle & \equiv \beta^2 r_{\rm vir}^{-2}
\end{align}
\rev{It can be shown that for an NFW profile (S2023),
\begin{align}
	\alpha(c)&=\sqrt{\frac{c\left(-3-3 c / 2+c^2 / 2\right)+3(1+c) \ln (1+c)}{c^2(-c+(1+c) \ln (1+c))}}\\
    \beta(c) &\approx \sqrt{\frac{c^2 \ln \left(r_{\mathrm{s}} / r_{\mathrm{c}}\right)+c^2 / 2-1 / 2}{\ln (1+c)-c /(1+c)}}
\end{align}
where $r_{\mathrm{c}}$ is the smallest radius that the profile extends to, which is assumed to be $0.01 r_{\mathrm{s}}$.} The binding energy $E_{\rm b}$ of the minihalo within the virial radius is parametrized by $\gamma$ as follows (S2023):
\begin{align}\label{eq:binding energy}
    E_{\rm b} = \gamma \frac{GM_{\rm vir}^2}{r_{\rm vir}}     
\end{align}
where $M_{\rm vir}$ is the mass of the minihalo contained within the virial radius. \rev{For an NFW profile, it can be shown that (S2023)
\begin{equation}
    \gamma(c)=\frac{c}{2} \frac{1-1 /(1+c)^2-2 \ln (1+c) /(1+c)}{[c /(1+c)-\ln (1+c)]^2}
\end{equation}
}
The energy injected per unit mass $\vert\Delta\varepsilon\vert$ into the minihalo is given by Refs.~\cite{spitzer1958disruption, greenGoodwin2007mini}:
\begin{equation}\label{eq:energy injected per unit mass (part 1)}
    \vert\Delta\varepsilon(r)\vert = \frac{\Delta E}{M_{\rm vir}} \frac{r^2}{\langle r^2 \rangle}
\end{equation}
where the $\Delta E$ is the net energy injected into the minihalo within the virial radius for a single encounter with a star of mass $m_{\ast}$, relative \rev{minihalo-star speed} $v_{\ast}$, and impact parameter $b$ is given by \cite{greenGoodwin2007mini}
\begin{equation}
    \label{eq:deltaE_general}
    \Delta E = \begin{cases} 
        \dfrac{4 \alpha^{2}(c)}{3} \dfrac{G^{2} m_{\ast}^{2} M_{\rm vir} r^{2}_{\rm vir}}{v_{\ast}^2} \dfrac{1}{b^4}\ , \qquad \text{$(b>b_{\rm s})$} \\
        \dfrac{4 \alpha^{2}(c)}{3} \dfrac{G^{2} m_{\ast}^{2} M_{\rm vir} r^{2}_{\rm vir}}{v_{\ast}^2} \dfrac{1}{b_{\rm s}^4}\ , \qquad \text{$(b\leq b_{\rm s})$}\,.
    \end{cases}
\end{equation}
\rev{Note} that this expression looks at two regimes of the impact parameter, while eqn.~(\ref{eq:input energy, final part})  assumed only large impact parameters. If we substitute $b=0$ in eqn.~(\ref{eq:input energy, final part}), the predicted energy injected diverges. This is because the distant-tide approximation breaks down for small impact \rev{parameters}. In reality, as $b$ becomes comparable to the size of the minihalo, the energy injection saturates, as was shown by ref.~\cite{greenGoodwin2007mini}. Here, $b_{\rm s} = f_{\rm b}\, (2\alpha/3\beta)^{1/2} r_{\rm vir}$ is the transition radius and $f_{\rm b}$ is an order-unity correction factor introduced by S2023.  From their simulations, S2023 finds that $f_{\rm b} = 6$.

From Eqs.~(\ref{eq:langle r^2 rangle}), (\ref{eq:binding energy}) and (\ref{eq:energy injected per unit mass (part 1)}), it follows that:
\begin{equation}\label{eq:energy injected per unit mass (part 2)}
    \vert\Delta\varepsilon(r)\vert = \Psi_0 \frac{\gamma}{\alpha^2} E_{\rm frac} \frac{r^2}{r_{\rm vir}^2}
\end{equation}
\begin{equation}\label{eq:normalized energy injected per unit mass (part 2) as a function of x}
    \vert\Delta\epsilon(x)\vert \equiv \frac{\vert\Delta\varepsilon(x)\vert}{\Psi_0}  = \frac{\gamma}{\alpha^2} E_{\rm frac} x^2
\end{equation}
where $\Psi_0 \equiv \frac{GM_{\rm vir}}{r_{\rm vir}}$, $E_{\rm frac} \equiv \frac{\Delta E}{E_{\rm b}}$, and $\vert\Delta\epsilon\vert$ is the normalized injected energy per unit mass into the minihalo. \rev{Plugging the expressions for $\Delta E$ from eqn.~(\ref{eq:deltaE_general}) and $E_{\rm b}$ from eqn.~(\ref{eq:binding energy}), it can be shown that (S2023)
\begin{equation}
    \label{eq:Efrac expression}
    E_{\rm frac} = \begin{cases} 
        \dfrac{\alpha^{2}(c)}{\pi \gamma(c)} \dfrac{G m_{\ast}^{2}}{v_{\ast}^2 \bar{\rho}_{\rm vir}} \dfrac{1}{b^4} \ , \qquad \text{$(b>b_{\rm s})$} \\
        \dfrac{\alpha^{2}(c)}{\pi \gamma(c)} \dfrac{G m_{\ast}^{2}}{v_{\ast}^2 \bar{\rho}_{\rm vir}} \dfrac{1}{b_{\rm s}^4} \ , \qquad \text{$(b\leq b_{\rm s})$}\,.
    \end{cases}
\end{equation}
}

The relative potential $\Psi$ is defined as $\Psi(r) \equiv -\Phi(r)$, where $\Phi(r)$ is the Newtonian gravitational potential. For an untruncated NFW minihalo of concentration parameter $c$, it can be shown that (see Appendix \ref{app:psi for untruncated NFW})
\begin{equation}\label{eq:Psi expression for NFW minihalo}
    \Psi(r) = \Psi_0\frac{1}{f_{\rm NFW}(c)} \frac{\ln\left(1+c\frac{r}{r_{\rm vir}}\right)}{\frac{r}{r_{\rm vir}}}
\end{equation}
\begin{equation}\label{eq:Psi expression for NFW minihalo as a function of x}
    \psi(x) \equiv \frac{\Psi(x)}{\Psi_0} = \frac{1}{f_{\rm NFW}(c)} \frac{\ln\left(1+cx\right)}{x}
\end{equation}
where $\psi$ is the normalized relative potential and
\begin{equation}
    \label{eq:fNFW}
    f_{\rm NFW}(x) \equiv \ln(1+x) - \frac{x}{1+x}.  
\end{equation}

\rev{
To compute the expression for the mass loss in a minihalo due to a stellar encounter, one needs to integrate the phase space distribution function $f(\varepsilon)$ of dark matter particles in the minihalo over the 6-dimensional phase space volume only for those particles that become unbound due to energy injection during the stellar encounter (K2021). To find such a condition for an unbound particle, we define the dark matter particle's specific relative (total) energy as
\begin{equation}\label{eq:varepsilon definition}
    \varepsilon(r,v) = \Psi(r) - \frac{v^2}{2} \ ,
\end{equation}
where $v$ is the speed of the particle. Note that for a bound particle, the potential energy term in Eq.~(\ref{eq:varepsilon definition}) is greater than the kinetic energy term, making $\varepsilon(r,v) \geq 0$. Conversely, if for a particle $\varepsilon(r,v) < 0$, we say that the particle is unbound. Now, $\Delta\varepsilon(r)$ is taken to be negative (K2021). Thus, a particle that is initially bound, becomes unbound during the stellar encounter if the following condition holds true:
\begin{align}\label{eq:condition for integration of phase space}
    &\varepsilon_{\rm after}(r,v) = \varepsilon_{\rm before}(r,v) + \Delta\varepsilon(r) < 0 \nonumber \\
    \implies &\varepsilon_{\rm before}(r,v) < -\Delta\varepsilon(r) =  \vert\Delta\varepsilon(r)\vert \ ,
\end{align}
where the subscripts \textit{before} and \textit{after} denote before and after the stellar encounter, for the same particle. Also, the equality in the second line applies because $\Delta\varepsilon(r)$ is negative. Thus, to compute the mass loss  $\Delta M$ within the virial radius of the minihalo, the phase space is integrated under the condition specified by eq.~(\ref{eq:condition for integration of phase space}). This then gives (K2021)
}
\begin{equation}\label{eq:mass loss NFW}
    \Delta M = 16\pi^2\int\limits_{r=0}^{r_{\rm vir}} \mathrm{d}r\ r^2 \int\limits_{\varepsilon=0}^{\min\left[\vert\Delta\varepsilon(r)\vert, \Psi(r)\right]} \mathrm{d}\varepsilon\sqrt{2(\Psi(r)-\varepsilon)}f(\varepsilon)
\end{equation}
\rev{Note that in K2021, the upper limit of the $\varepsilon$ integral is $\min\left[\Delta\varepsilon(r), \Psi(r)\right]$ instead of $\min\left[\vert\Delta\varepsilon(r)\vert, \Psi(r)\right]$. We believe this is not accurate since $\Delta\varepsilon(r) <0,\ \Psi(r)>0\ \forall\ r>0$. So, $\min\left[\Delta\varepsilon(r), \Psi(r)\right] = \Delta\varepsilon(r) \ \forall\ r>0$, making the $\min$ operator ineffective.}

Converting Eq.~(\ref{eq:mass loss NFW}) to a dimensionless form, we compute the survival fraction (SF) of the minihalo as (see Appendix \ref{app:survival fraction expression for NFW minihalo})
\begin{align}\label{eq:surival fraction expression dimensionless double integral}
    \text{SF} &\equiv 1 - \frac{\Delta M}{M_{\rm vir}} \nonumber\\
    &= 1 - \frac{4\pi c^3}{f_{\rm NFW}(c)}\int\limits_{x=0}^1\mathrm{d}x\ x^2\int\limits_{\epsilon=0}^{\min[\vert\Delta\epsilon(x)\vert, \psi(x)]} \mathrm{d}\epsilon\ \hat{f}(\epsilon)\sqrt{2(\psi(x)-\epsilon)}
\end{align}
where $\epsilon \equiv \frac{\varepsilon}{\Psi_0}$ is the normalized specific relative (total) energy and $\hat{f}(\epsilon) = \frac{\Psi_0^{3/2}}{\rho_{\rm s}} f(\varepsilon)$ (K2021) is the normalized phase space distribution function of the dark matter particles in the minihalo and it can be evaluated as follows (K2021): 
\begin{equation}\label{eq:f_hat}
    \hat{f}(\epsilon) = \frac{1}{\sqrt{8}\pi^2} \int\limits_{\psi=0}^\epsilon \frac{1}{\sqrt{\epsilon - \psi}} \frac{\mathrm{d}^2\varrho}{\mathrm{d}\psi^2} \mathrm{d}\mathrm{\psi}
\end{equation}
where $\varrho \equiv \frac{\rho}{\rho_s}$ is the normalized density of the minihalo. Thus, Eq.~(\ref{eq:surival fraction expression dimensionless double integral}) can be rewritten as a triple integral:
\begin{multline}\label{eq:surival fraction expression dimensionless triple integral}
    \text{SF} = 1 - \frac{4\pi c^3}{f_{\rm NFW}(c)}\int\limits_{x=0}^1 \int\limits_{\epsilon=0}^{\min[\vert\Delta\epsilon(x)\vert, \psi(x)]} \int\limits_{\psi^\prime=0}^\epsilon x^2 \frac{1}{\sqrt{8}\pi^2} \frac{1}{\sqrt{\epsilon - \psi^\prime}}\\ \frac{\mathrm{d}^2\varrho}{\mathrm{d}\psi^{\prime^2}}  \sqrt{2(\psi(x)-\epsilon)}\ \mathrm{d}\psi^\prime\ \mathrm{d}\epsilon\ \mathrm{d}x
\end{multline}

Using Eq.~(\ref{eq:surival fraction expression dimensionless triple integral}), one can evaluate mass loss by computing the survival fraction of the minihalo for a particular value of $E_{\rm frac}$ and concentration parameter. We used the \textit{Derivative}() function from Python's SymPy library to evaluate $\frac{\mathrm{d}^2\varrho}{\mathrm{d}\psi^{\prime^2}}$ as a function of $x$ and the \textit{solve}() function from the SymPy library to invert the expression for $\psi^\prime$ so that we could express $x$ as a function of $\psi^\prime$ and eventually express $\frac{\mathrm{d}^2\varrho}{\mathrm{d}\psi^{\prime^2}}$ as a function of $\psi^\prime$. We further used the \textit{nquad}() function from the SciPy library to numerically evaluate the triple integral in Eq.~(\ref{eq:surival fraction expression dimensionless triple integral}). The survival fraction of the minihalo is then plotted against the normalized total injected energy, $E_{\rm frac}$, for a fixed concentration parameter, $c$. Fig.~\ref{fig:SF vs Efrac for four values of concentration} shows this plot for $c = 10,30,100,500$.
\begin{figure}
    \includegraphics[width=\columnwidth]{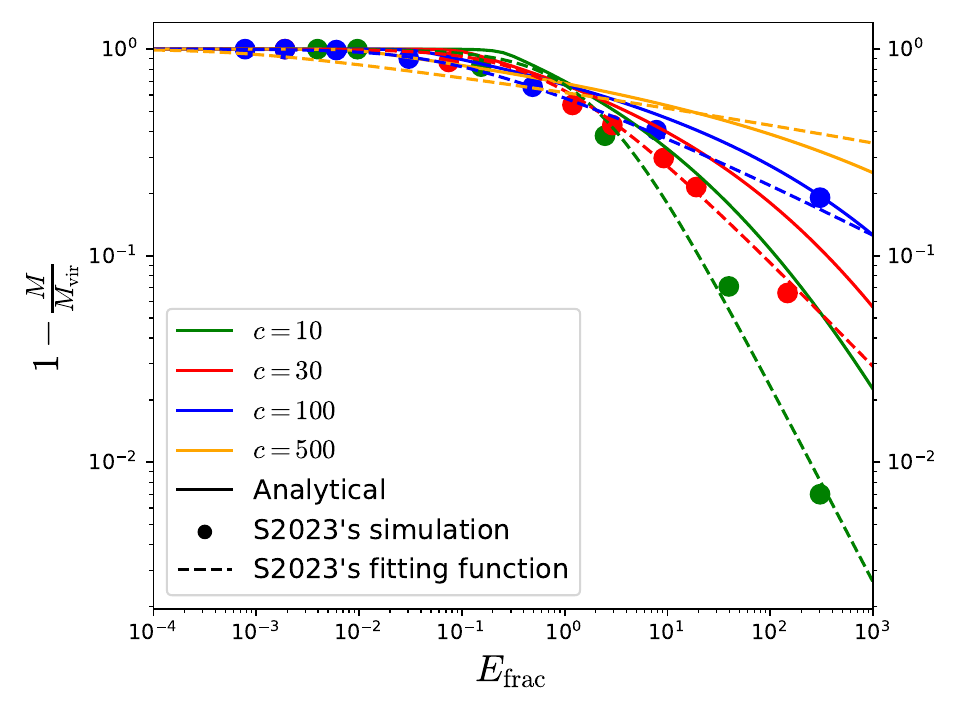}
    \caption[Survival fraction against $E_{\rm frac}$ for K2021 model]{Survival fraction (SF) as a function of the normalized total injected energy $E_{\rm frac}$ into the minihalo as a result of stellar interaction, for concentration parameters $c=10,30,100,500$. The solid curves are the output of our implementation of K2021's analytical approach.   
        The dots are numerical simulation data from S2023. The dashed curves are the empirical fitting functions used by S2023.}
    \label{fig:SF vs Efrac for four values of concentration}
\end{figure}
From Fig.~\ref{fig:SF vs Efrac for four values of concentration}, it is clear that the analytical method described so far does not approximate the simulated data very well except for the $c=100$ case. To improve on the analytical method, we introduce 
a sequential stripping model for the mass loss in the minihalo.

\section{The sequential stripping model}
\label{sec:The_sequential_stripping_model}
One of the flaws with the expression for $\psi(x)$ as given by Eq.~(\ref{eq:Psi expression for NFW minihalo as a function of x}) is that it assumes that when one takes a dark matter particle from position $x$ to $\infty$, all matter in regions greater than normalized radius $x$ remains intact. 

We introduce a model of dark matter particle ``unbinding," where we divide the minihalo into shells of infinitesimal thickness. During a stellar interaction, dark matter particles that will eventually be unbound will go to infinity. A dark matter particle in a particular shell that is going to infinity is not expected to feel a gravitational pull from a dark matter particle in an outer shell because we assume shell expansion is taking place in a spherically symmetric manner. Thus, we model this as first starting with the minihalo's outermost shell (not necessarily within the virial radius) and taking it to infinity. Then we take the next innermost shell to infinity, and so on. When we take a dark matter particle from a normalized radius $x$ to infinity, we assume that no matter exists in the region of radius $>x$ because the matter in this region has already been taken to infinity. We call this approach the ``sequential stripping model".

In Eq.~(\ref{eq:surival fraction expression dimensionless triple integral}), the upper limit of the $\epsilon$ \rev{integral} is a \textit{minimum} operation on two functions, $\psi(x)$ and $\vert\Delta\epsilon(x)\vert$. Setting $c=10$ for now, Fig.~\ref{fig:psi crossing over deltaEpsilon} shows $\psi(x)$ (according to Eq.~(\ref{eq:Psi expression for NFW minihalo as a function of x})) which is a decreasing function of $x$, and $\vert\Delta\epsilon(x)\vert$ (according to Eq.~(\ref{eq:normalized energy injected per unit mass (part 2) as a function of x})) which is a quadratically increasing function on $x$. The two curves intersect within or beyond the virial radius, and the normalized radius of the intersection is called the normalized crossover radius $x^*$.
\begin{figure}
    \includegraphics[width=\columnwidth]{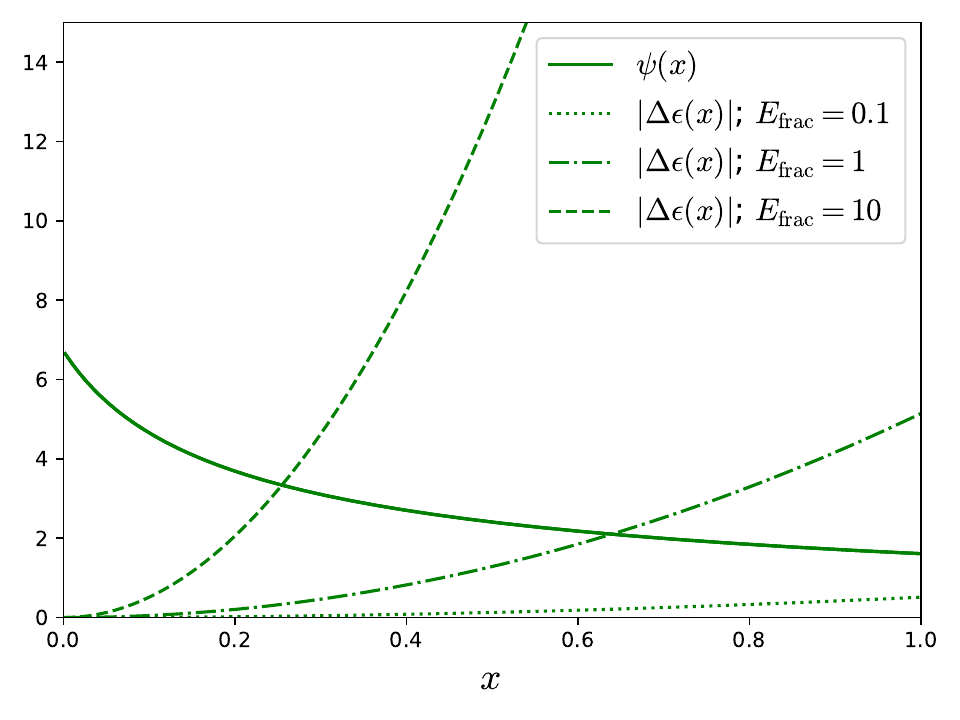}
    \caption[Normalized crossover radius from relative potential and normalized energy injected per unit mass]{The normalized injected energy per unit mass $\vert\Delta\epsilon\vert$ in a minihalo due to a stellar interaction is plotted as a function of the normalized radius $x$. In doing so, the total injected energy $E_{\rm frac}$ is set to 0.1, 1, 10. 
        The normalized relative potential $\psi$ is also plotted as a function of $x$. The $\psi(x)$ curve does not vary with $E_{\rm frac}$. In all cases for this figure, the concentration parameter $c=10$.  The value of $x$ at which $\vert\Delta\epsilon(x)\vert$ and $\psi(x)$ curves intersect is called the normalized crossover radius $x^*$.}
    \label{fig:psi crossing over deltaEpsilon}
\end{figure}
To make Eq.~(\ref{eq:surival fraction expression dimensionless triple integral}) more tractable, let's utilize the normalized crossover radius, $x^*$, which is mathematically defined as that value of $x$ such that:
\begin{equation}\label{eq:cross over radius condition}
    \vert\Delta\epsilon(x^*)\vert = \psi(x^*)
\end{equation}

As shown in Appendix~\ref{app:Evaluating the survival fraction using the sequential stripping model} the survival fraction with the sequential stripping model can be computed using
\begin{equation}\label{eq:SF expression in terms of mass fraction x < x^* and I_A copy}
    \text{SF} = \text{mass fraction}_{x < \min[x^*,1]} - \text{prefactor}\times I_{\rm A}
\end{equation}
where
\begin{equation}\label{eq:expression for mass fraction x < x^*, in terms of x copy}
    \text{mass fraction}_{x < \min[x^*,1]} = \frac{c^2}{f_{\rm NFW}(c)} \int\limits_{x=0}^{\min[x^*, 1]} \frac{x}{\left(1 + cx\right)^2} \mathrm{d}x,
\end{equation}
\begin{equation}\label{eq:prefactor copy}
    \text{prefactor} \equiv \frac{4\pi c^3}{f_{\rm NFW}(c)}
\end{equation}
and 
\begin{multline}\label{eq:I1 updated to sequential stripping model copy}
    I_{\rm A} \equiv \int\limits_{x=0}^{\min[x^*, 1]} \int\limits_{\epsilon=0}^{\vert\Delta\epsilon(x)\vert} \  \int\limits_{\psi_{\rm B}^\prime=0}^\epsilon x^2\frac{1}{\sqrt{8}\pi^2}\ \frac{1}{\sqrt{\epsilon - \psi_{\rm B}^\prime}} \frac{\mathrm{d}^2\varrho}{\mathrm{d}\psi_{\rm B}^{\prime^2}}\left(x^\prime\left(\psi_{\rm B}^\prime\right)\right)\\ \sqrt{2\left(\psi_{\rm A}(x)-\epsilon\right)} \ \mathrm{d}\psi_{\rm B}^\prime\ \mathrm{d}\epsilon\ \mathrm{d}x.
\end{multline}
The relative potentials in the above equation are given by 
\begin{align}\label{eq:psi1(x) copy}
    \psi_{\rm A}(x) = \frac{1}{f_{\rm NFW}(c)} \left[ \frac{\ln(1+cx)}{x} - \frac{c}{1+cx^*} \right]\rev{,} && x<x^*
\end{align}
and
\begin{align}\label{eq:psi2(x) copy}
    \psi_{\rm B}(x) = \frac{1}{f_{\rm NFW}(c)} \left[ \frac{\ln(1+cx)}{x} - \frac{c}{1+cx} \right]\rev{,} && x>x^*.
\end{align}

Fig.~\ref{fig:SF vs Efrac - sequential stripping model} shows how the survival fraction (solid curves) varies with the normalized injected energy $E_{\rm frac}$. The sequential stripping model gives a better fit to the simulation data points when compared to the approach used in Fig.~\ref{fig:SF vs Efrac for four values of concentration}.

\begin{figure*}
    \includegraphics[width=\textwidth]{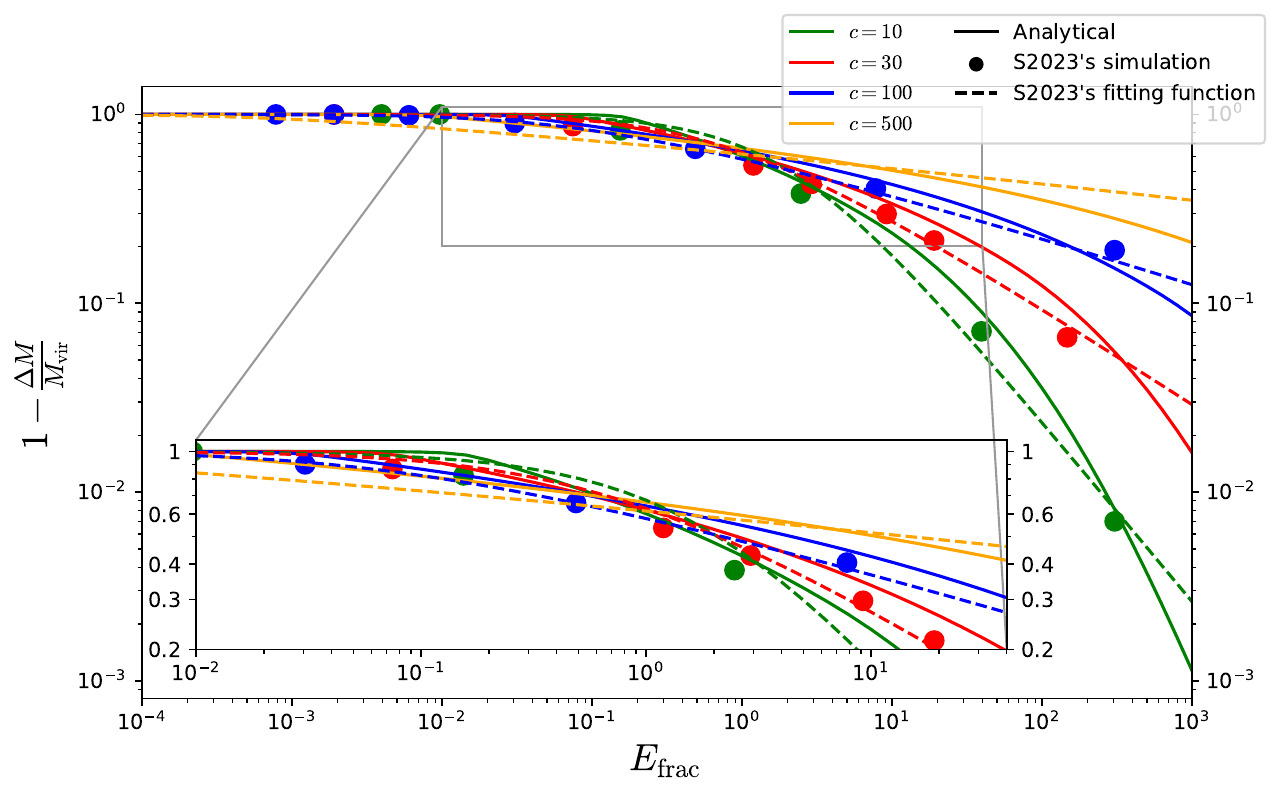}
    \caption[Survival fraction against $E_{\rm frac}$ for sequential stripping model, without relaxation]{ The same as Fig.~\ref{fig:SF vs Efrac for four values of concentration} except that
        the solid curves are the output of our analytical approach using the sequential stripping model of mass loss in the minihalo.
    }
    \label{fig:SF vs Efrac - sequential stripping model}
\end{figure*}

\subsection{The need to include relaxation}\label{sec:need for relaxation}
The S2023 simulation data shown in our  Fig.~\ref{fig:SF vs Efrac - sequential stripping model}, is computed after the remnant minihalo has undergone full relaxation following a stellar encounter. However, we have not accounted for this relaxation process for our analytical curves. But by juxtaposing our analytical curves with S2023's numerical data points, it is equivalent to assuming that, in our case, the remnant minihalo relaxes to the same initial NFW profile. This is a good approximation for small $E_{\rm frac}$ values. For example, in the limit $E_{\text{frac}} \rightarrow 0$, there is no actual stellar encounter, and the resulting (relaxed) minihalo is the same NFW minihalo we started with. Thus, for small finite $E_{\rm frac}$ values, the approximation that the remnant minihalo relaxes to the same NFW profile is a good one. Thus, we see that our analytical curves have a good match to the numerical data at low values of $E_{\rm frac}$. However, S2023 finds that, in general, the remnant minihalo relaxes to a broken power profile,
\begin{equation}
    \rho(r)=\frac{\rho_s}{\frac{r}{r_s}\left(1+\frac{r}{r_s}\right)^k},
\end{equation}
which has an outer logarithmic slope of $-(k+1)$ for large values of $r$ and a logarithmic slope of $-1$ for small values of $r$ \footnote{This is in contrast to Ref.~\cite{Delos:2019tsl} who finds that the profiles relax to a non-broken power law formula. However, Ref.~\cite{Delos:2019tsl} only fits numerical simulations up to $r\approx 2 r_s$ while S2023 fits them to $r\approx 20 r_s$.}. For our purposes, we assume that $k=3$, and this profile is known as the Hernquist density profile \cite{Hernquist:1990be}. Since we have argued that for small $E_{\rm frac}$, the remnant minihalo relaxes to an NFW profile, it is only for larger $E_{\rm frac}$ values that the remnant minihalo relaxes to a Hernquist profile. Since our analytical curves in Fig.~\ref{fig:SF vs Efrac - sequential stripping model} \rev{do not} account for this, there arises a discrepancy which becomes starker at higher values of $E_{\rm frac}$. This discrepancy leads to errors, which result in our analytical curves overshooting the numerical data points at larger $E_{\rm frac}$ values. We need to include the relaxation process in our calculations to account for this discrepancy.

\subsection{The Hernquist model}
In this section, we assume that after disruption, the minihalo relaxes to the Hernquist density profile, which is described by
\begin{equation}\label{eq:Hernquist profile in terms of r}
    \rho(r) = \frac{\rho_{\rm s}}{\frac{r}{r_{\rm s}} \left( 1 + \frac{r}{r_{\rm s}} \right)^3}
\end{equation}
\begin{equation}\label{eq:Hernquist profile in terms of x}
    \rho(x) = \frac{\rho_{\rm s}}{cx \left( 1 + cx \right)^3}
\end{equation}

Firstly, using the same sequential stripping model we used for the NFW minihalo, it can be shown that the normalized relative potential of an untruncated Hernquist minihalo is (see Appendix~\ref{app:psi for Hernquist minihalo})
\begin{align}\label{eq:psI_A(x) for Herquist Profile}
    \psi_{\rm A}(x) = (1+c)^2 \biggr[ \frac{x^* - x}{(1+cx^*)(1+cx)} + \frac{x^*}{(1+cx^*)^2} \biggr]\rev{,}&&   x<x^*
\end{align}
\begin{align}\label{eq:psI_B(x) for Herquist Profile}
    \psi_{\rm B}(x) = (1+c)^2 \frac{x}{(1+cx)^2}\rev{,}&&   x>x^*
\end{align}

To find $x^*$, we use the condition given in Eq.~(\ref{eq:cross over radius condition}) where $\vert\Delta\epsilon(x)\vert$ is given by Eq.~(\ref{eq:normalized energy injected per unit mass (part 2) as a function of x}). The $\alpha^2$ and $\gamma$ for a Hernquist minihalo are given by (see Appendix~\ref{app:alpha squared and beta for Herquist minihalo})
\begin{equation}\label{eq:alpha squared for Herquist minihalo}
    \alpha^2 = \frac{c(6 + 9c + 2c^2) - 6(1 + c)^2\ln(1 + c)}{c^4}
\end{equation}
\begin{equation}\label{eq:beta for Hernquist minihalo}
    \gamma = \frac{4+c}{6}
\end{equation}

\subsection{Density profile of the first-generation minihalo resulting from stellar interaction with an NFW minihalo}\label{sec:density profile of first-generation minihalo after single encounter}\label{sec:density profile of first-generation minihalo}
We would now like to plot the density profile of the first-generation minihalo resulting from a stellar interaction of an NFW minihalo. The first-generation minihalo is assumed to have a broken power law profile (S2023). We start by specifying the scale parameters of an unperturbed NFW minihalo. Using these, we look to compute the scale parameters of the first-generation broken power law profile minihalo. We use the subscript $s$ to denote the unperturbed NFW minihalo and subscript $1$ to denote the resulting first-generation minihalo.

We first note that the NFW minihalo in the region $x>x_{\rm s}^*$ is completely disrupted. In addition, there is a partial mass loss in the region $x<x_{\rm s}^*$ (see Appendix~\ref{app:Evaluating the survival fraction using the sequential stripping model}). One must note that $x_{\rm s}^*$ can be greater than 1. We first compute the total surviving mass of the NFW minihalo just after the stellar interaction. We then allow the remnant minihalo to fully relax to a broken power law profile of the form
\begin{equation}\label{eq:general broken powerlaw profile}
    \rho(r)=\frac{\rho_1}{\frac{r}{r_1}\left(1+\frac{r}{r_1}\right)^k}    
\end{equation}
where we are leaving $k$ unspecified for the moment, and later we will set it to 3.
We use the fact that the total surviving mass of the NFW minihalo after perturbation is equal to the total mass of the fully relaxed broken power law first-generation minihalo.
i.e.,
\begin{equation}\label{eq:mass condition for transition from NFW minihalo to first-generation minihalo}
    M_{\text {enc,s}}\left(x_{\rm s}^*\right)-\Delta M_{x_{\rm s}=0 \rightarrow x_{\rm s}^*}=\lim _{x_1 \rightarrow \infty} M_{\rm enc, 1}(x_1)
\end{equation}
where $M_{\text {enc,s}}\left(x_{\rm s}^*\right)$ is the mass of the NFW minihalo enclosed within the normalized crossover radius $x_{\rm s}^*$ and $\Delta M_{x_{\rm s}=0 \rightarrow x_{\rm s}^*}$ is the mass of the NFW minihalo lost within the normalized crossover radius $x_{\rm s}^*$. Also, $\lim\limits_{x_1 \rightarrow \infty} M_{\rm enc, 1}(x_1)$ is the total mass of the first-generation broken power law minihalo. The $x_{\rm s}$ and $x_1$ are ``local variables'' of the NFW minihalo and the first-generation minihalo, respectively. They are defined as
\begin{equation}\label{eq:x_s defintion}
    x_{\rm s}\equiv\frac{r}{r_{\text {vir,s}}}
\end{equation}
\begin{equation}\label{eq:x_1 defintion}
    x_1\equiv\frac{r}{r_{\text {vir,1}}}
\end{equation}
where $r_{\text {vir,s}}$ and $r_{\text {vir,1}}$ are the virial radii of the unperturbed NFW minihalo and first-generation minihalo, respectively.

We consider the disrupted halo to be  a broken power law of the form 
\begin{equation}
    \label{eq:broken power law}
    \rho_{k=2+\Delta}(r)=\frac{\rho_1}{\frac{r}{r_1}\left(1+\frac{r}{r_1}\right)^{2+\Delta}}
\end{equation}
for a parameter $\Delta > 0$.

We now look at Fig. 7 of S2023 and make the reasonable assumption that at small radii, the unperturbed NFW minihalo density profile and the first-generation Hernquist/broken power-law density profile are indistinguishable from each other. We write this as
\begin{align}\label{eq:small radius condition}
    \lim _{r \rightarrow 0} \rho_{\rm NFW}(r)&=\lim _{r \rightarrow 0} \rho_{k=2+\Delta}(r) \nonumber\\
    \lim _{r \rightarrow 0} \frac{\rho_{\rm s}}{\frac{r}{r_{\rm s}}\left(1+\frac{r}{r_{\rm s}}\right)^2}&=\lim _{r \rightarrow 0} \frac{\rho_1}{\frac{r}{r_1}\left(1+\frac{r}{r_1}\right)^{2+\Delta}} \nonumber\\
    \frac{\rho_{\rm s}}{\frac{r}{r_{\rm s}}}&=\frac{\rho_1}{\frac{r}{r_1}} \nonumber\\
    \Rightarrow \rho_{\rm s} r_{\rm s}&=\rho_1 r_1
\end{align}

As shown in Appendix~\ref{app:Computing expressions for the disrupted minihalo's parameters}
we can  compute $r_1$ in terms of $r_{\rm s}$ as follows
\begin{align}\label{eq:r_1 final expression - broken power law k=3.2 copy}
    r_1
    &=r_{\rm s} \times R_{\rm s}
\end{align}
where
\begin{equation}\label{eq:r_s defintion - broken power law k=3.2 copy}
    R_{\rm s} \equiv \sqrt{(\Delta + \Delta^2)\left[f_{\rm NFW}(c_{\rm s} x_{\rm s}^*)-4 \pi c_{\rm s}^3 I_{\rm s}\right]}
\end{equation}
and 
\begin{equation}\label{eq:c_s definition copy}
    c_{\rm s} \equiv \frac{r_{\rm vir,s}}{r_{\rm s}}.
\end{equation}
\begin{multline}\label{eq:I_s definition in main text}
    I_{\rm s} = \int\limits_{x=0}^{x_{\rm s}^*} \int\limits_{\epsilon=0}^{\vert\Delta\epsilon(x)\vert} \  \int\limits_{\psi_{\rm B}^\prime=0}^\epsilon x^2\frac{1}{\sqrt{8}\pi^2} \frac{1}{\sqrt{\epsilon - \psi_{\rm B}^\prime}} \frac{\mathrm{d}^2\varrho}{\mathrm{d}\psi_{\rm B}^{\prime^2}}\left(x^\prime\left(\psi_{\rm B}^\prime\right)\right)
    \sqrt{2\left(\psi_{\rm A}(x)-\epsilon\right)} \ \mathrm{d}\psi_{\rm B}^\prime\ \mathrm{d}\epsilon\ \mathrm{d}x
\end{multline}
It is also shown in the same appendix that 
\begin{equation}\label{eq:rho_1 final condition copy}
    \rho_1 = \frac{\rho_{\rm s}}{R_{\rm s}}
\end{equation}
Thus, if the scale parameters $\rho_{\rm s}$, $r_{\rm s}$ of the NFW profile are given, Eqs.~(\ref{eq:r_1 final expression - broken power law k=3.2 copy}) and (\ref{eq:rho_1 final condition copy}) give the scale parameters $\rho_1$, $r_1$ of the resulting first-generation broken power law minihalo.

Using N-body simulations, S2023 finds that when an NFW minihalo participates in a stellar interaction with impact parameter $b=2 \times 10^{-5} \mathrm{kpc}$, the resulting first-generation minihalo will have a broken power law profile with $k=3.2$ or $\Delta = 1.2$. In such a case,
\begin{equation}
    R_{\rm s} = \sqrt{\frac{66}{25}\left[f_{\rm NFW}(c_{\rm s} x_{\rm s}^*)-4 \pi c_{\rm s}^3 I_{\rm s}\right]}
\end{equation}
On the other hand, according to the  N-body simulations of S2023, if the impact parameter is $b=5 \times 10^{-5} \mathrm{kpc}$, the resulting first-generation minihalo will have a broken power law profile with $k=3.3$ or $\Delta = 1.3$. Then,
\begin{equation}
    R_{\rm s} = \sqrt{\frac{299}{100}\left[f_{\rm NFW}(c_{\rm s} x_{\rm s}^*)-4 \pi c_{\rm s}^3 I_{\rm s}\right]}
\end{equation}
Finally, assuming for simplicity, that the first-generation minihalo had a Hernquist profile ($k = 3$ or $\Delta = 1$),
\begin{equation}\label{eq:r_s definition}
    R_{\rm s} = \sqrt{2f_{\rm NFW}(c_{\rm s} x_{\rm s}^*) - 8 \pi c_{\rm s}^3 I_{\rm s}}
\end{equation}

\rev{Note that the value of the impact parameter $b$ was only introduced here to determine the value of $k$ for the broken power law profile that the NFW profile minihalo relaxes to, following a stellar encounter. As such the quanitity that characterises the stellar encounter is the normalized energy injection parameter $E_{\rm frac}$. The impact parameter is only important in so far as to determine the value of $E_{\rm frac}$ using eqn.~(\ref{eq:Efrac expression}).}

\rev{The virial mass of the unperturbed NFW minihalo impacts the value of $R_{\rm s}$ and consequently the survival fraction because the virial mass is a function of the scale parameters $\rho_{\rm s}$ and $r_{\rm s}$ of the NFW minihalo. To see how the virial mass explicitly determines the survival fraction,}  instead of specifying the unperturbed NFW minihalo by its two scale parameters, we would like to specify it by two other quantities: its concentration parameter and its virial mass. So, in order to compute the scale parameters of the first-generation minihalo, we need to first compute the scale parameters of the NFW minihalo from its concentration and virial mass.

We first use the definition of the virial radius of the NFW minihalo. The virial radius is that radius at which the average density $\bar{\rho}_{\rm vir}$ enclosed within the virial radius is 200 times the critical density $\rho_{\rm crit}$ of the universe. Therefore,
\begin{align}\label{eq:Virial radius condition}
    \frac{\text{Mass within virial radius}}{\text{Volume within virial radius}} &= 200\rho_{\rm crit} \nonumber\\\
    \frac{\int\limits_{r=0}^{r_{\rm vir, s}} \rho_{\rm NFW}(r) \times 4 \pi r^2 d r}{\frac{4 \pi}{3} r_{\rm vir, s}^3}&=200 \rho_{\text {crit }}
\end{align}
Let
\begin{equation}\label{eq:x_s definition}
    x_s \equiv \frac{r}{r_{\text {vir,s}}}\ \Rightarrow\ r=x_{\rm s} r_{\text {vir,s }}
\end{equation}
Substituting Eq.~(\ref{eq:x_s definition}) in Eq.~(\ref{eq:Virial radius condition})
\begin{equation}\label{eq:Virial radius condition, intermediate 2}
    \int\limits_{x_{\rm s}=0}^1 \rho_{\rm NFW}\left(x_{\rm s}\right) x_{\rm s}^2 \mathrm{d} x_{\rm s}=\frac{200}{3} \rho_{\text {crit}}
\end{equation}
Substituting Eq.~(\ref{eqn:NFW as a function of x}) in Eq.~(\ref{eq:Virial radius condition, intermediate 2}) and performing the integration with respect to $x_{\rm s}$ in the L.H.S. of Eq.~(\ref{eq:Virial radius condition, intermediate 2}), we get
\begin{equation*}
    \frac{\rho_{\rm s}}{c^3}\left[\ln (1+c)-\frac{c}{1+c}\right]=\frac{200}{3} \rho_{\rm crit}
\end{equation*}
\begin{equation}\label{eq:rho_s in terms of concentration for NFW profile}
    \Rightarrow \rho_{\rm s}=\frac{200}{3} \frac{c^3}{f_{\rm NFW}(c)} \rho_{\rm crit} 
\end{equation}
Cosmological observations \citep{Planck2018} fix the value of $\rho_{\rm crit}$ as
\begin{equation}\label{eq:rho_crit value}
    \rho_{\text {crit}}=1.3483 \times10^{-7} \mathrm{M}_\odot / \mathrm{pc}^3
\end{equation}
Thus, specifying the concentration of the NFW minihalo fixes its scale density through Eq.~(\ref{eq:rho_s in terms of concentration for NFW profile}).

Next, the average density within the virial radius of the NFW minihalo is expressed in terms of its virial mass and virial radius as follows
\begin{equation}\label{eq:average virial density definition}
    \bar{\rho}_{\text {vir}}=\frac{M_{\text {vir,s}}}{\frac{4 \pi}{3} r_{\text {vir,s}}^3}    
\end{equation}

Substituting Eq.~(\ref{eq:c_s definition copy}) in Eq.~(\ref{eq:average virial density definition}) and solving for $r_{\rm s}$, we get
\begin{equation}\label{eq:r_s in terms of of c_s and M_vir,s - pre-final expression}
    r_{\rm s}=\frac{1}{c_{\rm s}}\left(\frac{3 M_{\text {vir,s}}}{4 \pi \bar{\rho}_{\text {vir}}}\right)^{1 / 3}
\end{equation}
Noting that by definition
\begin{equation}
    \bar{\rho}_{\rm vir}=200 \rho_{\text {crit}}    
\end{equation}

Therefore, Eq.~(\ref{eq:r_s in terms of of c_s and M_vir,s - pre-final expression}) becomes
\begin{equation}\label{eq:r_s in terms of of c_s and M_vir,s - final expression}
    r_{\rm s}=\frac{1}{c_{\rm s}}\left(\frac{3 M_{\rm vir,s}}{800 \pi \rho_{\text {crit}}}\right)^{1 / 3} 
\end{equation}
Thus, given the concentration and virial mass of the NFW minihalo, its scale parameters can be found using Eqs.~(\ref{eq:rho_s in terms of concentration for NFW profile}) and (\ref{eq:r_s in terms of of c_s and M_vir,s - final expression}).
\begin{figure*}[btp]
    \centering
        \includegraphics[width=\columnwidth, height=0.47\textheight,keepaspectratio]{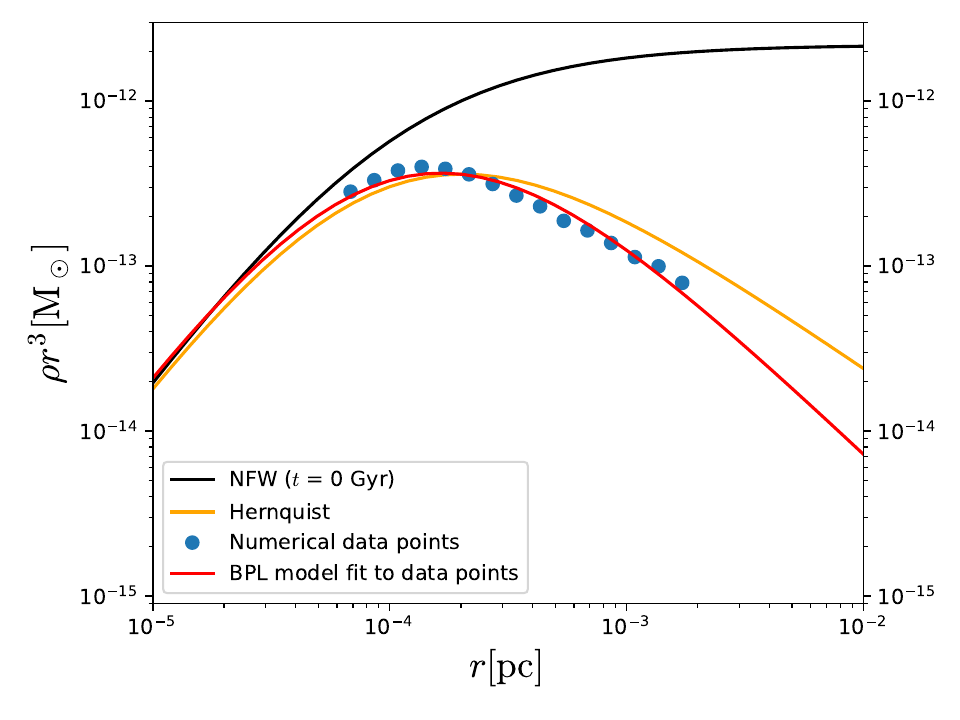}
    \medskip

        \includegraphics[width=\columnwidth, height=0.47\textheight,keepaspectratio]{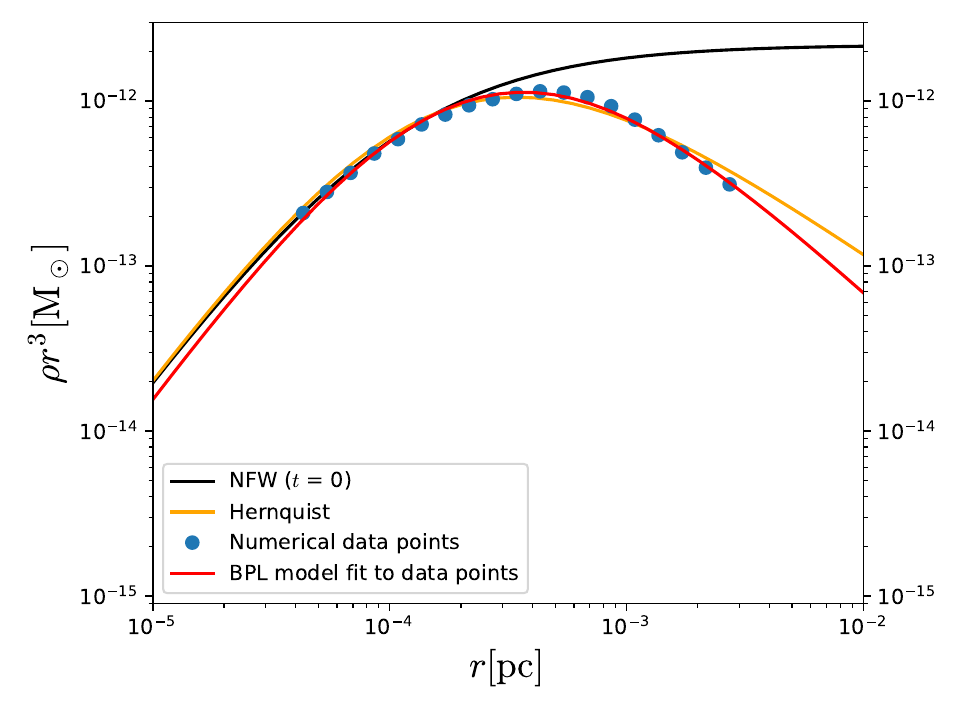}
    
    \caption[Density profiles of unperturbed NFW minihalo and resulting relaxed Hernquist minihalo]{The top and bottom panels show the density profile of an NFW minihalo (black curve) which has a stellar encounter with impact parameters $b = 2\times10^{-5}\rm kpc$ and $b = 5\times10^{-5}\rm kpc$ respectively. The NFW minihalo has an initial concentration $c_{\rm s}=100$ and virial mass $M_{\rm vir} = 10^{-10}\rm M_\odot$. The orange curve shows the case where the remnant minihalo has relaxed to a Hernquist density profile, which is a broken power law (BPL) profile with $k=3$. The \rev{blue} dots are numerical data points of the resulting density profile stabilized at $t = 2.5$ Gyr post-stellar interaction taken from S2023. We performed a curve fit of these data points to a broken power law profile and obtained its three parameters. The broken power law profile is shown as the red curves.}
    \label{fig:density profile of first-generation minihalo}
\end{figure*}
Given the concentration and virial mass of the unperturbed NFW minihalo, we are now able to plot the density profile of the first-generation minihalo. We use $c_{\rm s}=100$ and $M_{\text {vir,s}}=10^{-10} \rm M_\odot$. The top (bottom) panel of Fig.~\ref{fig:density profile of first-generation minihalo} shows the case where the impact parameter $b=2 \times 10^{-5} \mathrm{kpc}$ ($b=5 \times 10^{-5} \mathrm{kpc}$). The black curve shows the density profile of the unperturbed NFW minihalo while the orange curve shows the case where the fully relaxed first-generation minihalo is assumed to have a Hernquist profile $(k=3)$. For both impact parameter cases, S2023 numerically shows how the density profile of the resulting minihalo changes and stabilizes over time. We chose the data points on that portion of the density profile curve at $t$ = 2.5 Gyr where the profile had stabilized, and used these data points to find the model parameters of the resulting broken power law of the form given in Eq.~(\ref{eq:general broken powerlaw profile}). \rev{Note that the associated dynamical time of the minihalo in S2023's simulations was $t_{\rm dyn}(z=0) = 2.2\ \rm Gyr$.} These data points are shown in \rev{blue} in both panels of Fig.~\ref{fig:density profile of first-generation minihalo}. We did a least squares fit of the data points and obtained the optimal parameters $r_1$, $\rho_1$ and $k$. These parameters are given in Table~\ref{tab:BPL model parameters and survival fractions}. The resulting broken power law profiles are plotted as the red curves in Fig.~\ref{fig:density profile of first-generation minihalo}. We notice that the broken power law profile merges with the NFW profile at small radii. This further reinforces our assumption that the parent and relaxed child minihalos are indistinguishable at small radii during a stellar interaction. Next, we calculated the virial radius $r_{\rm vir,s}$ of the unperturbed NFW minihalo using Eq.~(\ref{eq:average virial density definition}). We found that $r_{\rm vir,s} = 9.6\times10^{-3}$pc. We then calculated the masses of the Hernquist and broken power law profiles inside this virial radius $r_{\rm vir,s}$. The ratio of the above mass to the virial mass of the NFW profile then gives us the survival fraction of the Hernquist or broken power-law profiles. Alternatively, we also calculate the value of normalized injected energy $E_{\rm frac}$ using the impact parameter with Eq.~(\ref{eq:Efrac expression}) when $b>b_{\rm s}$. Knowing $E_{\rm frac}$, we calculate the survival fraction according to S2023's empirical response function. These survival fractions are presented in Table~\ref{tab:BPL model parameters and survival fractions}. As can be seen, there is about a 0.03 difference in the SF value between the broken power law case and the SF from S2023's empirical response function. This is indicative of the level of systematic error in determining the SF. As can be seen, the difference between the SF from the Hernquist and the broken power law profiles is around 0.01, which indicates that within the level of systematic error, the Hernquist profile could be used instead of the broken power law. 

\begin{table}
    \begin{tabular}{|l|c|c|}
        \hline
        b &	$2\times10^{-5}$kpc & $5\times10^{-5}$kpc \\
        \hline
        $r_1$ & $1.14\times10^{-4}$pc & $2.47\times10^{-4}$pc \\
        \hline
        $\rho_1$ & $2.44$M$_\odot$/pc$^3$ & $0.716$M$_\odot$/pc$^3$\\
        \hline
        $k$  & 3.38 & 3.34\\
        \hline
        SF from Hernquist & 0.15 & 0.433 \\
        \hline
        SF from BPL & 0.137 & 0.423 \\
        \hline
        \rev{$E_{\rm frac}$ computed from eqn.~(\ref{eq:Efrac expression})} & \rev{305} & \rev{7.81} \\
        \hline
        SF from S2023 response function & 0.172 & 0.393 \\
        \hline
    \end{tabular}
    \caption
    [Parameters of the relaxed broken power law profile are tabulated for two different impact parameters. Also tabulated are the survival fractions.]
    {
        \label{tab:BPL model parameters and survival fractions}
        An NFW profile minihalo of concentration $c_{\rm s} \rev{= 100}$, virial mass \rev{$M_{\rm vir,s} = 10^{-10} M_\odot$ and $r_{\rm vir,s} = 9.6\times10^{-3} \rm pc$} undergoes a stellar interaction with two different impact parameters $b=2 \times 10^{-5} \mathrm{kpc}$ and $b=5 \times 10^{-5} \mathrm{kpc}$. In each case, we fit the numerical data points in Fig.~\ref{fig:density profile of first-generation minihalo} to obtain the parameters of the resulting broken power law (BPL) profile. The scale radius $r_1$, scale density $\rho_1$, and power parameter $k$ of the BPL profile are presented above. Moreover, the survival fractions using the Hernquist and BPL profiles in Fig.~\ref{fig:density profile of first-generation minihalo} as well as using S2023's empirical response function are presented.
    }
\end{table}

\subsection{Incorporating relaxation to a Hernquist profile}
We now try to improve upon Fig.~\ref{fig:SF vs Efrac - sequential stripping model}, this time incorporating the relaxation process to a Hernquist profile. We assume we are given the concentration $c_{\rm s}$ of the unperturbed NFW minihalo. Then, Eq.~(\ref{eq:rho_s in terms of concentration for NFW profile}) gives us the scale density $\rho_{\rm s}$ of the NFW minihalo. Having calculated $\rho_{\rm s}$, Eq.~(\ref{eq:rho_1 final condition copy}) gives us the scale density $\rho_1$ of the relaxed first-generation Hernquist minihalo, where $R_{\rm s}$ is given by Eq.~(\ref{eq:r_s definition})

Our next task is to compute the concentration $c_1$ of the first-generation Hernquist minihalo and get an equation analogous to Eq.~(\ref{eq:rho_s in terms of concentration for NFW profile}) but for the Hernquist profile. We start with the definition of the virial radius as before. This then leads us to an equation similar to Eq.~(\ref{eq:Virial radius condition, intermediate 2}) but for the first-generation Hernquist minihalo, finally resulting in an equation relating the concentration $c_1$ and scale density $\rho_1$ as follows (see Appendix~\ref{app:computing the concentration of Hernquist minihalo, given its scale density}): 

\begin{equation}\label{eq:relating concentration and scale radius for Hernquist profile}
    \frac{1}{2 c_1\left(1+c_1\right)^2}=\frac{200}{3}\frac{\rho_{\text {crit}}}{\rho_1}
\end{equation}
The survival fraction after relaxation is given by (see Appendix~\ref{app:Computing the expression for survival fraction with relaxation})

\begin{align}\label{eq:survival fraction of NFW minihalo incorporating relaxation}
    \text {SF} &\equiv \frac{M_{\text {enc,1}}\left(x_1^{r_{\rm vir,s}}\right)}{M_{\text {vir,s}}} \nonumber\\
    & =\frac{1}{2} R_{\rm s}^2 \frac{f_{\rm Hern}(c_1 x_1^{r_{\text{vir,s}}})}{f_{\rm NFW}(c_{\rm s})} 
\end{align}
where
\begin{equation}\label{eq:f_Hern definition}
    f_{\rm Hern}(x) = \frac{x^2}{(1 + x)^2}
\end{equation}
\begin{equation}
    x_1^{r_{\rm vir,s}} \equiv \frac{r_{\rm vir,s}}{r_{\rm vir,1}}
\end{equation}
is the virial radius of the unperturbed NFW minihalo expressed in terms of the ``local" normalized radial distance variable of the first-generation Hernquist minihalo. $M_{\rm enc,1}$ is the mass enclosed for a Hernquist profile.
$x_1^{r_{\text{vir,s}}}$ can be written as (see Appendix~\ref{app:Computing the expression for survival fraction with relaxation})
\rev{
\begin{align}\label{eq:x_1^rvirs}
    x_1^{r_{\text{vir,s}}} &= \frac{c_{\rm s}}{c_1} \frac{1}{R_{\rm s}} \nonumber\\
    \implies c_1 x_1^{r_{\text{vir,s}}} &= \frac{c_{\rm s}}{R_{\rm s}} \ .
\end{align}
Plugging in eqn.~(\ref{eq:x_1^rvirs}) into eqn.~(\ref{eq:survival fraction of NFW minihalo incorporating relaxation}), we get
\begin{equation}\label{eq:survival fraction of NFW minihalo incorporating relaxation, final expression}
    \text {SF}  =\frac{1}{2} R_{\rm s}^2 \frac{f_{\rm Hern}(c_{\rm s}/R_{\rm s})}{f_{\rm NFW}(c_{\rm s})} 
\end{equation}
}

\begin{figure*}
    \includegraphics[width=\textwidth]{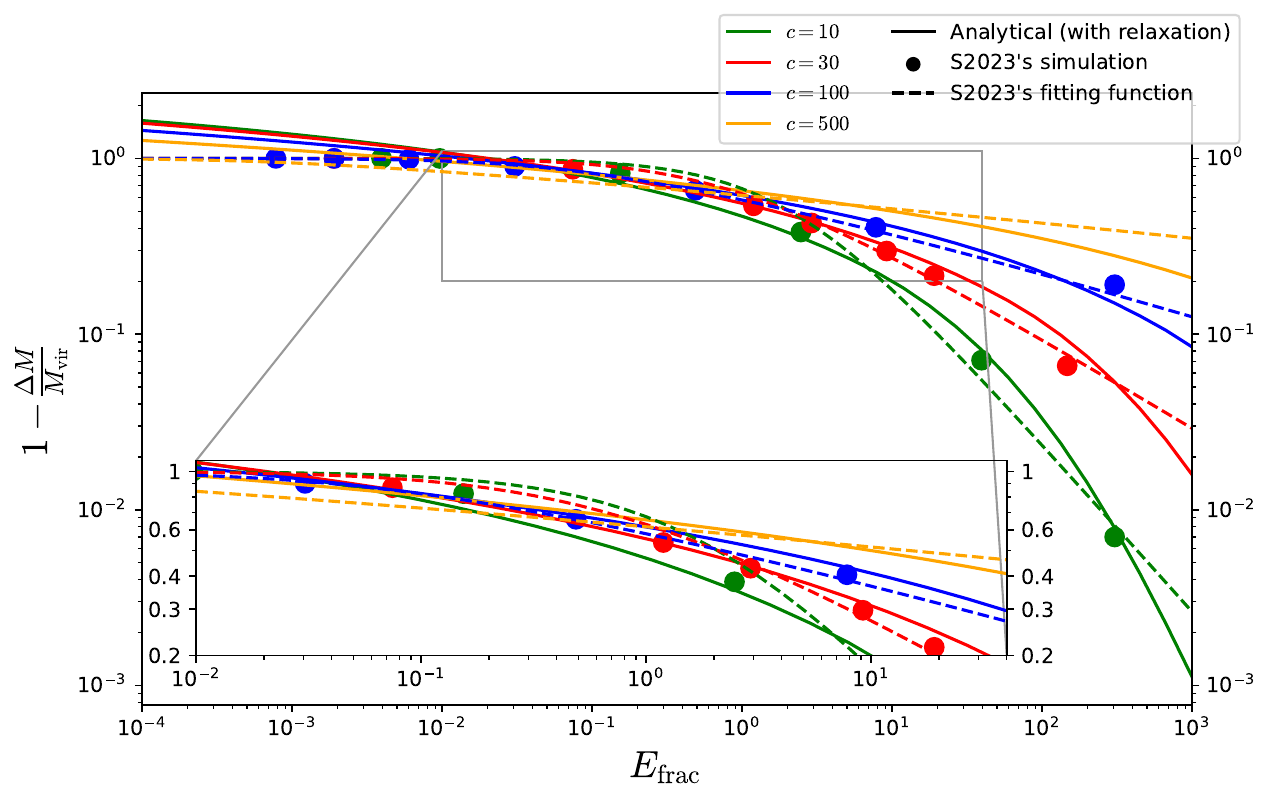}
    \caption[Survival fraction against $E_{\rm frac}$ for sequential stripping model, with relaxation to a Hernquist profile]{
        The same as Fig.~\ref{fig:SF vs Efrac - sequential stripping model} except that the solid curves are the output of our analytical approach using the sequential stripping model of mass loss in the minihalo and taking into account relaxation of the remnant minihalo to a Hernquist profile.
    }
    \label{fig:SF vs Efrac - including relaxation}
\end{figure*}
Using this procedure, the survival fraction can be computed against $E_{\rm frac}$. Fig.~\ref{fig:SF vs Efrac - including relaxation} shows the results. The solid lines are our analytical curves. The circular dots are numerical data points from S2023. The dashed curves are S2023's curve fits to the numerical data. Our analytical curves closely match the numerical data points for larger values of $E_{\rm frac}$. This is because it is a reasonable assumption that the remnant minihalo relaxes to a Hernquist profile in this regime. However, our analytical curves overshoot the numerical data \rev{and even increase to values greater than unity} for smaller values of $E_{\rm frac}$. This is because, as discussed in Section~\ref{sec:need for relaxation}, for smaller $E_{\rm frac}$ values, it is a better approximation to assume that the remnant minihalo relaxes to the same NFW profile, not the Hernquist profile. \rev{To mathematically see why the survival fraction increases above unity for such small $E_{\rm frac}$ values, we direct the reader to Fig.~\ref{fig:psi crossing over deltaEpsilon} which shows that as $E_{\rm frac}$ decreases, the $|\Delta \epsilon(x)|$ curve becomes more and more flat. Thus, the crossover radius increases. As $E_{\rm frac} \to 0$, $\lim\limits_{E_{\rm frac} \to 0}|\Delta \epsilon(x)| = 0\ \forall\ x\geq0$. Thus, the crossover radius $\lim\limits_{E_{\rm frac} \to 0}x_{\rm s}^* = \infty$. $R_{\rm s}$ is given by eqn.~(\ref{eq:r_s defintion - broken power law k=3.2 copy}). Now, eqn.~(\ref{eq:fNFW}) shows the expression for $f_{\rm NFW}(x)$. Thus,
\begin{align}
	\lim\limits_{x \to \infty}	f_{\rm NFW}(x) &= \lim\limits_{x \to \infty} \left[\ln(1+x) - \frac{x}{1+x} \right]= \infty \nonumber\\
	\implies \lim\limits_{x_{\rm s}^* \to \infty} f_{\rm NFW}(c_{\rm s} x_{\rm s}^*) &= \infty
\end{align}
Eqn.~(\ref{eq:I_s definition in main text}) shows the expression for $I_{\rm s}$. Thus,
\begin{equation}
	\lim\limits_{|\Delta \epsilon(x)| \to 0}I_{\rm s} = 0 \ .
\end{equation}
Thus, 
\begin{equation}
	\lim\limits_{E_{\rm frac} \to 0} R_{\rm s} = \infty \ .
\end{equation}
Next, using eqns.~(\ref{eq:survival fraction of NFW minihalo incorporating relaxation, final expression}) and (\ref{eq:f_Hern definition}), we can rewrite the expression for the survival fraction as
\begin{equation}
    \text{SF}  = \frac{1}{2f_{\rm NFW}(c_{\rm s})} \frac{c_{\rm s}^2}{\left(1 + \frac{c_{\rm s}}{R_{\rm s}}\right)^2} \ .
\end{equation}
Thus,
\begin{equation}
    \lim\limits_{E_{\rm frac} \to 0}\text{SF} = \lim\limits_{R_{\rm s} \to \infty}\text{SF} = \frac{c_{\rm s}^2}{2f_{\rm NFW}(c_{\rm s})} \approx 1379  > 1
\end{equation}
Thus, mathematically, this model produces survival fraction values of greater than unity for low $E_{\rm frac}$ values. Conceptually, in the mass conservation condition presented in eqn.~(\ref{eq:mass condition for transition from NFW minihalo to first-generation minihalo}), the L.H.S. of the equation, which dictates the surviving mass of the NFW minihalo within $x_{\rm s}^*$, increases without bounds since our NFW minihalo is untruncated. This then enforces the R.H.S. of the equation, which dictates the mass of the relaxed Hernquist minihalo to also grow without bounds, enforcing the scale radius $r_1 \to \infty$ via eqn.~(\ref{eq:r_1 final expression - broken power law k=3.2 copy}) to accommodate this ever-increasing mass. Since we always compute the survival fraction within the physical virial radius $r_{\rm vir,s}$ of the unperturbed NFW minihalo, there is no guarantee that the mass of the relaxed Hernquist minihalo within the same $r_{\rm vir,s}$ will be  $\leq M_{\rm vir, s}$, which is required to have a survival fraction of less than unity.
}

\subsection{The switching procedure}
Fig.~\ref{fig:SF vs Efrac - sequential stripping model} shows that the sequential stripping model without considering relaxation produces a good match to the numerical data at low values of $E_{\rm frac}$ but overshoots the data at high $E_{\rm frac}$ values. On the other hand, Fig.~\ref{fig:SF vs Efrac - including relaxation} shows that the sequential stripping model incorporating relaxation to a Hernquist profile overshoots the numerical data at low values of $E_{\rm frac}$ but is a good match at high $E_{\rm frac}$ values. We have previously discussed the physical intuitions for both these models. We now introduce a switching procedure where, for any given $E_{\rm frac}$ value, we compute the survival fraction both without and with considering relaxation to a Hernquist profile, and we choose the method that gives us the lesser value of survival fraction. Implementing this procedure in Python, we find that the algorithm selects the method without relaxation for low values of $E_{\rm frac}$. As $E_{\rm frac}$ is increased, the algorithm switches to the method with relaxation at some (unenforced) value of $E_{\rm frac}$.
\begin{figure*}
    \includegraphics[width=\textwidth]{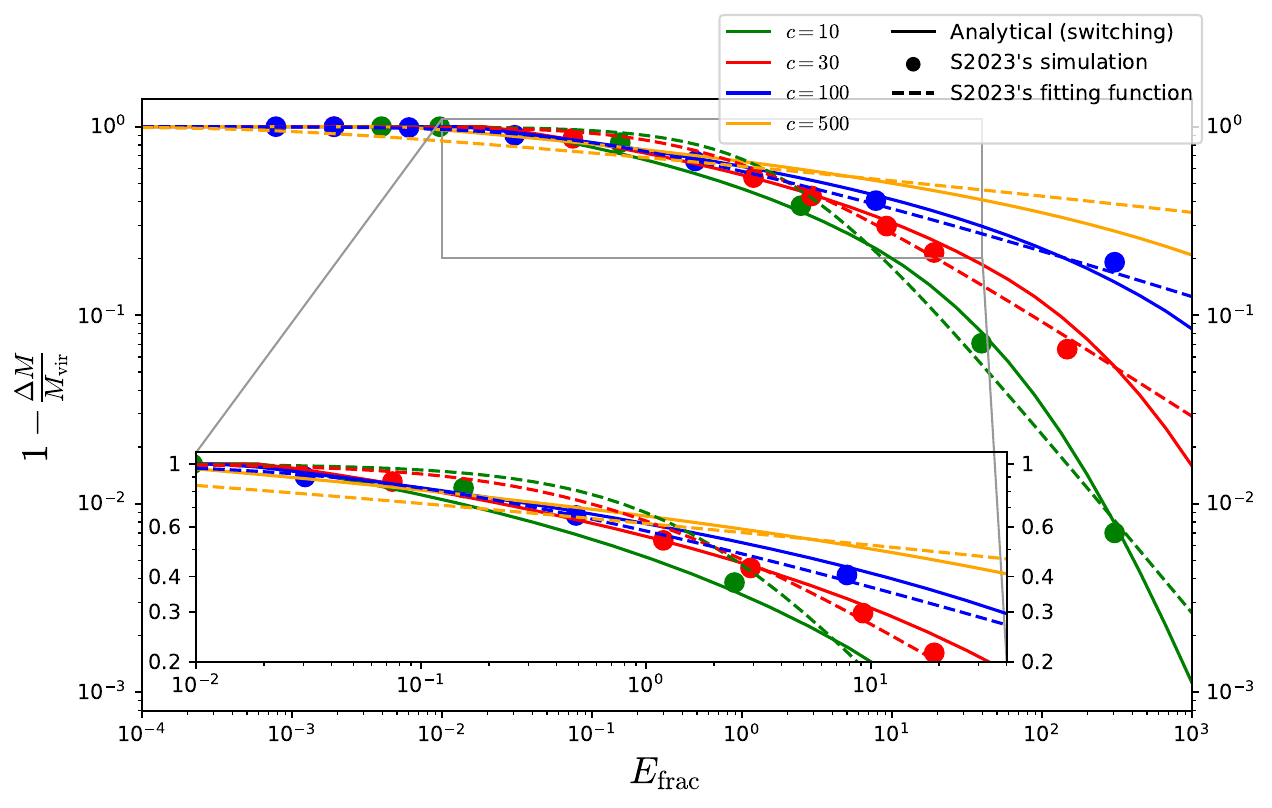}
    \caption[Survival fraction against $E_{\rm frac}$ for sequential stripping model with switching procedure]{
        The same as Fig.~\ref{fig:SF vs Efrac for four values of concentration} except that the solid curves are the output of our analytical approach using the sequential stripping model of mass loss in the minihalo and incorporating the switching procedure.
    }
    \label{fig:SF vs Efrac - switching}
\end{figure*}
Fig.~\ref{fig:SF vs Efrac - switching} shows the survival fraction plotted against $E_{\rm frac}$ using the switching procedure.  The switching procedure provides a good match to the numerical data for all regimes of $E_{\rm frac}$ values considered.

\begin{figure}
    \centering
    \includegraphics[width=\columnwidth]{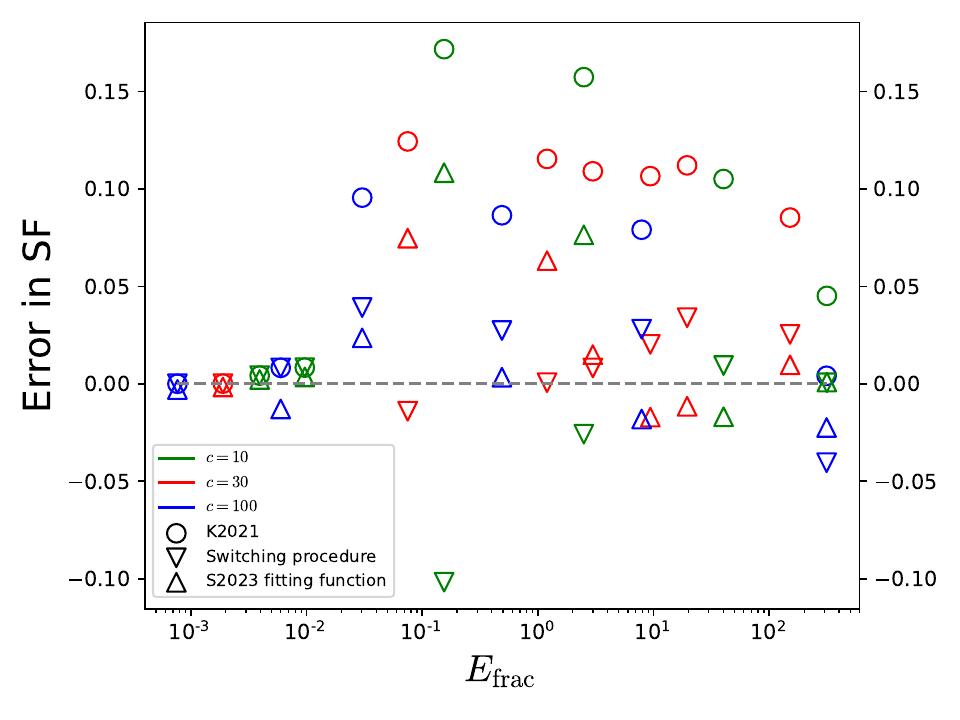}
    \caption[Error of the K2021 model, our switching model and S2023's fitting function]{The error in a model is defined as (model $-$ data). Plotted are the errors in the output of Ref.~\cite{K2020}'s (K2021) analytical approach, our analytical
        switching procedure and S2023's empirical fitting functions. The numerical data is taken from S2023's N-body simulations. Errors are plotted for concentration parameters $c=10,30,100$.
    }
    \label{fig:error}
\end{figure}
\begin{figure}
    \centering
    \includegraphics[width=\columnwidth]{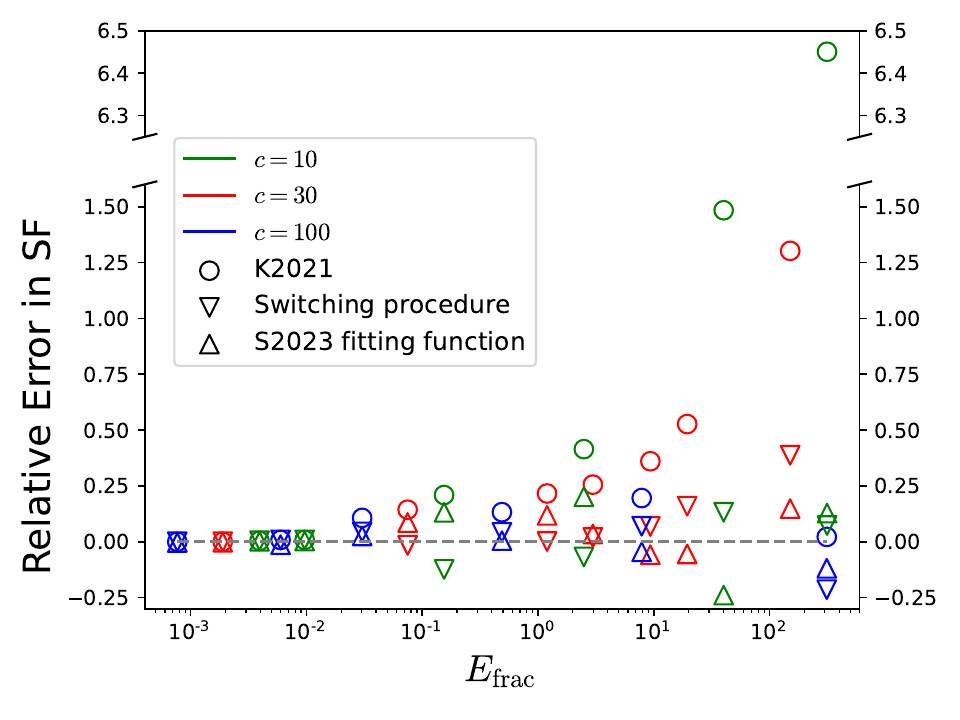}
    \caption[Relative error of the K2021 model, our switching model and S2023's fitting function]{The same as Fig.~\ref{fig:error} except for the relative error, which is defined as (model $-$ data)/data, rather than the error.
    }
    \label{fig:relative error}
\end{figure}
Figs.~\ref{fig:error} and \ref{fig:relative error} show the error and relative error respectively in the survival fraction vs $E_{\rm frac}$ curves for K2021's approach, our analytical 
switching procedure and S2023's fitting functions.

As can be seen, the smallest errors are shown by the switching procedure and S2023's semi-analytic fitting functions.

\section{Multiple stellar encounters of an NFW minihalo}
\label{sec:Multiple_stellar_encounters_of_an_NFW_minihalo}
Here, we consider the scenario of multiple successive stellar interactions with an NFW minihalo. After each stellar interaction, the remnant minihalo is allowed to fully relax before the next stellar interaction is applied. S2023 states that after a stellar encounter, the NFW minihalo will relax to a broken power law profile with a $k$ dependent on the impact parameter of the stellar interaction. In general, $k \sim 3$. We will assume $k=3$ for our analytical calculations, i.e., a Hernquist profile. We also assume that when a Hernquist profile is perturbed, the remnant minihalo will relax to a Hernquist profile with a different concentration and overall mass.

The NFW and Hernquist density profiles are two-parameter models. However, it turns out that we need only one piece of information to calculate the survival fraction (or mass loss fraction) due to a stellar encounter. Here, we use the concentration of the minihalo as that piece of information. As we will see later, the concentration of a two-parameter minihalo, like the NFW and Hernquist profiles, uniquely determines the scale density of that minihalo. So, we cannot compute the exact value of the scale radius as that would need a second piece of information. However, we can compute the ratio of the parent and child minihalos' scale radii. This will allow us to calculate the scale density of the child minihalo. Knowing the scale density of the child minihalo gives us its concentration parameter. We then iterate this process to get the concentrations (and hence survival fractions) of successive generations of minihalos.

We use the subscript $s$ to denote the unperturbed NFW minihalo. We use the numeral $n$ to denote the $n^{\text {th }}$ generation Hernquist minihalo, where the first-generation Hernquist minihalo is the child of the NFW minihalo and the $n^{\text {th }}$ generation Hernquist minihalo is the child of the $(n-1)^{\text {th }}$ generation Hernquist minihalo.

To compare the survival fractions of successive generations of minihalos, we will define the survival fraction of any minihalo as

\begin{equation}
    \text{SF}_n \equiv 1-\frac{\Delta M}{M_{\text {vir,s}}}    
\end{equation}
where $n = s, 1, 2, 3, \cdots$ and $\Delta M$ is the total mass lost within the virial radius of the unperturbed NFW minihalo. This means that our region of interest for the purposes of evaluating mass loss is always inside a 
sphere of radius $r_{\rm vir,s}$.

We start the procedure by specifying the concentration $c_{\rm s}$ of the NFW minihalo. To better understand how some of the equations in this section are derived, see Appendix~\ref{app:multiple stellar encounters}.

\noindent\underline{Step I}: Compute the survival fraction of the NFW minihalo

We use the switching procedure to find out the survival fraction of the NFW minihalo. We will restrict our incremental $E_{\rm frac}$ to be high enough such that the switching procedure forces the remnant minihalo to relax to a Hernquist profile. Knowing $c_{\rm s}$, we can calculate the scale density $\rho_{\rm s}$ of the NFW minihalo using Eq.~(\ref{eq:rho_s in terms of concentration for NFW profile}). Next we calculate $R_{\rm s}$ using Eq.~(\ref{eq:r_s definition}). We can then calculate the scale density $\rho_1$ of the first-generation Hernquist minihalo using Eq.~(\ref{eq:rho_1 final condition copy}). We can then calculate concentration $c_1$ of the first-generation Hernquist minihalo by solving for Eq.~(\ref{eq:relating concentration and scale radius for Hernquist profile}). The NFW minihalo's survival fraction is then given by Eq.~(\ref{eq:survival fraction of NFW minihalo incorporating relaxation}).

\noindent\underline{Step II}: Compute the survival fraction of the $n^{\rm th}$ generation Hernquist minihalo ($n \geq 1$)

In the $(n-1)^{\text{th}}$ step, we would have calculated the scale density $\rho_n$ and concentration $c_n$ of the $n^{\rm th}$ generation Hernquist minihalo. We next consider the conservation of mass condition for the transition from the $n^{\text{th}}$ to $(n+1)^{\text{th}}$ generation Hernquist minihalo.

\begin{equation}\label{eq:mass condition for transition from n^th to (n+1)^th generation Hernquist minihalo}
    M_{\text {enc,}n}\left(x_n^*\right)-\Delta M_{x_n=0 \rightarrow x_n^*}=\lim _{x_{n+1} \rightarrow \infty} M_{\text {enc,}n+1}\left(x_{n+1}\right)
\end{equation}
Evaluating each of the terms in Eq.~(\ref{eq:mass condition for transition from n^th to (n+1)^th generation Hernquist minihalo}), it can be shown that the ratio of the scale radii of the $(n+1)^{\text{th}}$ to the $n^{\text{th}}$ generation Hernquist minihalos is (see Appendix~\ref{app:rn+1_by_rn and rho_n+1 derivation})

\begin{equation}
    \frac{r_{n+1}}{r_n}=R_n
\end{equation}
where
\begin{equation}\label{eq:R_n definition}
    R_n \equiv \sqrt{f_{\rm Hern}(c_n x_n^*)-8 \pi c_n^3 I_n}
\end{equation}

\begin{multline}
    I_n \equiv \int\limits_{x_n=0}^{x_n^*}\int\limits_{\epsilon=0}^{\vert\Delta\epsilon(x_n)\vert} \int\limits_{\psi_{\rm B}^\prime=0}^\epsilon \frac{1}{\sqrt{8}\pi^2} x_n^2 \sqrt{2(\psi_{\rm A}(x_n)-\epsilon)} \\
    \times\frac{1}{\sqrt{\epsilon - \psi_{\rm B}^\prime}} \frac{\mathrm{d}^2\varrho}{\mathrm{d}\psi_{\rm B}^{\prime^2}} \left(x_n^\prime(\psi_{\rm B}^\prime)\right) \mathrm{d}\mathrm{\psi_{\rm B}^\prime}\mathrm{d}\epsilon \mathrm{d}x_n
\end{multline}
We can now calculate the scale density $\rho_{n+1}$ of the $(n+1)^{\text{th}}$ generation Hernquist minihalo using the following relationship (see Appendix~\ref{app:rn+1_by_rn and rho_n+1 derivation}):

\begin{equation}
    \rho_{n+1}=\frac{\rho_n}{R_n}
\end{equation}
Next, we compute the concentration $c_{n+1}$ of the $(n+1)^{\text{th}}$ generation Hernquist minihalo by adapting Eq.~(\ref{eq:relating concentration and scale radius for Hernquist profile}) and solving for $c_{n+1}$ in the expression below:

\begin{equation}
    \frac{1}{2 c_{n+1}\left(1+c_{n+1}\right)^2}=\frac{200}{3} \frac{\rho_{\rm crit}}{\rho_{n+1}}
\end{equation}
The survival fraction of the $n^{\text{th}}$ generation Hernquist minihalo is given by (See Appendix~\ref{app:SF_n derivation})
\begin{equation}
    \text{SF}_n=\frac{1}{2}\left(R_n R_{n-1} \ldots R_1 R_{\rm s}\right)^2 \frac{f_{\rm Hern}(c_{n+1} x_{n+1}^{r_{\text{vir,s}}})}{f_{\rm NFW}(c_{\rm s})}
\end{equation}
where
\begin{equation}
    x_{n+1}^{r_{\text{vir,s}}}=\frac{c_{\rm s}}{c_{n+1}} \frac{1}{R_n R_{n-1} \cdots R_1 R_{\rm s}}
\end{equation}
Step II is applied for $n=1$, then $n=2$, and so on. This completes the theoretical procedure for computing survival fractions for multiple stellar encounters of an NFW minihalo.
\begin{figure}
    \includegraphics[width=\columnwidth]{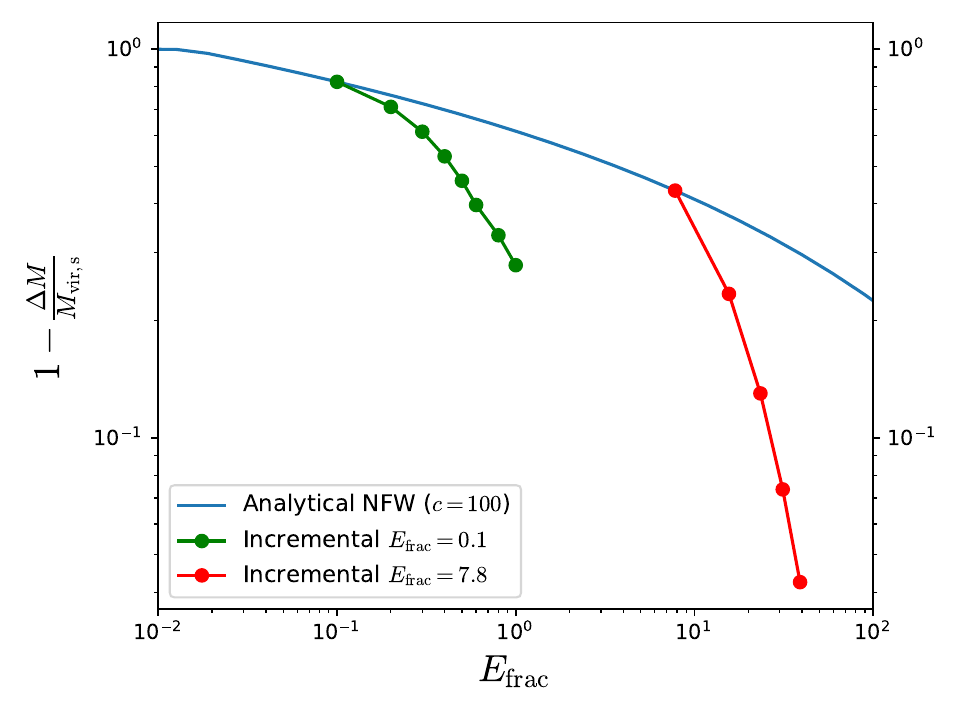}
    \caption[Survival fraction against $E_{\rm frac}$ for multiple stellar encounters]{The survival fraction of an NFW minihalo is plotted against the total normalized injected energy $E_{\rm frac}$. The blue curve shows the case of a single stellar encounter. The green (red) curves represent the multiple stellar encounter scenario where an incremental $E_{\rm frac} = 0.1 (7.8)$ is used repeatedly. \rev{Note that for the green curve, the last two incremental $E_{\rm frac}$ were 0.2.} When the incremental $E_{\rm frac}$ is higher (red curve), the deviation from the single energy injection case (blue curve) is higher as well.}
    \label{fig:survival fractions for multiple encounters - fixed Efrac}
\end{figure}
We now set $c_{\rm s} = 100$ and evaluate the survival fractions of the NFW minihalo and successive generations of Hernquist minihalos using the above mentioned procedure. Fig.~\ref{fig:survival fractions for multiple encounters - fixed Efrac} shows the survival fractions resulting from multiple energy injections into the minihalo due to successive stellar encounters. The blue curve represents the survival fraction vs $E_{\rm frac}$ due to single stellar encounter scenarios. The \rev{green (red)} curves represent multiple encounters, each characterized by an incremental \rev{$E_{\rm frac} = 0.1\ ( 7.8)$}, respectively. However, the green curve's last two incremental energy injections are characterized by an $E_{\rm frac} = 0.2$.

It can be seen that multiple encounters generate more mass loss than a single-shot encounter case of the same cumulative energy injection $E_{\rm frac}$ because of the relaxation occurring in between stellar encounters.

In Fig.~\ref{fig:survival fractions for multiple encounters - fixed Efrac}, we have fixed the incremental $E_{\rm frac}$ between successive encounters. \rev{For a large enough impact parameter, the first expression in eqn.~(\ref{eq:Efrac expression}) gives us the normalized energy injection parameter $E_{\rm frac}$.} $\alpha^2$ and $\gamma$ are functions of $c$, which depends on the density profile of the minihalo - specifically, it depends on the scale density of the minihalo. Between successive encounters of the minihalo, its density profile changes. Thus, the ratio $\frac{\alpha^2}{\gamma}$ changes. Assuming we fix $m_*$, $v_*$ and $\bar{\rho}_{\rm vir}$ between encounters, fixing $E_{\rm frac}$ necessarily means that the impact parameter $b$ changes between successive encounters.
\begin{figure}
    \includegraphics[width=\columnwidth]{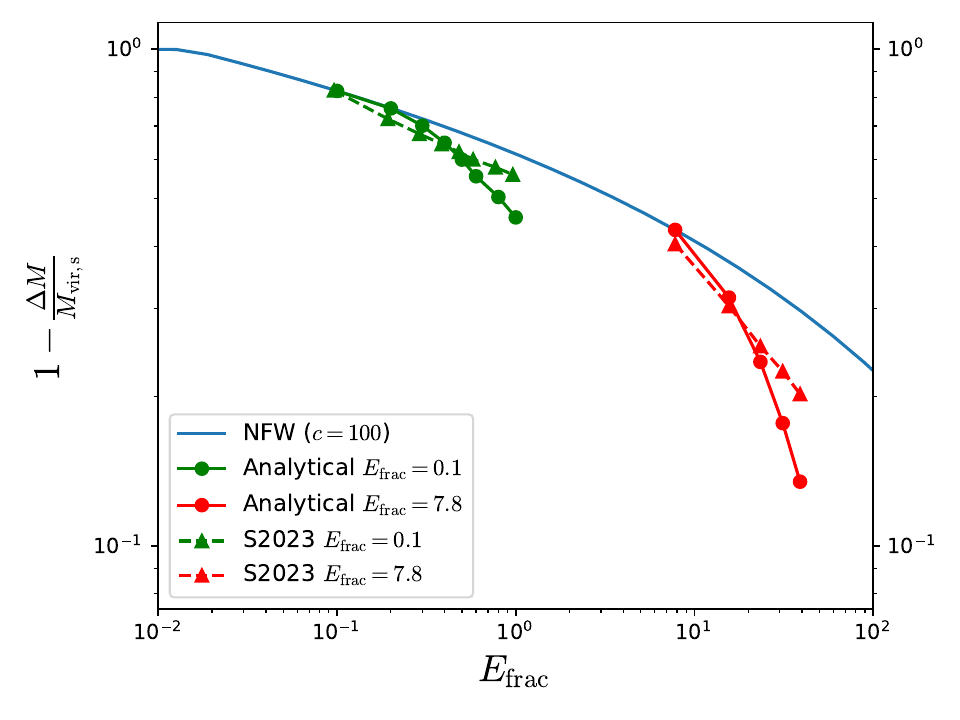}
    \caption[Survival fraction against $E_{\rm frac}$ for multiple stellar encounters: compare to S2023 results]{The survival fraction of an NFW minihalo is plotted against the total normalized injected energy $E_{\rm frac}$. The blue curve shows the case of a single stellar encounter. The green (red) curves represent the multiple stellar encounter scenario where an incremental $E_{\rm frac} = 0.1 (7.8)$ is used repeatedly. \rev{Note that for the green curve, the last two incremental $E_{\rm frac}$ were 0.2.} The solid curves with circular markers represent the output of our analytical method, while the dashed curves with triangular markers represent the output of S2023's N-body simulations.}
    \label{fig:survival fractions for multiple encounters - fixed Efrac and fixed b}
\end{figure}
In Fig.~16 of S2023, they have a graph where they use their N-body results to evaluate the survival fractions of an NFW minihalo for a multiple stellar encounter scenario. We would like to compare our results with those of S2023. They use incremental $E_{\rm frac} = 0.1, 0.1, 0.1, 0.1, 0.1, 0.1, 0.2, 0.2$ in one case and $E_{\rm frac} = 7.8, 7.8, 7.8, 7.8, 7.8$ in a second case. We will also use these values of incremental $E_{\rm frac}$ in our analytical calculations. S2023 evaluates \rev{an equivalent value of} $b$ from Eq.~(\ref{eq:Efrac expression}) for an NFW profile, given knowledge of the concentration of the NFW minihalo and the $E_{\rm frac}$ value. S2023 then fixes $b$ between successive encounters while simultaneously assuming that $E_{\rm frac}$ is held constant \rev{$(0.1 , 0.2 , 7.8)$}~\cite{Shen_personal_communication_2023}.
This approach is inaccurate since the density profile of the minihalo changes between encounters, thus changing the ratio $\frac{\alpha^2}{\gamma}$. Thus, both $E_{\rm frac}$ and $b$ can't be fixed between encounters. As we mentioned, S2023 fixes $b$ between successive encounters. To compare the output of our analytical procedure to S2023's results, we emulate this by first evaluating \rev{an equivalent value of} $b$ from Eq.~(\ref{eq:Efrac expression}) for the NFW profile, with a known $E_{\rm frac}$ \rev{$(0.1 , 0.2 , 7.8)$} and then fix that value of $b$ and evaluate the actual $E_{\rm frac}$ using Eq.~(\ref{eq:Efrac expression}) for the remaining encounters. Using the actual value of $E_{\rm frac}$ for each encounter, we analytically compute the survival fraction for that encounter. However, when plotting the results in a graph, we use \rev{$E_{\rm frac} = 0.1,0.2,7.8$} for each encounter instead of the actual value of $E_{\rm frac}$. This allows us to compare the output of our analytical procedure to the numerical simulations of S2023. Fig.~\ref{fig:survival fractions for multiple encounters - fixed Efrac and fixed b} shows the corresponding results. The blue curve represents the single encounter case. The green (red) curves represent the multiple encounter scenario with incremental $E_{\rm frac}=0.1 (7.8)$. The solid curves with circular markers represent our analytical method, while the dashed curves with triangular markers represent S2023's numerical simulations. There is a fair amount of agreement between our results and S2023's. However, the slight deviation in the two results is likely because, as is apparent from Fig.~\ref{fig:SF vs Efrac - switching}, our analytical method of computing survival fractions gives a slightly different answer compared to S2023's numerical simulations. Moreover, we assume that the successive generations of minihalos have Hernquist density profiles. However, in reality, those minihalos will have a broken power law profile with the $k$ parameter close to, but not exactly, 3 (which would be the Hernquist profile). \rev{We would like to emphasise that Fig.~\ref{fig:survival fractions for multiple encounters - fixed Efrac and fixed b} should be viewed with some reservation. This is because in attempting to compare our new analytical method with Fig.~16 of S2023, we have fixed the mean density of every generation minihalo in the multiple encounter case as a constant ($= 200\rho_{\rm crit}$) when using the expression for $E_{\rm frac}$ in eqn.~(\ref{eq:Efrac expression}). Strictly speaking, this is only accurate when considering the unperturbed NFW minihalo. Since our analytical method fixes the region of interest in which we evaluate the survival fraction to be the physical virial radius $r_{\rm vir, s}$ of the unperturbed NFW minihalo, we must really evaluate the mean density of all future generation minihalos within this same $r_{\rm vir, s}$. However, by using the mass density as $200\rho_{\rm crit}$ even for the future generation minihalos, we have effectively redefined our physical region of interest to smaller and smaller physical radii every stellar encounter we consider in the sequence, thus giving rise to slightly inaccurate values for the survival fractions. Nevertheless, this caveat only applies for Fig.~\ref{fig:survival fractions for multiple encounters - fixed Efrac and fixed b} and isn't applicable to the rest of the results in this thesis.}

\section{Empirical method of accounting for multiple encounters}
Inspired by ref.~\cite{Stucker2023}, we propose the following empirical method 
for evaluating the effective single energy injection from multiple stellar encounters:
\begin{equation}
    \label{eq:multiple encounters}
    E_{\rm frac,eff} = \left(  \sum_i E_{{\rm frac},i}^{p/2} \right)^{2/p}
\end{equation}
where $p$ is a parameter to be determined. 
For $p=2$, the value of $E_{\rm frac,eff} $ would correspond to the sum of all individual $E_{{\rm frac},i}$. For $p<2$, multiple energy injections would have an enhanced effect. \rev{The larger the value of $p$, the more is the value of $E_{\rm frac,eff}$ influenced by the strongest energy injection in the sequence.}

Ref.~\cite{Stucker2023} finds that $p=1.2$ for a prompt cusp which has a density that differs from the NFW form, following a steep $r^{-1.5}$ density profile between outer boundary set by the curvature of the initial density peak and inner core determined by the physical nature of the dark matter. 
S2023 effectively assumed $p=2$. As our results in the previous section indicate, $p<2$ would provide a better fit.   
A similar finding was obtained by ref.~\cite{Delos:2019tsl}. However,
due to the different methods and assumptions used there, it is difficult to make a precise comparison between our results.

We performed a least squares fit to find the optimal value of $p$ for the cases we investigated in Figs.~\ref{fig:survival fractions for multiple encounters - fixed Efrac}. The results are shown in Fig.~\ref{fig:survival fractions for multiple encounters - fixed Efrac - best fit p}. Our best-fit values of $p=0.8$ and $p=0.6$ 
are consistent with our earlier findings that successive encounters are more destructive than a single effective encounter, with the effective fractional energy equal to the sum of the effective fractional energies of each actual encounter.

\begin{figure}
    \includegraphics[width=\columnwidth, height=0.47\textheight,keepaspectratio]{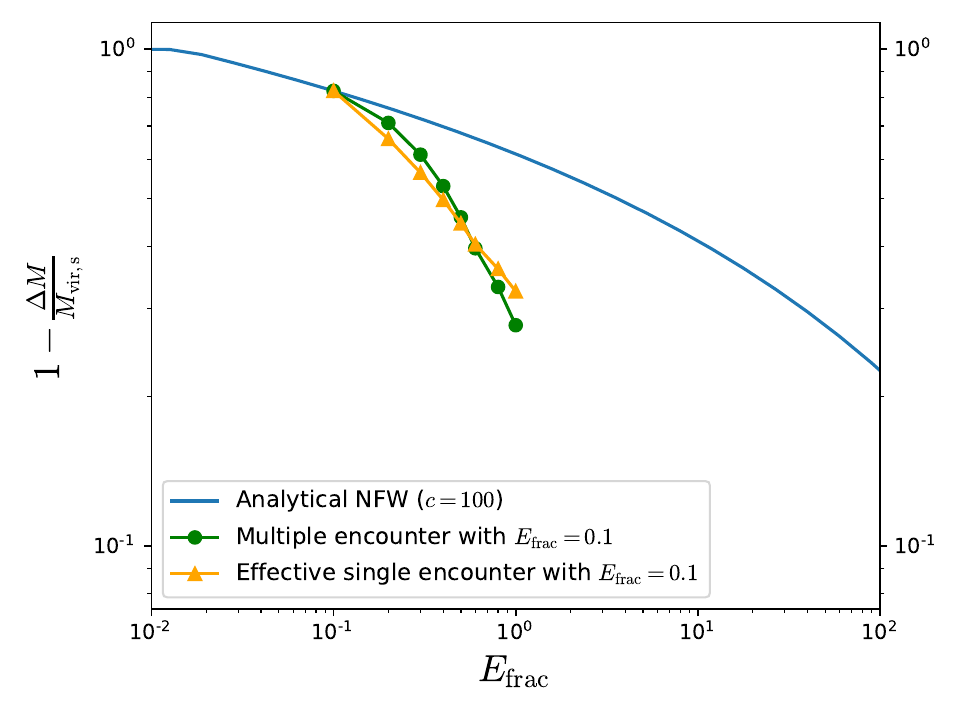}	\includegraphics[width=\columnwidth, height=0.47\textheight,keepaspectratio]{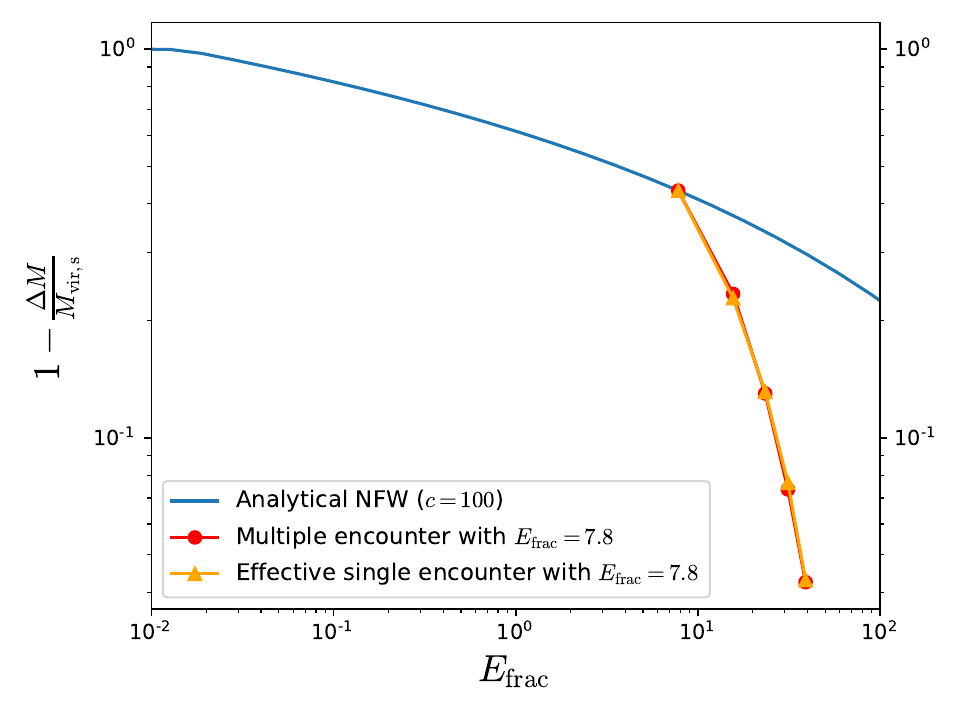}
    \caption[Curve-fitting the formula for effective energy injection in the multiple encounter scenario using the results of our model]{The survival fraction (SF) of an NFW minihalo is plotted against the total normalized energy injection $E_{\rm frac}$. The blue curve shows the case of a single stellar encounter. The green (red) curves represent the multiple stellar encounter scenario where an incremental $E_{\rm frac} = 0.1 (7.8)$ is used repeatedly. In the top and bottom panels, the orange curve is the best fit $p$ using Eq. \ref{eq:multiple encounters} for individual energy injections of $E_{\rm frac}=0.1$  and 7.8, respectively. The respective best-fit values of $p$ were 0.8 and 0.6. 
        \label{fig:survival fractions for multiple encounters - fixed Efrac - best fit p}}
\end{figure}

\section{Discussion and Conclusions}
\label{sec:Discussion_and_Conclusions}

In extending the K2021 method for evaluating minihalo mass loss due to stellar encounters, we have introduced a sequential stripping model. This model conceptually divides the minihalo into infinitesimally thin shells, sequentially focusing from the outermost to the innermost. However, this sequential approach does not imply a temporal sequence in the stripping process. Instead, it is a methodological tool for analysis. In reality, during a stellar encounter, the stripping of shells occurs simultaneously, regardless of their distance from the minihalo center.

Building on this, it is crucial to note that our model also accounts for changes in the minihalo's profile after a stellar encounter. This aspect becomes significant if the energy injection during the encounter is sufficiently large, altering the minihalo's structure and dynamics.

As shown in  Figs.~\ref{fig:error} and \ref{fig:relative error}, our new method provides a significantly better fit to S2023's N-body simulation results
compared to the K2021 method. This is particularly noticeable for $E_{\rm frac}\gtrsim 10^{-2}$, which is relevant for modeling the mass loss incurred by multiple passes through the Milky Way disk (S2023).

A significant finding of our research centers on the treatment of minihalos undergoing multiple stellar encounters. 
Note that this is for the case where the minihalo has had time to stabilize after each encounter.  As discussed in S2023, for example, this scenario may be appropriate for successive cases of the minihalo passing through the Galactic disk. Although many encounters may occur during a single passing through of the disk, these will be in such rapid succession that they can be considered one encounter where the fractional energies have been summed up to give the effective fractional energy of one encounter.

Contrary to the results presented by S2023, our analysis suggests that sequential stellar interactions lead to a more pronounced mass loss in minihalos when they are allowed to fully relax between encounters. This finding will have implications for our understanding of minihalo survival and evolution in dense stellar environments. S2023 found that by $z=0$ at the Solar System location, around 60\% of the mass 
in minihalos has survived stellar disruption from the Milky Way disk. However, as we have shown, they assume the fractional energy injections for each passage through the disk can be added to make a single effective fractional energy injection. Our results indicate that this underestimates the effects of sequential energy injections. 

\chapter[Disruption of Axion Minihalos in the Milky Way Environment]{Disruption of Axion Minihalos by Multiple Stellar Encounters in the Milky Way Environment}
\label{ch:journal_paper_2}

\graphicspath{{Journal_Paper_2/Figs/}}

\section{Introduction}
\label{sec:Introduction for journal paper 2}

This chapter extends the results of Chapter~\ref{ch:journal_paper_1} to the population of minihalos in the Milky Way Galaxy. As mentioned in Chapter~\ref{ch:journal_paper_1}, S2023 examined the disruption of axion minihalos due to stellar encounters, employing a linear addition of energy injections from multiple encounters. However, the results of Chapter~\ref{ch:journal_paper_1} indicate that this approach may underestimate the cumulative effects when minihalos have time to relax between encounters.

In this chapter, we account for the dynamical timescales of minihalos to determine whether they can relax between encounters, leading to a nonlinear addition of energy injections when appropriate. By generating a population of minihalo orbits using Monte Carlo simulations and evolving them within a model of the Galactic potential, we compute the stellar-disrupted mass function of minihalos more precisely.

This chapter follows our published article \cite{dsouza2025enhanced}. The chapter is organised as follows. In Sec.~\ref{sec: pre-infall mass function}, we discuss the pre-infall mass function of axion minihalos formed from isocurvature perturbations. Sec.~\ref{sec: Formation halos from adiabatic perturbations} describes the formation of larger adiabatic halos from adiabatic perturbations. Here, we also derive the undisrupted mass function of minihalos within adiabatic halos. The mass-concentration relationship for minihalos is presented in Sec.~\ref{sec: Mass-Concentration relationship}. In Sec.~\ref{sec: Accounting for multiple stellar encounters}, we detail our method for accounting for multiple stellar encounters, considering the minihalo's ability to relax between encounters.

We then describe our Monte Carlo simulations in Secs.~\ref{sec: Monte Carlo sampling of orbits in singular isothermal sphere} to \ref{sec: Monte Carlo Simulations to determine the stellar-disrupted mass function of minihalos}. Finally, we present our results on the stellar-disrupted mass function and discuss the implications for axion dark matter detection in Sec.~\ref{sec: computing the survival fraction of the minihalo} and give our conclusions in Sec.~\ref{sec: conclusions}.

\section{Pre-infall mass function}\label{sec: pre-infall mass function}
	 The axion isocurvature perturbations gravitationally collapse to form axion miniclusters around matter-radiation equality. These initially formed axion miniclusters undergo hierarchical mergers to form larger minihalos. The comoving number density of minihalos in a given mass range is quantified by the \textit{mass function} of minihalos.
	Ref.~\cite{X2021}, hereafter referred to as \citelink{X2021}{X2021}, performed numerical simulations to generate the mass function of such minihalos. They later fitted a modified  Sheth-Tormen formula \cite{ShethTormen99} to match the results of their simulations. We use this formula for the mass function of axion minihalos.
	
	The isocurvature growth function tells us how the isocurvature density fluctuations of axions evolve with redshift. We use the Code for Anisotropies in the Microwave Background (\texttt{CAMB}; \cite{lewis2000efficient, howlett2012cmb}) package in Python to determine this growth function. We assume a flat $\Lambda$CDM cosmology with $\Omega_{\rm m} = 0.2814$, $\Omega_\Lambda = 0.7186$, scalar spectral index $n_{\rm s} = 0.9667$, and $h = 0.697$. These are the values that \citelink{S2023}{S2023} assumed and they are consistent with the Planck 2018 Results \cite{Planck2018}. We also set $\Omega_{\rm r} = 8.6113\times10^{-5}$, $\Omega_{\rm b}h^2 = 0.0240$, $\Omega_{\rm m}h^2=0.1404$, and $\sigma_{\rm 8} = 0.796$ from the Planck 2018 results.
	We set $T_{\rm CMB} = 2.7255\mathrm{K}$ from Ref.~\cite{Fixsen2009}.
	Using \texttt{CAMB}, we can set up the initial conditions of isocurvature perturbations for cold dark matter. We initially calculate the value of the growth function relative to redshift $z=100$. \rev{Note that during the radiation-dominated epoch, the growth function has $k$-dependence, and the power spectrum changes shape as redshift changes. On the other hand, the growth function becomes $k$-independent during the matter- and dark energy-dominated epochs. Thus, we choose our reference redshift to be in one of these two epochs. For convenience, we have chosen the reference redshift of $z=100$, which is deep in the matter-dominated epoch.}
	Thus, we first calculate the power spectrum at the desired redshift and also at $z=100$. We then calculate the square of the growth function as the ratio of the value of the power spectrum at the desired redshift to the value of the power spectrum at $z=100$. When selecting the value of the power spectrum, we look at the power spectrum corresponding to small length scales (or high $k$) because it is in this regime that the power spectrum becomes scale-independent. 
	In our code, we consider the $k$ values up to $2h\,\mathrm{Mpc}^{-1}$.

	We use \texttt{CAMB} to evolve the isocurvature perturbations. See Appendix \ref{app: renormalizing the CAMB growth function} for details.
	Knowing how the isocurvature perturbations of axion dark matter evolve, we can now calculate the mass function $\frac{\mathrm{d} n_0}{\mathrm{d} M}\left(M, z\right)$ at a given redshift ($z$) and mass ($M$) using a modified Sheth-Tormen formalism. See Appendix~\ref{app: pre-infall mass function} for further details.

\section{Formation of halos from adiabatic perturbations}
\label{sec: Formation halos from adiabatic perturbations}
Sec.~\ref{sec: pre-infall mass function} detailed how axion minihalos are formed from isocurvature perturbations. On the other hand, there exist adiabatic perturbations in the axion density field in the primordial universe as well.
These perturbations collapse to form larger halos, which we refer to as {\em adiabatic halos\/}. Note that  \citelink{X2021}{X2021} and \citelink{S2023}{S2023} refer to them as {\em CDM halos\/}.
These halos generally form much later than the axion minihalos and can be host to galaxies and galaxy clusters.

We use the \texttt{hmf} package \cite{hmf}
to compute the collapse fraction of adiabatic halos, $f_{\rm adiab}$, as a function of redshift. 
We generate the redshift-dependent mass function $\mathrm{d}n^{\rm adiab}/\mathrm{d}M$  using the Press-Schecter formula \cite{PressSchecter74} and set the growth model to ``CambGrowth" and the transfer model to ``CAMB" in the \texttt{hmf} package. This mass function incorporates both the baryonic and cold dark matter. When the minihalos are in the adiabatic halos, they are predicted to freeze in their evolution due to the high virial velocities. This will only happen for adiabatic halos that are substantially more massive than the minihalos.
To account for this, \citelink{S2023}{S2023} states that they impose a minimum adiabatic mass halo of  $M_{\rm min} = 10^{-2} M_\odot$. They base their computation on the results of \citelink{X2021}{X2021}, which states that they impose $M_{\rm min} = 10^{-3} M_\odot$. However, 
they 
actually used  $M_{\rm min} = 10^2 M_\odot$ 
\cite{Xiao2023Communication}.    
To ascertain how sensitive the result is to this choice, we used two different values of $M_{\rm min}$ in all our calculations. \rev{We use $M_{\rm min} = 10^2 M_\odot$ to compare our results to S2023. On the other hand, we also use $M_{\rm min} = 10^{-2}   M_\odot$ because in the Monte Carlo simulation that follows in Sec.~\ref{sec: Monte Carlo Simulations to determine the stellar-disrupted mass function of minihalos}, the upper mass bound of the undisrupted mass function of minihalos was $10^{-3} M_\odot$. We wanted the lightest adiabatic halo to be heavier than the heaviest minihalo we consider because the adiabatic halo hosts the minihalos. Thus, the choice of $M_{\rm min} = 10^{-2}   M_\odot$. Note that the optimal value of $M_{\rm min}$ would somewhat depend on the axion mass through how suppressed the undisrupted mass function of minihalos is at high minihalo masses. In eqn.~(\ref{eq: characteristic mass of minihalo as a function of axion mass}), the characteristic mass scale $M_0$ of the minihalos is given as a function of the axion mass $m_a$. Smaller axion masses will increase $M_0$. This increases the variance $\sigma^2(M,z)$ in eqn.~(\ref{eq: sigma definition}) for a given minihalo mass $M$. This decreases $\nu(M,z)$ in eqn.~(\ref{eq: nu definition}). This means a minihalo mass of $M$ is less rare, or more abundant, for lower axion masses. Thus, the undisrupted mass function of minihalos shifts to higher minihalo masses, and we would need to increase the minimum mass of the adiabatic halo $M_{\rm min}$ in an $m_a$-dependent way. However, such a mass-dependent $M_{\rm min}$ would not significantly affect our calculations because most of the support of the mass function is around $10^{-8}M_\odot$ (as is presented in Fig.~\ref{fig: mass_function}) which is well below our fixed choice of $M_{\rm min} = 10^{-2}M_\odot$.
}

Next, we choose an upper bound for the adiabatic halos of  $M_{\rm max}=10^{20} M_\odot$ because the mass function of adiabatic halos as generated by the \texttt{hmf} package is sufficiently suppressed for masses $> 10^{20} M_\odot$, and increasing the value of $M_{\rm max}$ does not affect the value of the collapse fraction. We  calculate the collapse fraction 
using
\begin{equation}\label{eq: collapse fraction}
	f_{\rm adiab}(z) = \frac{1}{\bar{\rho}_{\rm m}} \int_{M_{\rm min}}^{M_{\rm max}} M \frac{\mathrm{d}n^{\rm adiab}}{\mathrm{d}M}(M, z) \mathrm{d}M \ ,
\end{equation}
where $M$ is the mass of the adiabatic halo and $\bar{\rho}_{\rm m}$ is the average comoving mass-density of matter (which is also equal to the average physical mass-density of matter today). We numerically perform the integral in Eq.~(\ref{eq: collapse fraction}) using Simpson's rule from the \texttt{SciPy} package.

\begin{figure}
	\includegraphics[width=\columnwidth]{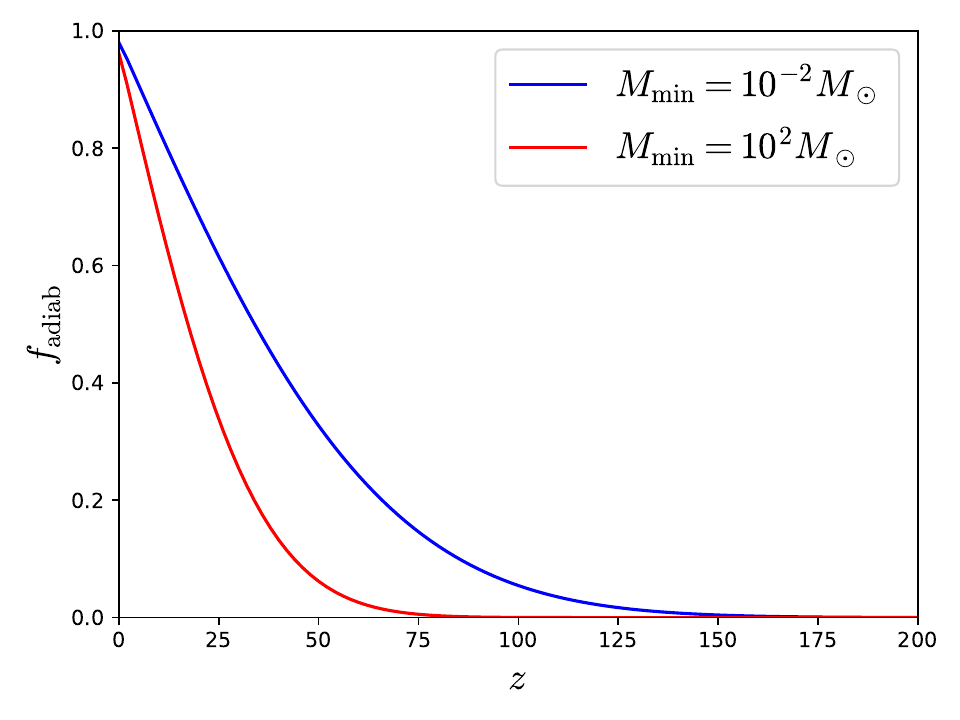}
	\caption[Collapse fraction of adiabatic halos plotted against redshift]{The collapse fraction of adiabatic halos, with masses between $M_{\rm min}$ and $M_{\rm max}=10^{20}M_\odot$, plotted against redshift. %
	}
	\label{fig: collapse fraction}
\end{figure}

We compute the collapse fraction using Eq.~(\ref{eq: collapse fraction}) for different values of redshift. Fig.~\ref{fig: collapse fraction} shows how the collapse fraction of adiabatic halos changes with redshift. 

We further numerically compute the derivative $\mathrm{d}f_{\rm adiab} / \mathrm{d}z$.
We then create an interpolation object that takes as input the redshift and outputs the value of $\mathrm{d}f_{\rm adiab} / \mathrm{d}z$ corresponding to that redshift.

\citelink{S2023}{S2023} gives the expression of the undisrupted mass function of minihalos $\mathrm{d} n_{\mathrm{f}} / \mathrm{d} M$ inside the adiabatic halos at redshift $z=0$ as: 
\begin{equation}\label{eq: undisrupted mass function}
	\frac{\mathrm{d} n_{\mathrm{f}}}{\mathrm{d} M}(M)=\int_{z_{\mathrm{eq}}}^0 \mathrm{d} z \frac{\mathrm{d} f_{\rm adiab}}{\mathrm{d} z}\left(z\right) \frac{\mathrm{d} n_0}{\mathrm{d} M}\left(M, z\right),
\end{equation}
where $\mathrm{d} n_0 / \mathrm{d} M$ is the pre-infall mass function of minihalos, $z$ is the infall redshift of the minihalos, $z_{\rm eq}$ is the redshift of matter-radiation equality.

\section{Mass-Concentration relationship}\label{sec: Mass-Concentration relationship}
We consider minihalos that have the spherically symmetric NFW density profile. Note that when the axion miniclusters initially collapse from isocurvature density fluctuations, they do not have an NFW profile. However, by the time the adiabatic halos form, these primordial miniclusters will undergo hierarchical merging to give rise to NFW density minihalos (\citelink{X2021}{X2021}). 
In this chapter, when we talk about the mass of a minihalo, we are referring to its virial mass.

For a given minihalo, the concentration, virial mass, and the redshift at which they are evaluated are related to each other \cite{lee2021probing, bullock2001profiles}. 
\citelink{S2023}{S2023} has made available \cite{ShenCode} (and presented in Fig.~1 in their article) a tabulated relationship between these quantities for three different axion masses ($m_{\rm a} = 1.25, 25, 500 \mu\mathrm{eV}$). We adopt this relationship in our calculations.

\rev{
The mass-concentration relationship is axion mass-dependent. The reason for this is that the axion mass $m_a$ sets the size of the Hubble horizon at the time $t_{\rm osc}$ at which the oscillations in the axion field $\theta$ arise, through:
\begin{equation}
	H(T_{\rm osc}) \sim m_a
\end{equation}
The Hubble horizon at $t_{\rm osc}$ sets the characteristic mass scale $M_0$ through eqn.~(\ref{eq: characteristic mass of minihalo as a function of axion mass}). $M_0$ determines the variance of the power spectrum of isocurvature perturbations of the axion through eqn.~(\ref{eq: sigma definition}). The variance then determines the rareness $\nu$ of the minihalo through eqn.~(\ref{eq: nu definition}). In the Press-Schecter theory, the redshift of collapse $z_c$ is defined as
\begin{equation}\label{eq:PS condition for redshift of collapse}
	\nu(M, z_c) \equiv \frac{\delta_{\mathrm{c}}^2}{\sigma^2(M, z_c)}= \nu_* \sim \mathcal{O}(1)
\end{equation}
Eqn.~(\ref{eq:PS condition for redshift of collapse}) fixes the value of $\sigma(M, z_c)$. We can write $\sigma(M, z_c) \equiv \sigma(M,z_c; M_0)$. Thus, for a given minihalo mass $M$, changing $M_o$ changes $z_c$. Thus, the collapse redshift $z_c$ is axion mass-dependent. Then Ref.~\cite{bullock2001profiles} gives the relation 
\begin{equation}
	c(M) \propto \frac{1+z_c(M)}{1+z_i} \ ,
\end{equation}
where $z_i$ is the redshift of infall of the minihalo into the host adiabatic halo. Thus, the mass-concentration relationships have an axion mass dependence.
}

\section{Accounting for multiple stellar encounters}
\label{sec: Accounting for multiple stellar encounters}
In Chapter \ref{ch:journal_paper_1}, we addressed the mass loss incurred by a minihalo during an interaction with a star. In this chapter, we would like to extend that analysis to the mass loss incurred by a population of minihalos in the  
Milky Way galaxy.

As mentioned in Chapter \ref{ch:journal_paper_1}, each time a minihalo has a stellar encounter, there is an associated energy injection, which we quantify as 
\begin{equation}\label{eq: E_frac definition}
	E_{\rm frac} \equiv \frac{\Delta E}{E_{\rm bind}}\ ,
\end{equation}
\rev{where the expression for $E_{\rm frac}$ in terms of the parameters of the star, minihalo and the encounter itself is given in eqn.~(\ref{eq:Efrac expression}).}

Let's say there are $N$ stellar encounters by a minihalo. Thus, we have $N$ single encounter events, which we would now like to approximate as an effective single encounter event with an effective energy injection parameter $E_{\rm frac, eff}$, such that the survival fraction of the minihalo, in either case, is the same. Equation (\ref{eq:multiple encounters}) is a formula for such an $E_{\rm frac, eff}$.

When two stellar encounters happen in quick succession such that the minihalo does not have enough time to gravitationally relax in between encounters, the minihalo is unable to tell if it has been subjected to two stellar encounters of known energy injection parameters or a single stellar encounter with a higher energy injection parameter. In such a case, what does get added up linearly is the total energy $\Delta E$ injected into the minihalo within its virial radius. Thus, the effective energy injection parameter corresponding to these two single stellar encounters is
\begin{align}\label{eq: effective E_frac for two encounters when relaxation is not allowed in between encounters}
	E_{\rm frac,eff} &= \frac{\Delta E_1 + \Delta E_2}{E_{\rm bind}} \nonumber\\
	&= E_{\rm frac,1} + E_{\rm frac,2} \ ,
\end{align}
where $E_{\rm frac,i} \equiv \Delta E_i / E_{\rm bind}$, for $i=1,2$. Note that in the first equality of Eq.~(\ref{eq: effective E_frac for two encounters when relaxation is not allowed in between encounters}), we did not have a binding energy term separately for each encounter. This is because since the minihalo's density profile does not have time to significantly change in between encounters, its density profile just before the start of either encounter is approximately the same. If the density profile is the same, so should the binding energy of the minihalo. It is precisely this fact that leads us to the conclusion that the energy injection parameters are added linearly when the minihalo does not have time to change in between encounters. Comparing Eq.~(\ref{eq: effective E_frac for two encounters when relaxation is not allowed in between encounters}) to Eq.~(\ref{eq:multiple encounters}), we see that $p=2$ when the minihalo's density profile does not have enough time to relax in between encounters.

On the other hand, consider the scenario where we have two stellar encounters with a large amount of time in between encounters. The minihalo is able to completely gravitationally relax in between the encounters. Now, we are no longer able to use the method in Eq.~(\ref{eq: effective E_frac for two encounters when relaxation is not allowed in between encounters}) because the binding energy of the minihalo just before each encounter is different. The amount by which it is different will depend on the energy injected into the minihalo during the first encounter as this will change the density profile of the relaxed minihalo and hence the binding energy of the minihalo just before the second encounter. To address this, in Chapter \ref{ch:journal_paper_1}, we parameter-fitted the value of $p$ in  Eq.~(\ref{eq:multiple encounters}) to various such multiple encounter cases. We found that $p \lesssim 1$ in this case. The smaller the value of $p$ in Eq.~(\ref{eq:multiple encounters}), the larger is the value of $E_{\text {frac}, \text {eff}}$ for fixed $E_{\text {frac}, i}$. To be on the conservative side, we choose $p=1$. \rev{When we say \textit{conservative}, we mean that we are under-predicting the effect that the gravitational relaxation of minihalos in between stellar encounters has on the survival fraction, compared to the S2023 result that did not consider this gravitational relaxation effect.} Thus, using Eq.~(\ref{eq:multiple encounters}) for our two encounter case with complete minihalo relaxation in between encounters, the effective energy injection parameter is given by
\begin{equation}
	E_{\rm frac,eff} =  \left( E_{\rm frac,1}^{1/2} + E_{\rm frac,2}^{1/2} \right)^2
\end{equation}

To add up the energy injection parameters corresponding to two consecutive stellar encounters, we must choose whether we want to add them with $p=1$ or $p=2$. This would depend on how time elapses between the two encounters. To be accurate, we must define what it means to have a small versus a large time in between the two encounters. To do this, we first define the dynamical time of the minihalo. This is a quantitative measure of how fast the minihalo is able to relax. The dynamical time chosen could also be called the crossing time. It is defined as the time taken by a small test particle that is released from the surface of the minihalo under the influence of the gravitational potential of the minihalo, to reach the center of the minihalo. This effectively tells us how fast perturbations on the surface of the minihalo propagate through its volume. For simplicity, we consider a homogeneous sphere (instead of the NFW profile of the minihalo) to calculate this crossing time because the crossing time is independent of the radius from which the particle is released in this case. Thus, the dynamical time of the minihalo is given by \cite{binneyTremaine}
\begin{equation}\label{eq: dynamical time}
	t_{\mathrm{dyn}}=\sqrt{\frac{3 \pi}{16 G \bar{\rho}_{\mathrm{vir}}}}
\end{equation}

We compare the time between consecutive stellar encounters to the dynamical time of the minihalo. If the time between consecutive encounters is smaller than the dynamical time, then the minihalo does not have enough time to relax and we add up the energy injection parameters  linearly with $p=2$. On the other hand, if the time between consecutive stellar encounters is more than the dynamical time, then the minihalo has enough time to relax and we add up the energy injection parameters non-linearly with $p=1$. We refer to this approach as the {\em hybrid method\/}.

\section{Generating a population of minihalo orbits in the Milky Way galaxy}\label{sec: Monte Carlo sampling of orbits in singular isothermal sphere}

The energy injected into a minihalo during its lifetime in the galaxy can vary depending on the actual orbit of the minihalo since the stellar density varies with position in the galaxy. Thus, we look into generating a large number of orbits and calculate the energy injected into minihalos for each of those orbits. In modeling the population of minihalos, we assume that the minihalos are distributed in the galaxy according to the density profile of a singular isothermal sphere given by the spherically symmetric density profile and potential (\citelink{S2023}{S2023}):   
\begin{equation}
\label{eq:singular isothermal sphere}
	\rho(r)=\frac{V_{\mathrm{C}}^2}{4 \pi G r^2}, \quad \Phi(r)=V_{\mathrm{C}}^2 \ln \left(\frac{r }{ r_0}\right),
\end{equation}
where $V_{\mathrm{C}} = 200\, \mathrm{km / s}$ is the constant circular velocity of the singular isothermal sphere - any minihalo in a circular orbit around the galaxy's center will have this speed, no matter what radius it orbits at:  this emulates the flat rotation curves in our galaxy at sufficiently large radii. \rev{The value of $V_{\mathrm{C}} = 200\, \mathrm{km / s}$ was chosen since S2023 employed this value, and we wanted to compare our results to theirs.} Also, $r_0 = 10\, \mathrm{kpc}$ is the reference radius of zero potential. In actuality, we expect the minihalos to be distributed around the galaxy's center according to an NFW profile. However, the phase space distribution of an NFW profile does not have an analytical form, making it difficult to randomly draw orbits with a given total energy and angular momentum. \rev{Note that when employing the sampling method of Ref.~\cite{van1999substructure}, as we have done, one needs an analytical expression for the phase space distribution function of the chosen underlying density profile. This is because in this sampling method, one computes an analytical expression for the joint probability distribution function (PDF) of the energy and angular momentum of the orbit using the analytical expression for the phase space distribution function. It is then from this joint PDF that the energies and angular momenta of the sample orbit are drawn. In principle, even if one doesn't have an analytical expression for the phase space distribution function, one can approximate the coefficients of an analytic curve that is fitted to the observational or numerical data, and then use this expression to compute the joint PDF of the energy and angular momentum. In the absence of either method, one has to resort to other methods, like Eddington inversion, using the density profile itself.} Thus, as done, for example, by \citelink{S2023}{S2023}, we use the singular isothermal sphere.

We use the Monte Carlo sampling procedure outlined in the appendix of Ref.~\cite{van1999substructure} to sample orbits from a singular isothermal sphere density distribution. We sample such that the minihalo is present in the solar neighborhood today.
This is done by setting the galactocentric radius of observation to be $r_{\rm obs}=8\, \mathrm{kpc}$. 
We assume the phase space distribution function is isotropic. Hence, the term $h(\eta)$ in that article is set to 1. \footnote{\rev{There is a typo in the Monte Carlo procedure outlined in the appendix of Ref.~\cite{van1999substructure}. The condition $\mathcal{R}_4 > P_{\text{comp }}\left(\eta_{\text{try }}\right) / P\left(\eta_{\text{try }}\right)$ should be $\mathcal{R}_4  >   P\left(\eta_{\text{try }}\right) /  P_{\text{comp }}\left(\eta_{\text{try }}\right) $
\cite{vandenBoschCommunication}.}}
We evaluate the total energy per unit mass $E$ of a randomly drawn orbit as: 
\begin{equation}
	E = \Phi(r_{\rm obs}) - V_{\mathrm{C}}^2 \ln(1 - \mathcal{R}_1)\ ,
\end{equation}
where $\mathcal{R}_1$ is a random number in the interval [0, 1] - see Ref.~\cite{van1999substructure} for details. 
The angular momentum per unit mass $L$ of the randomly drawn orbit is given by: 
\begin{equation}
	L = \eta r_{\rm C}(E) V_{\mathrm{C}}\ ,
\end{equation}
where $\eta$ is called the orbital circularity (related to the orbital eccentricity) and $r_{\rm C}(E)$ is the radius of a circular orbit that has given energy per unit mass $E$. It is given by (\citelink{S2023}{S2023}): 
\begin{equation}
	r_{\rm C}(E)=r_0 \exp \left(\frac{E}{V_{\mathrm{C}}^2}-\frac{1}{2}\right)\, .
\end{equation}

From here on out, we will refer to $E$ and $L$ as just the energy and angular momentum, respectively, of the orbit, but note that they are actually the energy per unit mass and angular momentum per unit mass, respectively.

\section{Initial conditions to evolve an orbit}\label{sec: initial conditions to evolve an orbit}
In the previous section, we sampled orbits from a singular isothermal sphere density distribution such that the minihalos in those orbits are present in the solar neighborhood today. We found the energies and angular momenta of those orbits. We now have to evolve those orbits backward in time to find out the past positions of those minihalos since the stellar density distribution depends on the position, and we want to estimate the mass disruption effects of the stellar population on each minihalo. To do this, we need to convert the energy $E$ and angular momentum magnitude $L$ to the initial state parameters of the minihalo. The state parameters here are the position and velocity vectors of the minihalo with respect to the galactic frame. 
We use the definition of angular momentum of the orbit as: 
\begin{equation}\label{eq: definition of L}
	{ L} = { r}_{\rm obs} \, { v}_{\mathrm{init,}\perp}\ ,
\end{equation}
where $v_{\mathrm{init,}\perp}$ is the magnitude of the initial velocity vector of the minihalo projected onto the plane perpendicular to the radial direction (the line connecting the Galactic center and the minihalo) at the initial position of the minihalo. Using Eq.~(\ref{eq: definition of L}), we can compute the value of $v_{\mathrm{init,}\perp}$. Next, we compute the magnitude of the initial velocity vector using the definition of the total energy $E$ as: 
\begin{equation}\label{eq: definition of E}
	E = \Phi(r_{\rm obs}) + \frac{1}{2}v_{\rm init}^2\ ,
\end{equation}
where $v_{\rm init}$ is the magnitude of the initial velocity vector of the minihalo. Eq.~(\ref{eq: definition of E}) allows us to compute the value of $v_{\rm init}$. Next, we use the vector addition relation: 
\begin{equation}\label{eq: velocity vector addition}
	v_{\rm init}^2 = v_{\mathrm{init,}\parallel}^2 + v_{\mathrm{init,}\perp}^2\ ,
\end{equation}
where $v_{\mathrm{init,}\parallel}$ is the magnitude of the radial component of the initial velocity vector of the minihalo. Using Eq.~(\ref{eq: velocity vector addition}), we can calculate the value of $v_{\mathrm{init,}\parallel}$.

We still need two more pieces of information to fix the initial velocity vector. We first fix the galactic Cartesian coordinate system such that the Sun is located along the positive $X$-axis and the Galactic disk is in the $X-Y$ plane as shown in Fig.~\ref{fig:geometry}. Then, the vector corresponding to $v_{\mathrm{init,}\parallel}$ will be along the $X$-axis. Next, the vector corresponding to $v_{\mathrm{init,}\perp}$ lies in a plane parallel to the $Y$-$Z$ plane of the galactic coordinate system. Assuming this vector makes an angle $\theta$ with the positive $Z$-axis, we can decompose the vector along the $Z$-axis as $v_{\mathrm{init,}\perp} \cos(\theta)$ and along the $Y$-axis as $v_{\mathrm{init,}\perp} \sin(\theta)$. Now come two key steps. We choose the value of $\theta$ randomly in the interval $[0, 2\pi)$. The value of $\theta$ determines the inclination angle between the plane of orbit and the Galactic disk (which lies along the $X-Y$ plane). Furthermore, we choose the direction of the vector corresponding to $v_{\mathrm{init,}\parallel}$ randomly to be either along the positive or negative $X$-axis. With these two steps, we uniquely determine the orbit. Since our minihalo is randomly chosen to move clockwise or counter-clockwise, we are free to evolve our minihalo backward or forward in time. 
\rev{This is because, since we have a time-independent galactic potential, there exists time-reversal symmetry.}
Evolving our minihalo backward in time with a given initial velocity vector is equivalent to evolving our minihalo forward in time with the initial velocity vector flipped in its direction, i.e., $t \to -t$ corresponds to $\vec{v}_{\mathrm{init}} \to -\vec{v}_{\mathrm{init}}$. In our code, we choose to evolve our minihalo forward in time using the singular isothermal potential given in Eq.~\ref{eq:singular isothermal sphere}. The details of how we did this are given in Appendix~\ref{app: Code to evolve an orbit}.
\begin{figure}
    \centering
\begin{tikzpicture}
        \begin{scope}[viewport={160}{15}, very thin, rotate around z=-45,scale=2]

            \draw[dashed] (\ToXYZr{0.5}{90}{45}) -- (\ToXYZr{0.78}{90}{45}); 
            \draw[dashed,color=gray, opacity=0.9] (\ToXYZr{0.78}{90}{45}) -- (\ToXYZr{1}{90}{45}); 
            \draw[dashed,->] (\ToXYZr{1}{90}{45}) -- (\ToXYZr{1.7}{90}{45}) node[anchor=north]{$X$}; 
            
            \draw[dashed,->] (\ToXYZr{0.5}{90}{45}) -- (-0.56568542, 1.27279221, 0) node[anchor=north]{$Y$};
            \draw[dashed,->] (\ToXYZr{0.5}{90}{45}) -- (0.35355339, 0.35355339, 0.75 ) node[anchor=south]{$Z$};
            \node at (0.36769553, 0.36769553, -0.05) {$\mathcal{O}$};

            \coordinate (A) at (0.28284271, 1.13137085, 0.375);  
            \coordinate (B) at (0.28284271,  1.13137085, -0.375);  
            \coordinate (C) at (1.13137085,  0.28284271, -0.375);  
            \coordinate (D) at (1.13137085, 0.28284271, 0.375);  
            
            \draw[thick] (A) -- (B) -- (C) -- (D) -- cycle;

            \node at (\ToXYZr{1}{90}{45}) {\Large $\star$};

            \coordinate (Sun) at (\ToXYZr{1}{90}{45});
            

            \coordinate (velocity_vector) at (0.80961941, 1.30459415, 0.35); 
            \coordinate (velocity_vector_parallel) at (1.05710678, 1.05710678, 0.);
            \coordinate (velocity_vector_perpendicular) at (0.45961941, 0.95459415, 0.35);
            
            \coordinate (vertical_dashed_line) at (0.70710678, 0.70710678, 0.35);

            \draw[thick, ->, blue] (Sun) -- (velocity_vector_parallel) node[anchor=north, xshift=3pt, yshift=0pt] {$\vec{v}_{\mathrm{init,}\parallel}$};
            \draw[thick, ->, darkgreen] (Sun) -- (velocity_vector_perpendicular) node[anchor=north, xshift=6pt, yshift=-10pt] {$\vec{v}_{\mathrm{init,}\perp}$};
            \draw[thick, ->, red] (Sun) -- (velocity_vector) node[anchor=south, xshift=-6pt, yshift=-2pt] {$\vec{v}_{\rm init}$};
            
            \draw[dash pattern=on 0.6pt off 1pt, line width=0.6pt, black] (velocity_vector) -- (velocity_vector_parallel);
            \draw[dash pattern=on 0.6pt off 1pt, line width=0.6pt, black] (velocity_vector) -- (velocity_vector_perpendicular);

            \draw[dashed, black] (Sun) -- (vertical_dashed_line);
            
            \coordinate (top_of_plane) at (\ToXYZr{1.3}{90}{45}); 

            \draw[dotted] (\ToXYZr{1}{90}{45}) -- (top_of_plane);

            \pic[draw, <-, angle radius=4mm, angle eccentricity=1.2]{angle=velocity_vector_perpendicular--Sun--vertical_dashed_line};
            \node at ($(Sun)!0.7!(velocity_vector_perpendicular)$) [xshift=-2mm, yshift=-1mm] {$\theta$};
        \end{scope}
    \end{tikzpicture}
    \caption[Orientation of a minihalo's initial velocity vector in the Galactic frame]{Orientation of a minihalo's initial velocity components. The Cartesian coordinate system is centered on the Galactic center ($\cal O$), with the Sun ($\star$) located along the positive $X$-axis and the Galactic disk in the $X-Y$ plane. The velocity vector \( \vec{v}_{\text{init}} \) is decomposed into a radial (\( \vec{v}_{\text{init}, \parallel} \)) component and a perpendicular (\( \vec{v}_{\text{init}, \perp} \)) component which lies on a plane perpendicular to the $X$-axis. The angle \( \theta \) determines the inclination of the perpendicular velocity relative to the $Z$-axis.}
    \label{fig:geometry}
\end{figure}
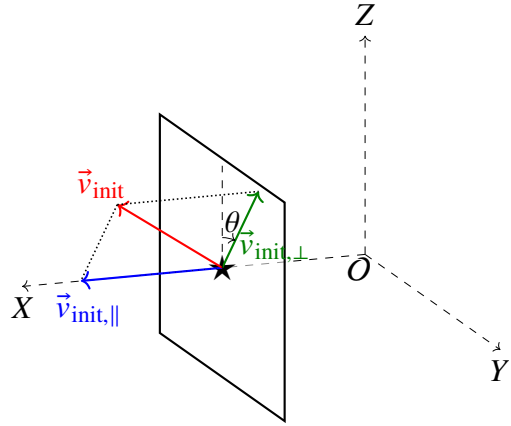

\section[Monte Carlo Simulations for stellar-disrupted mass function]{Monte Carlo Simulations to determine the stellar-disrupted mass function of minihalos}\label{sec: Monte Carlo Simulations to determine the stellar-disrupted mass function of minihalos}

In Section~\ref{sec: Formation halos from adiabatic perturbations}, we theoretically outlined how to determine the undisrupted mass function of minihalos. In this section, we follow a Monte-Carlo approach similar to that taken by \citelink{S2023}{S2023} to determine both the undisrupted mass function and the stellar-disrupted mass function of minihalos. However, here we take into account the potentially more destructive nature of multiple stellar encounters discussed in Sec.~\ref{sec: Accounting for multiple stellar encounters}. 

\subsection{Mass-Redshift Grid}\label{sec: mass-redshift grid}
We start off by creating a two-dimensional grid of virial mass ($M$) of minihalos and the infall redshift ($z$) at which the minihalos fell into their respective adiabatic halos. Thus, each grid point is associated with an ordered pair of $(M_i , z_j)$. We assign a randomly generated orbit to each grid point using the procedure in Section~\ref{sec: Monte Carlo sampling of orbits in singular isothermal sphere}. We choose the values of $M_i$ to be logarithmically spaced in the interval $[10^{-14}, 10^{-3}] M_\odot$. We choose the values of $z_j$ to be logarithmically spaced in $(1 + z)$ such that $z$ is in the range $[0, 150]$. We select 1,000 values of $M_i $ and 1,001 values of $z_j$ in those ranges.

 We approximate the integral in Eq.~(\ref{eq: undisrupted mass function}) by
\begin{equation}
	\label{eq: weight sum}
	\left. \frac{\mathrm{d} n_{\mathrm{f}}}{\mathrm{d} M}\right|_{M_i}=\sum_{j} w_{ij},
\end{equation}
where
\begin{equation}
	\label{eq: weights}
	w_{ij} = \delta z_{j} \frac{\mathrm{d} f_{\mathrm{adiab}}}{\mathrm{d} z}\left(z_{j}\right) \frac{\mathrm{d} n_0}{\mathrm{d} M}\left(M_i, z_{j}\right)\ ,
\end{equation}
and $\delta z_{j}=z_{j+1}-z_j$. Note that the value of $\delta z_{j}$ will vary with the $j$ considered since the $z$ values are logarithmically spaced. We will have one less number of $\delta z_{j}$ values than the number of $z_j$ values. Thus, the weight $w_{ij}$ is evaluated using 1,000 values of $M_i$, and for each value of $M_i$, there will be 1,000 (not 1,001) values of $z_j$. Thus, we will have one million weights to evaluate.

\rev{The dynamical timescale of any minihalo in this 2D $(M_i , z_j)$ grid is given by eqn.~(\ref{eq: dynamical time}), where $\bar{\rho}_{\mathrm{vir}} \equiv 200 \rho_{\rm crit}(z_j)$. Thus, the dynamical times of the minihalos in our Monte Carlo simulation range from $\approx 2.2 \rm Gyr$ at $z=0$ to $\approx 2.2\times 10^{-3} \rm Gyr$ at $z=150$.}

\subsection{Energy Injection due to Stellar Interactions}

Each minihalo in our simulation will undergo some mass loss during its lifetime in our galaxy due to tidal interactions with the Galactic disk's stars. 
For a single pass through the Galactic disk, the time between individual stellar encounters will be much less than $t_{\rm dyn}$, so the energy injections can be added linearly, which means $p=2$ in Eq.~\ref{eq:multiple encounters}.
\citelink{S2023}{S2023} estimates the injected energy for a passage through the disk to be
\begin{equation}\label{eq: E_frac for continuous stellar distribution}
	E_{\rm frac} =\frac{G m_\kappa \Sigma_*}{\sigma_*^2+v_{\mathrm{mh}}^2} \frac{\alpha^2(c)}{\gamma(c) \bar{\rho}_{\mathrm{vir}}(z)} \frac{2}{b_{\mathrm{s}}^2(c)+2 b_{\mathrm{C}}^2(\Sigma_*)}\ .
\end{equation}

As discussed in Section~\ref{sec: Mass-Concentration relationship}, given $M$ and $z$ one can work out the value of $c$.
The characteristic mass associated with the present-day mass function of stars in our Milky Way galaxy is $m_\kappa \sim 0.6 M_\odot$. The stellar surface density $\Sigma_*$ is obtained by integrating the stellar volume density $\rho_*$ along the one-dimensional trajectory of the minihalo within the galaxy as follows: 
\begin{equation}\label{eq: Sigma_* general formula}
	\Sigma_* = \int\limits_{\rm traj,\ s.d.p.} \rho_* \mathrm{d}l = \int\limits_{\rm traj,\ s.d.p.}\rho_* v \, \mathrm{d}t \ ,
\end{equation}
where ``traj'' represents the trajectory of the minihalo and \rev{``s.d.p.'' refers to a single disk pass, since for our problem, we apply eqn.~(\ref{eq: Sigma_* general formula}) to only a single disk pass.} $\mathrm{d}l$ represents the infinitesimal distance travelled by the minihalo along its trajectory, $v$ represents the instantaneous velocity magnitude of the minihalo relative to the Galactic disk, which we will assume to have static density distribution over the lifetime of the galaxy, and $\mathrm{d}t$ represents the infinitesimal time increment. The internal (to the galaxy) one-dimensional velocity dispersion of stars in the galaxy is denoted $\sigma_*$. \rev{The stars are assumed to have a Maxwell-Boltzmann speed distribution function, with the stellar velocity distribution being Gaussian}. Note that although the stars are moving in the galaxy, they do so as to keep the stellar volume density static on scales significantly larger than the average inter-stellar distance within the galaxy.  The variance of the velocity of the minihalo relative to the rest frame of the galaxy is denoted $v_{\rm mh}^2$. 
The term $\sigma_*^2+v_{\mathrm{mh}}^2$ has a weak dependence on the minihalo orbits and its value is assumed to be $(250\, \mathrm{km/s})^2$ in Eq.~(\ref{eq: E_frac for continuous stellar distribution}). \rev{Note that since the stellar disk contains stars that are revolving around the galactic center, one might naively expect the relative minihalo-disk velocity to appear in some form in eqn.~(\ref{eq: E_frac for continuous stellar distribution}). This relative velocity will depend on the angle between the minihalo's velocity and the net rotation velocity of the stars in the disk. However, this angle does not appear in eqn.~(\ref{eq: E_frac for continuous stellar distribution}). This is because, in deriving the expression for energy injection due to a disk pass, although they don't mention it, S2023 effectively works in the local frame where the net rotation of the disk has velocity $\vec{v}_{\rm rot} = 0$. Thus, the Gaussian distribution function of the stellar velocities is given as
\begin{equation}
	f(\vec{v}_*) \propto  \exp \left(-\frac{v_*^2}{2 \sigma_*^2}\right) \ ,
\end{equation} 
instead of 
\begin{equation}\label{eq:stellar velocity distribution when v_rot is not 0}
	f(\vec{v}_*; \vec{v}_{\rm rot}) \propto  \exp \left(-\frac{|\vec{v}_* - \vec{v}_{\rm rot}|^2}{2 \sigma_*^2}\right) \ .
\end{equation} 
Since $ f(\vec{v}_*)$ depends only on the magnitude $v_*$, stellar velocities are isotropic in this local frame, and the distribution $f$ is characterised by a single 1-D speed dispersion $\sigma_*$. Let's say we work in a spherical coordinate system where the minihalo is moving along the $Z$-axis -- for the purpose of this calculation, this $Z$-axis is not necessarily the same as the $Z$-axis of the Galactic coordinate system. Then, let $\theta_{\rm *,mh}$ be the angle between the star's and the minihalo's velocity vectors. S2023 then performs the following integral
\begin{equation}\label{eq:stellar velocity integral}
    \int \mathrm{d}^3\vec{v}_* f(\vec{v}_*)\frac{1}{|\vec{v}_* - \vec{v}_{\rm mh}|^2} \ .
\end{equation}
Then, $\theta_{\rm *,mh}$ is essentially the polar angle in this spherical coordinate system, and is part of the volume element $\mathrm{d}^3\vec{v}_*$ in eqn.~(\ref{eq:stellar velocity integral}). Thus, it gets integrated out and doesn't show up in the final expression for the energy injection. More importantly, the angle between $\vec{v}_{\rm mh}$ and $\vec{v}_{\rm rot}$ does not appear in the final expression. However, if we work in a frame where $\vec{v}_{\rm rot}\neq 0$, and we define a new velocity variable $\vec{u} \equiv \vec{v}_* - \vec{v}_{\rm rot}$, the integral in eqn.~(\ref{eq:stellar velocity integral}) would become
\begin{equation}\label{eq:stellar velocity integral when v_rot is not 0}
\int \mathrm{d}^3\vec{u}\, f(\vec{u})\,\frac{1}{\left|\vec{u}-\left(\vec{v}_{\mathrm{mh}}-\vec{v}_{\mathrm{rot}}\right)\right|^{2}} \ .
\end{equation}
Now, the cosine of the angle between $\vec{v}_{\rm mh}$ and $\vec{v}_{\rm rot}$ will appear through the magnitude $\left|\vec{v}_{\mathrm{mh}}-\vec{v}_{\mathrm{rot}}\right|$ when expanding out the denominator in eqn.~(\ref{eq:stellar velocity integral when v_rot is not 0}) and will end up in some form in the final expression for the energy injection.
}

\rev{
In eqn.~(\ref{eq: E_frac for continuous stellar distribution}), the stellar disk has a smooth surface density $\Sigma_*$, which is a consequence of the underlying stellar volume density $\rho_*$ of the disk being smooth. This requires $\rho_*$ to be approximated as a fluid, not discrete particles. This continuum approximation is only valid if the \emph{expected} number of stars in the region of impact parameters that dominates the energy injection is large. As a rough estimate, consider a cylinder of radius $b$ centred on the minihalo trajectory, with the axis of the cylinder along the minihalo trajectory. The expected number of ``typical'' stars of mass $m_\kappa$ that give rise to the stellar surface density is
\begin{equation}
	N_{\rm eff}(b) \;\sim\; \frac{\Sigma_*}{m_\kappa}\,\pi b^2.
\end{equation}
When $N_{\rm eff} \gg 1$, the stellar distribution can be considered to be effectively smooth for the purposes of computing the expression of the energy injection. As a rough calculation, a cylinder of radius $b_c$ will on average contain only one star when the condition
\begin{equation}
	\pi b_c^2 \Sigma_* \sim m_\kappa \ ,
\end{equation}
is true. Now, if the transition radius $b_{\rm s} \ll b_c$, then in the vicinity of the minihalo, there is one average less than one star. Thus, $\rho_*$ cannot be approximated as a continuous fluid. This is the problem of what is usually called the shot noise. Hence the shot noise parameter $b_C$ was introduced by S2023 to account for the scenario when the stellar volume density of the disk cannot be approximated as a continuous fluid. Thus,
\begin{equation}\label{eq: b_c definition}
	b_{\rm C}\sim\sqrt{\frac{m_\kappa }{ \pi \Sigma_* }}\,.
\end{equation}
}

\subsection{Computing the Stellar Surface Density and Time-stamp of each Effective Disk Pass}

The Milky Way's disk consists of a thick disk and a thin disk. We model the stellar volume density in cylindrical coordinates with the Galactic center at the origin \cite{mcmillan2011mass}: 
\begin{equation}\label{eq: rho_* expression}
	\rho_*(R, Z)= \sum\limits_{d = {\rm thin, thick}} \frac{\Sigma_{\mathrm{d}, 0}}{2 Z_{\mathrm{d}}} \exp \left(-\frac{|Z|}{Z_{\mathrm{d}}}-\frac{R}{R_{\mathrm{d}}}\right) \ ,
\end{equation}
where $R, Z$ are the cylindrical coordinates of the galaxy.  The scale length is denoted $R_{\mathrm{d}}$, and $Z_{\mathrm{d}}$ is the scale height. They tell us how fast the stellar volume density falls off in the plane of the Galactic disk and perpendicular to the Galactic disk, respectively. The central surface density is denoted $\Sigma_{\mathrm{d}, 0}$. If we integrate out the $Z$-component of $\rho_*$, we will get the stellar surface density of the Milky Way as a function of $R$, and $\Sigma_{\mathrm{d}, 0}$ will be the stellar surface density at $R=0$. The parameters of Eq.~(\ref{eq: rho_* expression}) are given in Table~\ref{tab: paramters of rho_* expression}.
The $R$ in Eq.~(\ref{eq: rho_* expression}) is calculated as $R = \sqrt{X^2 + Y^2}$, where the $X$ and $Y$ axes define the plane of the Galactic disk. \rev{Note that we haven't included the stellar bulge in our disk model, which will be relevant for minihalos on highly eccentric radial orbits. According to the leftmost panel of Fig.~17 of S2023, the PDF($e$) $\to 0$ when $e \to 1$, where $e$ is the eccentricity of the orbit. So, radial orbits will be vastly under-represented in the population of minihalo orbits in our Monte Carlo simulation. So, including the bulge may not affect our results significantly.}

\begin{table}
	\begin{tabular}{|l|c|c|}
		\hline
		&	Thin disk & Thick disk \\
		\hline
		$\Sigma_{\mathrm{d}, 0}\ [M_\odot/\mathrm{pc}^2]$ & 816.6 & 209.5 \\
		\hline
		$R_{\mathrm{d}}\ [\mathrm{kpc}]$ & 2.9 & 3.31\\
		\hline
		$Z_{\mathrm{d}}\ [\mathrm{kpc}]$ & 0.3 & 0.9\\
		\hline
	\end{tabular}
	\caption
    [Parameters of the galactic stellar volume density model are tabulated for the thin and thick disks of the MW]
    {
		\label{tab: paramters of rho_* expression}
		The parameters of Eq.(\ref{eq: rho_* expression}) are presented for both the thin and thick Galactic disks of the Milky Way galaxy. The parameter $\Sigma_{\mathrm{d}, 0}$ is the central surface density, $R_{\mathrm{d}}$ is the scale length, and $Z_{\mathrm{d}}$ is the scale height.
	}
\end{table}

Given the initial state parameters (position and velocity vectors) of a minihalo, we use the \texttt{lbparticles} code \cite{forbes2025semi} (see Appendix~\ref{app: Code to evolve an orbit}) to evaluate the position and velocity of the minihalo at certain discrete times in the future. Knowledge of the position vector at any instant of time allows us to evaluate the local stellar volume density $\rho_*$ at that time via Eq~(\ref{eq: rho_* expression}). Thus, at any given time, the integrand $\rho_* v$ (in Eq.~(\ref{eq: Sigma_* general formula})) corresponding to that time can be evaluated. The top panel of Fig.~\ref{fig: integrand vs time - regular} shows a sample minihalo in orbit around the Galactic center. The integrand $\rho_* v$ is plotted against time. We notice that the plot has local maxima and minima. The local maxima are denoted by a circular red marker, and in general correspond to the instant when a minihalo passes through the Galactic disk ($Z=0$). They are local maxima because the stellar volume density is highest at $Z=0$ for any given $R$ in Eq.~(\ref{eq: rho_* expression}). We consider a local maximum to be the time stamp of a single disk pass by the minihalo. On the other hand, the local minima in Fig.~\ref{fig: integrand vs time - regular} are denoted by an orange diamond marker, and in general correspond to being locally the furthest away from a disk pass that the minihalo can be at. The bottom panel of Fig.~\ref{fig: integrand vs time - regular} also plots the value of the galactocentric $Z$ coordinate versus time for that same orbit. It can be seen that, to a good approximation, the term $\rho_* v$ achieves a local maximum when $Z=0$ and a local minimum when $Z$ has a local extremum. We consider a single disk pass as being from one local minimum of the $\rho_* v$ curve to the next consecutive local minimum, with a local maximum in between. \rev{This is the domain of ``traj, s.d.p.'' in the integral presented in eqn.~(\ref{eq: Sigma_* general formula}). The energy injection is then computed for every single disk pass.} One thing to note is that the different local maxima do not have the same value of $\rho_* v$, which is evident from Fig.~\ref{fig: integrand vs time - regular}. This is because, although the plane of the orbit remains fixed, the minihalo does not have closed orbits. Instead, the orbit precesses with time. This implies that the minihalo's multiple passes through the disk occur at different phases of the minihalo's orbit. Thus, the minihalo crosses the Galactic disk at different galactocentric radii $R$. Hence, the local stellar volume density $\rho_*$ at each disk pass will be different according to Eq.~(\ref{eq: rho_* expression}), creating different values for the integrand $\rho_* v$. Furthermore, it might be worth noting that the integrand $\rho_* v$ may not be a smooth function of time at the instant when $Z=0$. It is not smooth because $\rho_*$ is not smooth at $Z=0$ according to Eq.~(\ref{eq: rho_* expression}). Nonetheless, our procedure for finding the local maxima of $\rho_* v$ curve works because we use the discrete second difference in $\rho_* v$ values, and not the continuous second derivative.

\begin{figure}[htp]
	\centering
	\includegraphics[width=\columnwidth, height=0.47\textheight,keepaspectratio]{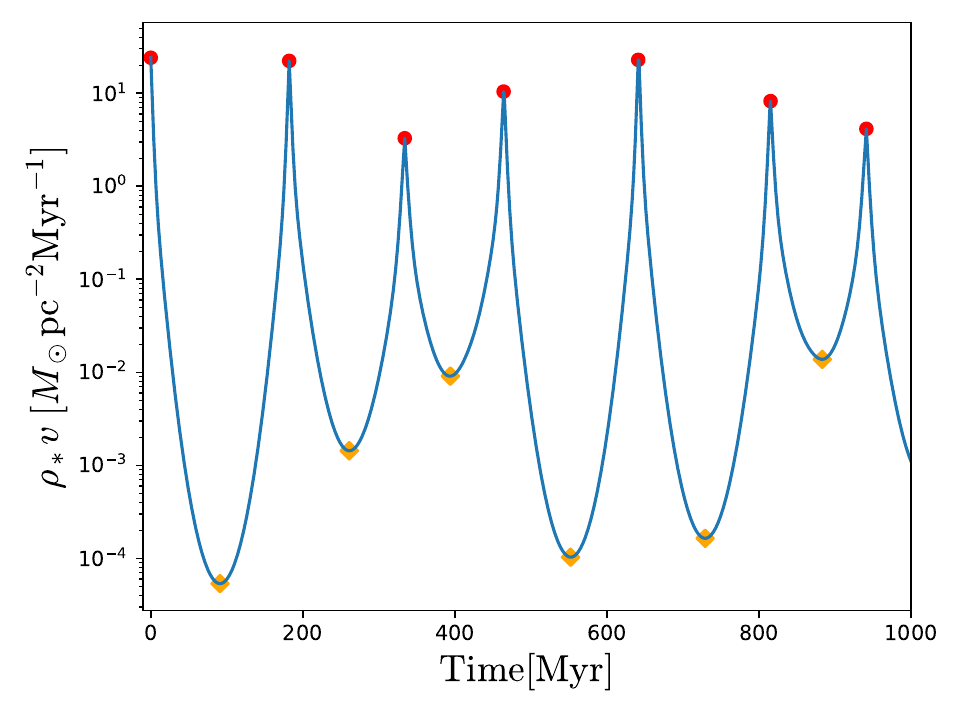}
	
	\hspace{0.5cm}\includegraphics[width=\columnwidth, height=0.47\textheight,keepaspectratio]{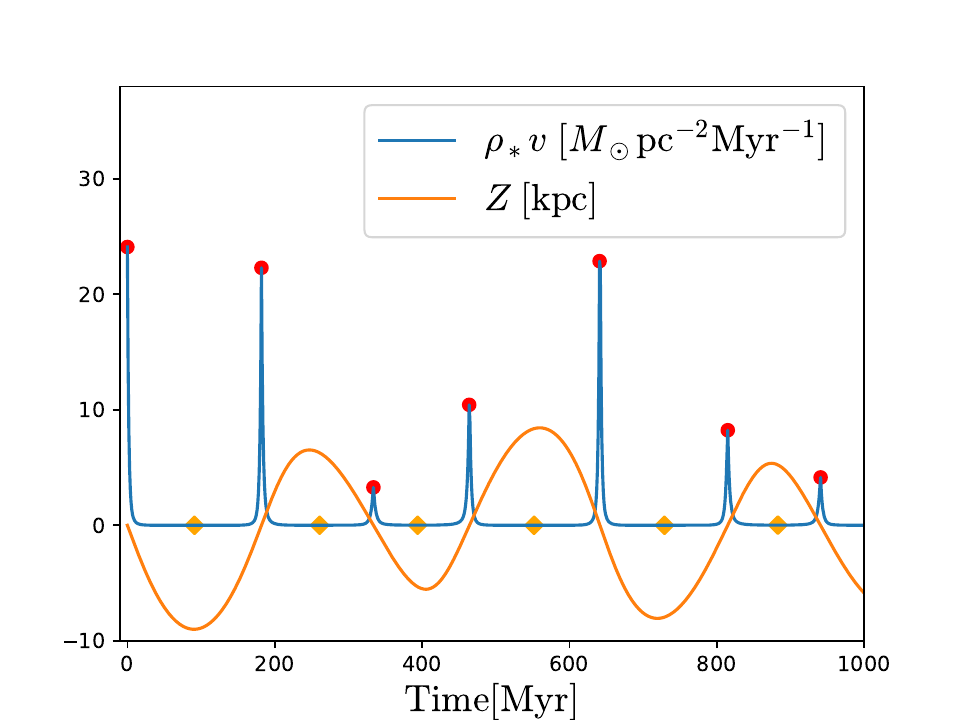}\hspace{-0.5cm}
	
	\caption[Stellar volume density plotted against time along a minihalo's trajectory in the MW galaxy]{The minihalo is in orbit around the Galactic center. In the top panel, the integrand term $\rho_* v$ is plotted against time elapsed where $\rho_*$ is the stellar volume density at the position of the minihalo at any given time, and $v$ is the instantaneous speed of the minihalo at that time. The local maxima are denoted by a circular red dot. These points correspond to when the minihalo crosses the Galactic disk. The local minima are denoted by an orange diamond marker. In the bottom panel, additional information about the galactic $Z$ coordinate is provided.}
	\label{fig: integrand vs time - regular}
\end{figure}

Looking more closely at the bottom panel of Fig.~\ref{fig: integrand vs time - regular}, we see that the local maximum of the $\rho_* v$ curve is coincident with the instant of time when $Z=0$. This is because as $Z$ goes from negative to positive values through $Z=0$, $|Z|$ achieves a (non-smooth) local minimum at $Z=0$. Thus, from Eq.~(\ref{eq: rho_* expression}), we can see that the term $\exp\left(-|Z|\right)$ achieves a (non-smooth) local maximum at $Z=0$, hence forcing $\rho_*$ and consequently $\rho_* v$ to achieve a (non-smooth) local maximum here. On the other hand, the local minimum of the $\rho_* v$ curve is a little offset from the instant of time when the local extremum of the $Z$ curve occurs, i.e., $\dot{Z}=0$. This is because the integrand $\rho_* v$ is not just a function of $Z$ but also a function of $R$ and $v$ (see Appendix~\ref{app: local minimum of rho_*v does not coincide with z_dot=0} for details).

In our \texttt{Python} code using the \texttt{NumPy} library, we start with an array of linearly spaced time values with a resolution of approximately 1 Myr. We then evaluate the integrand $\rho_* v$ at each value of time. We then perform the following operation on the integrand array:  \texttt{diff}(\texttt{sign}(\texttt{diff}($\rho_* v$ array))). The \texttt{diff}() operation computes the difference between neighboring values of the input array. The \texttt{sign}() operation takes any real number as input and outputs $-1$ if the input is negative, $+1$ if the input is positive, and $0$ if the input is zero. The above net operation is a discrete form of the second derivative of the integrand $\rho_* v$ with respect to time. It is easy to see (with an example) that the output array of the net operation is two elements shorter than the original integrands array. Most of the values of this output array will be zero (corresponding to monotonic portions of the integrands vs time plot). If we find a value of $+2$ in this output array, the $\rho_* v$ curve achieves a local minimum at the position corresponding to that entry. On the other hand, if we find a value of $-2$ in the output array, the time instant corresponding to that entry is a local maximum. Thus, we are able to find the local extrema of the integrands array efficiently.

\rev{
We note that the galactic system adopted in this chapter is not self-consistent. The minihalo orbits are sampled and evolved using the singular isothermal sphere model. On the other hand, when evaluating the energy injection, an axisymmetric disk is introduced. In principle, one could expect the presence of this disk to modify the trajectory of the minihalo slightly, especially at small galactocentric radii in the plane of the disk, as the stellar volume density in the disk here is high, and can dominate the singular isothermal sphere's potential. However, it is mostly minihalos on highly eccentric orbits that access such small galactocentric radii. As we mentioned earlier in this section, highly eccentric orbits are underrepresented in our Monte Carlo simulations. Thus, this inconsistency in modelling the galaxy may not affect our results significantly.
}

\subsection{Summing up the energy injection parameters for multiple Galactic disk passes}\label{sec: Summing up the energy injection parameters for multiple Galactic disk passes}

In Section~\ref{sec: mass-redshift grid}, we stated that in our Monte-Carlo simulations, we consider 1,000 values of $M$ and 1,000 values of $z$. Thus, we have one million ordered pairs or grid points of $(M, z)$. To each of these grid points, we assign one minihalo orbit that was generated using the procedure described in Section~\ref{sec: Monte Carlo sampling of orbits in singular isothermal sphere}. We also assign to each grid point a concentration parameter that depends on the $M$ and $z$ of that grid point according to Section~\ref{sec: Mass-Concentration relationship}. The goal is to calculate the effective energy injection parameter experienced by the orbit for each grid point during its lifetime in the galaxy. We then calculate the survival fraction of the minihalo given knowledge of this effective energy injection parameter and the concentration of the minihalo at $z$.

 Note that the time for which we calculate the effective energy injection parameter of the minihalo orbit corresponds to the lifetime of the orbit in the galaxy, which in turn corresponds to the infall redshift $z$. We must convert $z$ to its corresponding lookback time $T$ because the \texttt{lbparticles} code takes time (and not redshift) as input. We use the standard conversion formula (see, for example, Appendix~\ref{app: converting infall redshift to look back time}):
\begin{multline}\label{eq: look back time}
	T(z) = \frac{1}{H_0} \int_0^{z} \frac{1}{1+z^\prime} \\
	\times \left[\Omega_{\rm m}(1+z^\prime)^3 + \Omega_{\rm r}(1+z^\prime)^4 + \Omega_\Lambda\right]^{-1/2} \mathrm{d}z^\prime
\end{multline}

We numerically integrate Eq.~(\ref{eq: look back time}) using the \texttt{SciPy}'s \texttt{quad}() function in Python. We now can use the \texttt{lbparticles} code to evolve the orbit of the minihalo from the infall redshift until today.

\section{Computing the survival fraction of minihalos}\label{sec: computing the survival fraction of the minihalo}

In Chapter \ref{ch:journal_paper_1}, we presented a procedure to evaluate $\Delta M/M$, for an NFW minihalo, given $c$ and $E_{\rm frac}$
where $\Delta M$ is the mass loss due to the energy injection.
However, the ranges of concentration and $E_{\rm frac}$ values considered in that chapter were rather narrow compared to what is generated by the Monte-Carlo simulation in this chapter. Thus, we generate an interpolation object offline using Python's \texttt{RegularGridInterpolator} function from the \texttt{SciPy.Interpolate} package that takes as input the energy injection parameter and concentration, and outputs the survival fraction. To achieve this, we first create a log-spaced array of concentration values in the range $c \in [0.1, 10^5]$ and similarly for energy injection parameters values in the range $E_{\rm frac} \in [10^{-13}, 3\times10^6]$. We then use \texttt{NumPy}'s \texttt{meshgrid} function to generate all possible ordered pairs of $(E_{\rm frac}, c)$. Then, we use the procedure in Chapter \ref{ch:journal_paper_1} to compute the survival fraction corresponding to each ordered pair. This can be a time-consuming process, but it is done offline and only once. However, once the interpolation object is generated, it can compute the survival fraction rapidly as long as the input $E_{\rm frac}$ and $c$ are in the ranges that were originally used to generate the interpolation object in the first place.

We can integrate the mass function, $\frac{\mathrm{d}n_{f}}{\mathrm{d}M}$, to get the collapsed fraction $f$ which is the ratio of the number of axions that have collapsed into minihalos to the total number of axions. 
As done by \citelink{S2023}{S2023}, we choose the lower bound of this mass range to be $10^{-12}M_\odot$. 
Our conclusions about the ratio of the collapsed fraction with and without disruption will not be sensitive to making this lower bound even lower.
The upper bound of the mass range is chosen to be the same as the upper bound that we used for the Monte-Carlo simulation, i.e., $10^{-3}M_\odot$. This value is motivated by wanting our most massive minihalo to be substantially smaller than our least massive adiabatic halo. Therefore, the original collapsed fraction without stellar disruption is given by 
\begin{equation}\label{eq: expression for 'f'}
	f_{\rm ori} = \frac{1}{\bar{\rho}_{\rm c}}\int_{10^{-12}M_\odot}^{10^{-3}M_\odot} M \frac{\mathrm{d}n_{f}}{\mathrm{d}M}\ \mathrm{d}M \, .
\end{equation}
where $\bar{\rho}_{\rm c}$ is the comoving density of cold dark matter. 
Using Eqs.~(\ref{eq: weight sum}) and (\ref{eq: weights}), the above equation can be approximated by
\begin{align}\label{eq: expression for 'fundisrupted'b}
	f_{\rm ori} &= \frac{1}{\bar{\rho}_{\rm c}} \sum_{i,j} M_i          w_{ij} \delta M_i\nonumber \\
	&= \frac{1}{\bar{\rho}_{\rm c}}\sum_{i,j}       M_i^2   w_{ij} \delta \ln(M),
\end{align}
where $\delta M_i = M_{i+1} - M_i$ is the mass bin width and $\delta\ln(M)\approx\delta M_i/M_i$ is the constant bin width for $\ln(M)$.
We can then approximate the derivative with respect to the  mass as
\begin{equation}\label{eq: expression for 'dfundisrupt'}
	\begin{aligned}
		\left.\frac{{\rm d}f_{\rm ori}}{{\rm d\,ln}(M)}\right|_{M_k} &= \frac{1}{\bar{\rho}_{\rm c}} M_k^2\sum_j           w_{kj}\, .
	\end{aligned}
\end{equation}
For plotting purposes we use 
\begin{equation}\label{eq: mass function definition, df/dM - part 1}
	\frac{\mathrm{d}f}{\mathrm{d}\log_{10}(M)} = \ln(10) \frac{\mathrm{d}f}{\mathrm{d}\ln (M)} %
	\ .
\end{equation}

Combining Eqs.~(\ref{eq: mass function definition, df/dM - part 1}) and (\ref{eq: expression for 'dfundisrupt'}), we get: 
\begin{equation}\label{eq: S2023 mass function in terms of ordinary mass function}
	\left. \frac{\mathrm{d}f_{\rm ori}}{\mathrm{d}\log_{10}(M)}\right|_{M_k} =  \frac{\ln(10)}{\bar{\rho}_{\rm c}} M_k^2   \sum_j        w_{kj}\,  .
\end{equation}
This quantity is plotted in Fig.~\ref{fig: mass_function} as the black curve. It is a good match to the analogous dashed grey curve in the top panel of Fig.~10 of \citelink{S2023}{S2023}. 

\begin{figure}
	\includegraphics[width=\columnwidth]{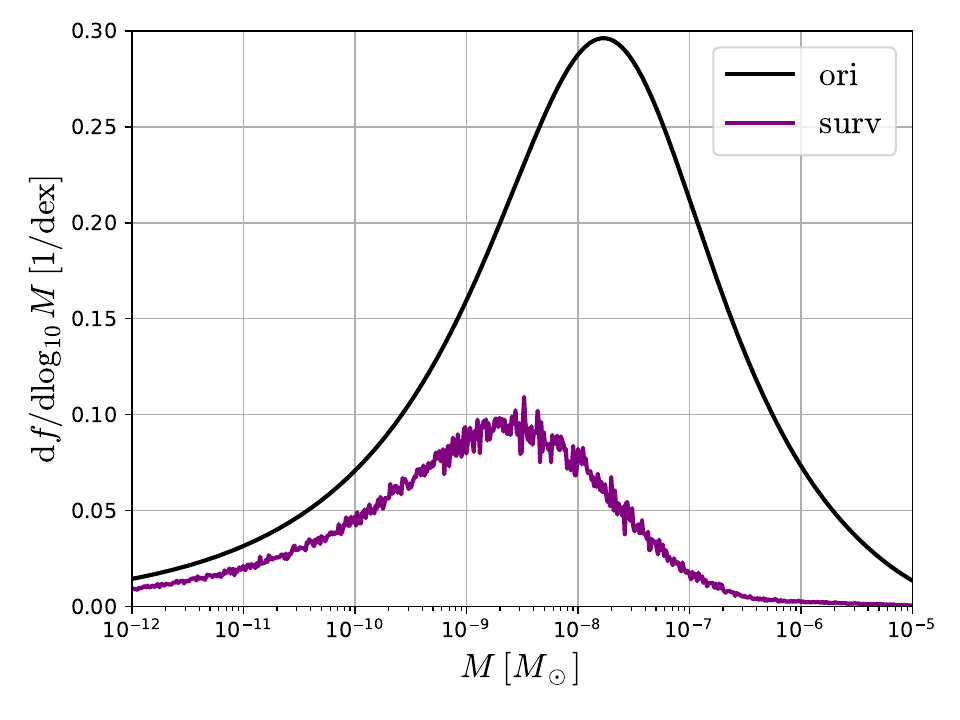}
\caption[The undisrupted and stellar-disrupted mass functions of minihalos in the MW galaxy]{\rev{$f$ is the fraction of galactic axion mass bound in minihalos. $\mathrm{d}f/\mathrm{d}\log_{10}M$ is the fraction of bound axion mass per unit minihalo mass interval $M$, and is used as a proxy for the usual mass function. An axion mass of $m_a=25$~$\mu$eV is assumed} . In this figure, we have assumed the minimum adiabatic halo mass is $10^2M_\odot$. The mass function without considering any disruption is shown as a black curve, while the mass function incorporating the effects of stellar disruption is shown as a purple curve. In generating the disrupted mass functions, we used the hybrid method to combine multiple energy injections.  }
	\label{fig: mass_function}
\end{figure}

To generate the mass function in the presence of stellar disruption, we first compute $\Delta M_{ij}$ for the minihalo orbit corresponding to each grid point.

Similarly to Eq.~(\ref{eq: expression for 'fundisrupted'b}) the surviving disrupted fraction of axions in minihalos is given by
\begin{align}\label{eq: expression for 'f'b}
	f_{\rm surv} =& \frac{1}{\bar{\rho}_{\rm c}} \sum_{i,j} (M_i - \Delta M_{ij})          w_{ij}\delta M_i\nonumber \\
	=& \frac{1}{\bar{\rho}_{\rm c}} \sum_{i,j} (M_i - \Delta M_{ij})          w_{ij}M_i\delta\ln(M)\nonumber \\
	=& \frac{\delta\ln(M)}{\bar{\rho}_{\rm c}} \times \nonumber\\
	&\sum_k \sum_{M_k\leq M_i-\Delta M_{ij}<M_{k+1}} (M_i - \Delta M_{ij})   M_i       w_{ij},
\end{align}
where the sum in the last line of the above equation is over all $i,j$ that satisfy $M_k\leq M_i-\Delta M_{ij}<M_{k+1}$.
We can then approximate the derivative with respect to the mass as
\begin{equation}\label{eq: expression for 'df'}
	\begin{aligned}
		\left.\frac{{\rm d}f_{\rm surv}}{{\rm d\,ln}(M)}\right|_{M_k} =& \frac{1}{\bar{\rho}_{\rm c}} \times \\
		&\sum_{M_k\leq M_i-\Delta M_{ij}<M_{k+1}} (M_i - \Delta M_{ij})M_i          w_{ij}\, .
	\end{aligned}
\end{equation}
Using equation (\ref{eq: mass function definition, df/dM - part 1}) with the above equation gives
\begin{equation}
	\begin{aligned}
		\label{eq: mass fraction step by step - part 3}
		\left.\frac{{\rm d}f_{\rm surv}}{{\rm d\,}\log_{10}(M)}\right|_{M_k} =& \frac{\ln(10)}{\bar{\rho}_{\rm c}}\times \\
		&\sum_{M_k\leq M_i-\Delta M_{ij}<M_{k+1}} (M_i - \Delta M_{ij})M_i          w_{ij}\, .
	\end{aligned}
\end{equation}

Eq.~(\ref{eq: mass fraction step by step - part 3})  gives us the alternative form of the stellar disrupted mass function of minihalos in the galaxy today. This mass function is presented in Fig.~\ref{fig: mass_function} as the purple curve. The stellar-disrupted mass function is significantly more suppressed than the undisrupted mass function. Moreover, the peak in the stellar-disrupted mass function has shifted towards lower mass relative to the peak of the undisrupted mass function. The reason for both these observations is that the minihalos lose mass when subjected to stellar interactions. Finally, the reason that the stellar-disrupted curve is noisy is that we have assigned a random minihalo orbit to each $(M, z)$ grid point. If the Monte-Carlo simulation is run multiple times independently, the exact values taken by this curve change but only up to the statistical noise induced by the noise in the energy and angular momentum of the generated orbits. As we will see later in the section, our final results will be insensitive to this noise. \rev{If one needs a smooth curve for the purposes of plotting, one can run the Monte Carlo simulations multiple times and perform an appropriate averaging procedure in each mass bin to wash out the statistical noise.}

The disrupted (purple) curve in Fig.~~\ref{fig: mass_function} can be compared to the red curve in Fig.~10 of \citelink{S2023}{S2023}. They also considered a smooth tidal disruption from the Milky Way potential. But as can be seen from the same figure, including that had a negligible effect once the stellar tidal effects were accounted for. Their disrupted curve has about twice the area of our one as they added the energy injections linearly, equivalent to our $p=2$ case. Their disrupted curve is also smoother than ours. The reason for this is that they used correction factors based on Monte Carlo averages. We couldn't employ that technique as it was incompatible with our hybrid method of adding multiple energy injections.


To get a measure of how much disruption has taken place, we evaluate 
\begin{equation}
	\frac{M_{\rm surv}}{M_{\rm ori}} = \frac{f_{\rm surv}}{f_{\rm ori}} 
\end{equation}
where $M_{\rm surv}$ is the amount of mass in minihalos 8~kpc from the Galactic center today, and $M_{\rm ori}$ would be the amount of mass in minihalos 8~kpc from the Galactic center today if stellar disruption had not taken place.

Using the minimum mass of adiabatic halos considered to be $10^{2} M_\odot$ for the Monte-Carlo simulation, we find that $M_{\rm surv} / M_{\rm ori} = 30\%$ for $m_a=25\mu \rm eV$. On the other hand, \citelink{S2023}{S2023} states that this value is $58\%$ using their method of correction factors and setting $p=2$.  This lost mass enters the inter-minihalo space called minivoids. Thus, the axion density in the minivoids increases. Despite the stellar disrupted mass function in Fig.~\ref{fig: mass_function} being noisy and its exact values changing slightly in between independent runs of the Monte-Carlo simulation, the resulting value of $M_{\rm surv} / M_{\rm ori}$ does not change in between runs at the level of $0.1\%$.

It is expected that the average inter-minihalo distance is significantly greater than the virial radius of a typical minihalo. Thus, the Earth is statistically more likely to be inside a minivoid than inside a minihalo (e.g., \cite{Eggemeier23,OHare2023}). Thus, the lower value of $M_{\rm surv} / M_{\rm ori}$ means that the local axion density at the Earth's position is likely to be higher than previously predicted. This increases the chances of axion dark matter direct detection using haloscopes relative to what had been previously estimated without considering the more destructive nature of multiple stellar encounters.

\begin{figure}[htp]
	\centering
	\includegraphics[width=\columnwidth, height=0.47\textheight,keepaspectratio]{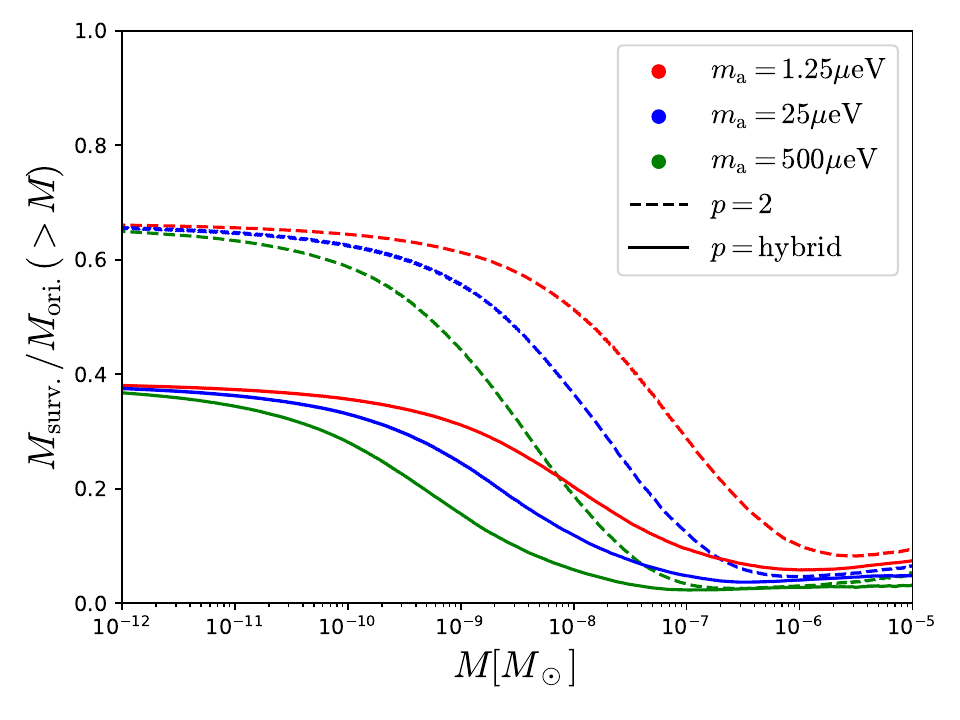}
	
	\hspace{0.5cm}\includegraphics[width=\columnwidth, height=0.47\textheight,keepaspectratio]{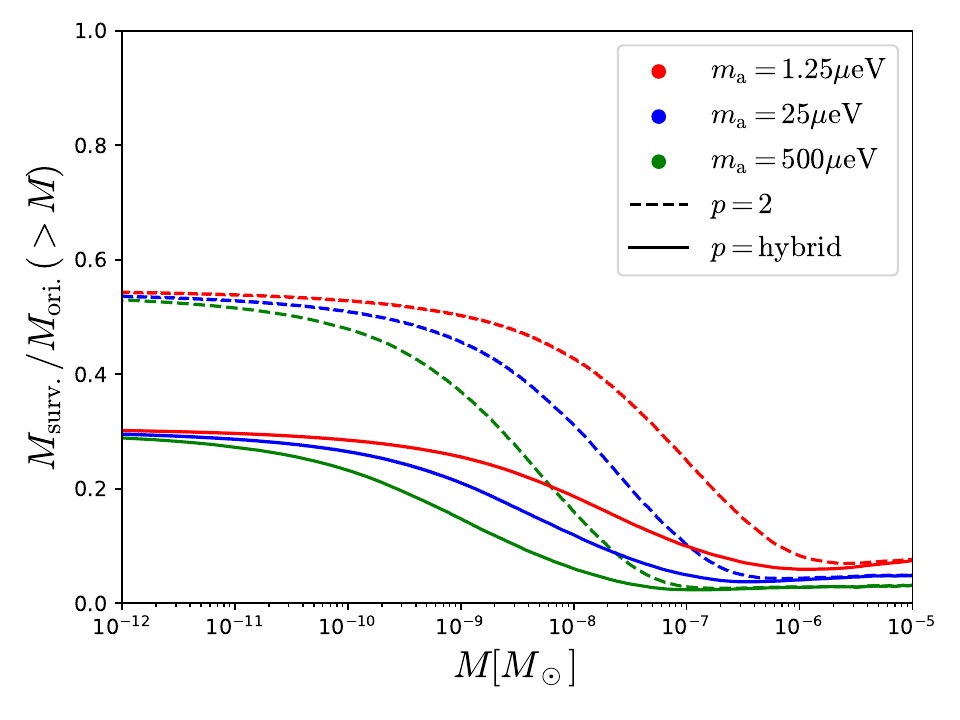}
	
	\caption[$M_{\rm surv} / M_{\rm ori} (>M)$ plotted against the minimum minihalo mass $M$]{The value of $M_{\rm surv} / M_{\rm ori} (>M)$ is plotted against the minimum minihalo mass $M$. The parameter $M_{\rm ori}$ would be the mass at $8$ kpc, of minihalos if there was no stellar or other tidal  disruption. While $M_{\rm surv}$ is the corresponding mass if stellar disruption is accounted for. The top and bottom panels use the lower bound in the mass of adiabatic halos to be equal to $10^{-2}$ and $10^2 M_\odot$ respectively. The results for three different axion masses $m_{\rm a}=1.25, 25, 500 \mu \text{eV}$ are presented as different colors. The dashed lines represent the $p=2$ method of linearly adding up the energy injection parameters corresponding to consecutive disk passes by the minihalo. The solid lines represent the ``hybrid'' method of adding up the energy injection parameters.}
	\label{fig: M_surv by M_ori vs M}
\end{figure}

In Fig.~\ref{fig: M_surv by M_ori vs M}, we have calculated the values of $M_{\rm surv} / M_{\rm ori}$ 
as a function of the lower bound of the 
integration in Eq.~(\ref{eq: expression for 'f'}). 
We see from this figure that 
the $p=\text{``hybrid''}$ method results in lesser mass surviving in minihalos compared to the $p=2$ method. We also see from the figure that when we set that adiabatic minimum halo mass $M_{\rm min} = 10^2 M_\odot$, we find that lesser mass in minihalos survives compared to setting $M_{\rm min} = 10^{-2}M_\odot$. This is because, as can be seen from Fig.~\ref{fig: collapse fraction}, the $M_{\rm min}=10^2 M_{\odot}$ case will freeze at a lower redshift.
From Fig.~1 of \citelink{S20204}{S2023}, we can see that this implies the concentration of the minihalos in the $M_{\rm min}=10^2M_{\odot}$ case will be higher.

To ascertain this, we want to generate the joint probability density function (PDF) of $E_{\rm frac}$ and $c$ for the entire physical population of minihalos in our simulation. We first consider our two-dimensional grid points of ordered pairs of $(M_i, z_j)$ and their corresponding weight $w_{ij}$. For each grid point, we compute the concentration $c_{ij}$ that will be assigned to that grid point. As discussed in Sections~\ref{sec: Monte Carlo sampling of orbits in singular isothermal sphere} to \ref{sec: Monte Carlo Simulations to determine the stellar-disrupted mass function of minihalos}, each grid point has a random orbit which we can use to calculate the total energy injection parameter $E_{{\rm frac},ij}$ that the assigned orbit experiences during its lifetime in the galaxy, which corresponds to infall redshift $z_j$. Thus, to generate the joint PDF of $E_{\rm frac}$ and $c$ for the physical population of minihalos, we first create a logarithmically spaced one-dimensional array each for the possible $E_{{\rm frac},k}$ and $c_l$. We then evaluate the binned PDF as follows:
\begin{equation}
\begin{aligned}
& P_{k l} \propto \sum_{\substack{ E_{{\rm frac},k} \leq E_{{\rm frac},ij} \rev{<} E_{{\rm frac},k}+\Delta E_{{\rm frac},k}\\ c_l \leq c_{ij} \rev{<} c_l+\Delta c_l}} w_{i j} \\
& 
\end{aligned}
\end{equation}
where the sum is over all $i$ and $j$ values that fulfill the specified conditions. The proportionality constant can be determined so that $\sum_{k,l}P_{kl}=1.$ It should be noted that $P_{kl}$ will be grainy, the extent to which depends on the resolution of the two-dimensional $(E_{\rm frac}, c)$ bins.

\begin{figure}
        \vspace{0.1 cm}
	\includegraphics[width=\columnwidth]{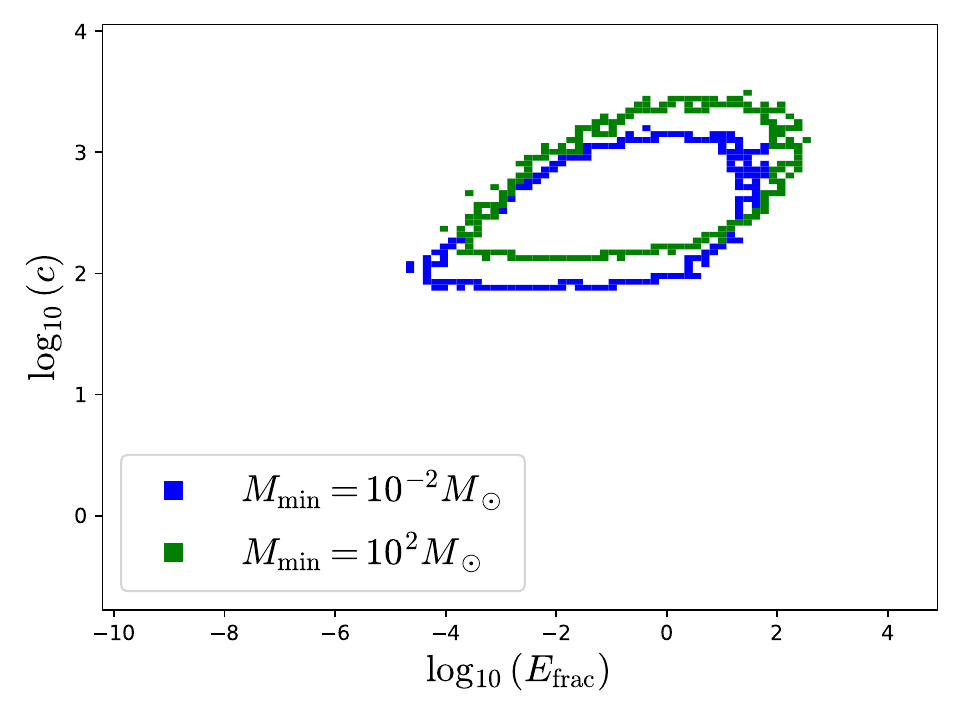}
	\caption[95\% contour lines of the joint PDF of total $E_{\rm frac}$ and concentration $c$ for the physical population of minihalos in the MW galaxy]{The  95\% contour lines of the joint probability density function (PDF) of total energy injection parameter $E_{\rm frac}$ and concentration $c$ for the entire physical population of minihalos is presented for both the $M_{\rm min}=10^{-2}M_\odot$ and $M_{\rm min}=10^{2}M_\odot$ cases. }
	\label{fig: contour plot of joint PDf of Efrac and c}
\end{figure}

 In Fig.~\ref{fig: contour plot of joint PDf of Efrac and c}, we present the  95\% confidence interval of the joint PDF of $E_{\rm frac}$ and $c$ for both the $M_{\rm min}=10^{-2}M_{\odot}$ and $M_{\rm min}=10^{2}M_{\odot}$ cases.
Each confidence interval was generated by finding the threshold value $P_{\rm thresh}$
such that 
\begin{equation}
    \sum_{P_{kl}>P_{\rm thresh}}P_{kl}\approx 0.95.
\end{equation}
for the corresponding $P_{kl}$.

In our Monte-Carlo simulation, we find that $E_{\rm frac} \in [10^{-10}, 10^4]$ and $c \in [10^{-1}, 10^4]$ approximately. Our analytical method of calculating the survival fraction given the values of $E_{\rm frac}$ and $c$ is accurate over this regime. However, as can be seen in Fig.~\ref{fig: contour plot of joint PDf of Efrac and c}, the vast majority of the physical minihalo population has $E_{\rm frac} \in [10^{-4}, 10^2]$ and $c \in [100, 3000]$ approximately. \citelink{S2023}{S2023} performed numerical simulations and generated their response function to work more or less in this narrower range. Thus, when we use \citelink{S2023}{S2023}'s response function to generate the survival fraction given $E_{\rm frac}$ and $c$ while still using the other aspects of our Monte-Carlo simulation, we find that the value of $M_{\rm surv} / M_{\rm ori} (M>10^{-12}M_\odot)$ differs from our result by (on average) 0.1\% to 1\%.

\rev{As can be seen from Fig.~\ref{fig: contour plot of joint PDf of Efrac and c}, most of the minihalos will have an effective $E_{\rm frac}\lesssim 1$. Also, in the $M_{\rm min}=10^2M_{\odot}$ case, the minihalos have higher concentrations than in the $M_{\rm min}=10^{-2}M_\odot$ case. As shown from Fig.~4 of \citelink{S2023}{S2023}, the higher concentration minihalos will suffer a greater mass loss in the regime $E_{\rm frac}\lesssim 1$. This, and the increased $E_{\rm frac}$ in the $M_{\rm min}=10^2M_{\odot}$ case, explains why the $M_{\rm min}=10^2M_{\odot}$ case is found to have a greater mass loss in minihalos in comparison to the $M_{\rm min}=10^{-2}M_\odot$ case. }

We see that as we increase the mass $M$ in Fig.~\ref{fig: M_surv by M_ori vs M} from $M=10^{-12}M_\odot$, the value of $M_{\rm surv} / M_{\rm ori}$ decreases. The reason for this trend becomes clearer when we look at Fig.~\ref{fig: mass_function}, where the undisrupted mass function rises more sharply relative to the stellar-disrupted mass function. This causes $M_{\rm surv} / M_{\rm ori}$ to decrease. However, in Fig.~\ref{fig: M_surv by M_ori vs M}, when we go to high masses of the order of $\gtrsim 10^{-7} M_\odot$, we see that the value of $M_{\rm surv} / M_{\rm ori}$ starts to increase. This is because in Fig.~\ref{fig: mass_function}, the value of the stellar-disrupted mass function begins to level off around $M \gtrsim 10^{-7} M_\odot$ while the undisrupted mass function still keeps dropping. This causes the value of $M_{\rm surv} / M_{\rm ori}$ to increase.

\begin{figure}
	\includegraphics[width=\columnwidth]{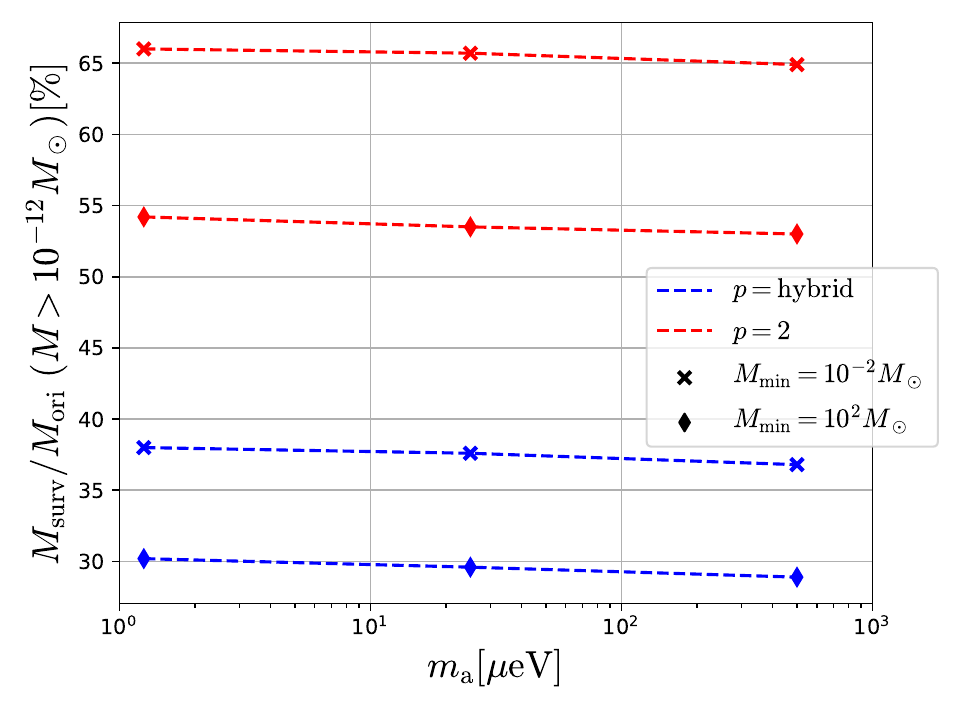}
	\caption
    [$M_{\rm surv} / M_{\rm ori}(M>10^{-12}M_\odot)$ plotted against axion mass $m_{\rm a}$]
    {
    A plot of $M_{\rm surv} / M_{\rm ori}(M>10^{-12}M_\odot)$ against discrete values of the axion mass $m_{\rm a}=1.25, 25, 500 \mu \text{eV}$. The cross and diamond markers represent the case where the lower bound $M_{\rm min}$ on the adiabatic halos is $10^{-2}$ and $10^2 M_\odot$, respectively. The red and blue lines represent the $p=2$ and hybrid method, respectively, of summing up the energy injection parameters of consecutive Galactic disk passes.}
	\label{fig: M_surv by M_ori greater than 1e-12 MSolar vs axion mass}
\end{figure}

The dashed curves in the bottom panel of Fig.~\ref{fig: M_surv by M_ori vs M} show reasonable agreement with the corresponding orange curves in Fig.~14 of \citelink{S2023}{S2023}. The minor differences are likely due to methodological distinctions: \citelink{S2023}{S2023} generated probability density functions (PDFs) from their Monte Carlo simulations and subsequently derived correction factors from those PDFs, whereas we directly implemented the Monte Carlo simulations without relying on intermediate corrections.

\rev{
We note that although one might potentially expect isolated minihalos of masses $< 10^{-12}M_\odot$, including such small mass minihalos won't change the value of $M_{\rm surv} / M_{\rm ori}(>M)$ when $M< 10^{-12}M_\odot$ because, as is seen in Fig.~\ref{fig: M_surv by M_ori vs M}, the curve already asymptotes when go to low enough masses, and certainly by $M = 10^{-12}M_\odot$. Thus, including such small minihalo masses won't change our result for the fraction of axion mass contained in bound minihalos at the Solar radius today.
}

Next, we combine both the panels of Fig.~\ref{fig: M_surv by M_ori vs M} and only present the value of $M_{\rm surv} / M_{\rm ori}$ where the lower bound of the integral in Eq.~(\ref{eq: expression for 'f'}) is set back to $10^{-12}M_\odot$. This is presented in Fig.~\ref{fig: M_surv by M_ori greater than 1e-12 MSolar vs axion mass} where we plot $M_{\rm surv} / M_{\rm ori} (M>10^{-12}M_\odot)$ against axion mass $m_{\rm a}$ for discrete axion masses $m_{\rm a} = 1.25, 25, 500 \mu \text{eV}$. 
This information is also presented in Table~\ref{table:mass ratios}. We checked the Monte-Carlo simulation convergence
was insensitive to small changes in 
the maximum mass of the minihalos, maximum infall redshift, and number of sample points of mass and infall redshifts. \textcolor{black}{Finally, note that in reality, we expect the dynamical time of a minihalo to increase slightly after each disk pass due to a change in the density profile of the minihalo. To be computationally efficient, we have not considered this effect. However, in Appendix~\ref{app:updating dynamical time}, using a back-of-the-envelope calculation, we have estimated that the true value of $M_{\rm surv}/M_{\rm ori}$ will not increase by more than a few percent compared to the values presented in Table~\ref{table:mass ratios} if we did update the dynamical time.}

\begin{table}
	\begin{tabular}{|l|c|l|l|}
		\hline
		$M_{\rm min} [M_\odot]$&	p & $m_{\rm a} [\mu \rm eV]$ & $M_{\rm surv} / M_{\rm ori}(M>10^{-12}M_\odot)[\%]$  \\
		\hline
		$10^{-2}$ &  2 & 1.25  & 66  \\
		\hline
		$10^{-2}$ &  2 &  25 & 65.7  \\
		\hline
		$10^{-2}$ &  2 &  500 &  64.9 \\
		\hline
		$10^{-2}$ & hybrid  & 1.25  &  38 \\
		\hline
		$10^{-2}$ & hybrid &  25 &  37.6 \\
		\hline
		$10^{-2}$ &  hybrid & 500  &  36.8 \\
		\hline
		$10^2$ & 2  & 1.25  &  54.2 \\
		\hline
		$10^2$ & 2  &  25 &  53.5 \\
		\hline
		$10^2$ & 2  &  500 &  53 \\
		\hline
		$10^2$ & hybrid  &  1.25 &  30.2 \\
		\hline
		$10^2$ & hybrid  &  25 &  29.6 \\
		\hline
		$10^2$ & hybrid  & 500  & 28.9  \\
		\hline
	\end{tabular}
	\caption
    [$M_{\rm surv} / M_{\rm ori}(M>10^{-12}M_\odot)$ tabulated for different values of minimum adiabatic halo mass, $p$-values, and axion mass $m_{\rm a}$]
    {
    A table of $M_{\rm surv} / M_{\rm ori}(M>10^{-12}M_\odot)$. 
    See Fig.~\ref{fig: M_surv by M_ori greater than 1e-12 MSolar vs axion mass} for a plot of these values.
		\label{table:mass ratios}
	}
\end{table}

\section{Conclusions}
\label{sec: conclusions}

In this work, we have investigated the disruption of axion minihalos due to stellar encounters in the Milky Way galaxy. We extended previous analyses by incorporating a more accurate treatment of multiple stellar encounters, taking into account whether minihalos have sufficient time to relax between encounters based on their dynamical timescales. By generating a population of minihalo orbits using Monte Carlo simulations and evolving them within a model of the Galactic potential, we computed the stellar-disrupted mass function of minihalos.

Our results indicate that the cumulative effect of stellar interactions is more destructive to minihalos than previously estimated. Specifically, we find that the surviving mass fraction $M_{\rm surv} / M_{\rm ori}$ of minihalos is significantly reduced when accounting for the proper addition of energy injections from multiple stellar encounters. For example, when using a minimum mass of adiabatic halos of $M_{\rm min} = 10^2\, M_\odot$, we find that only about $30\%$ of the original mass in minihalos survives, compared to previous estimates of around $60\%$. This reduction is due to the increased mass loss when minihalos have time to relax between encounters, leading to a nonlinear addition of energy injections.

The suppression of the stellar-disrupted mass function, as illustrated in Fig.~\ref{fig: mass_function}, has important implications for the distribution of axion dark matter in the Galaxy. With a larger fraction of axion dark matter residing in the inter-minihalo space (minivoids), the local axion density at the Earth's position may be higher than previously predicted. This enhancement increases the prospects for the detection of axions via haloscopes relative to the case where the more destructive nature of multiple stellar encounters was not accounted for.

Furthermore, our analysis highlights the importance of accurately modeling the cumulative effects of stellar encounters on minihalos. By considering the dynamical timescales of minihalos and adopting a hybrid method for summing energy injections (as opposed to the linear addition method with $p=2$), we provide a more realistic estimate of minihalo survival.


In conclusion, our findings suggest that stellar disruption plays a significant role in shaping the minihalo mass function and the distribution of axion dark matter in the Milky Way. Accurately accounting for these effects is crucial for interpreting observational data and guiding the search for dark matter.


\chapter{Summary and Future Work}
\label{chap:summary}

\graphicspath{{summary/Figs/}}


This chapter provides an integrative narrative that threads together the main discoveries reported in Chapters \ref{ch:journal_paper_1} and \ref{ch:journal_paper_2}. 

\section{Key insights from minihalos undergoing multiple stellar encounters}
Chapter~\ref{ch:journal_paper_1} undertakes a detailed exploration of the mass loss incurred by minihalos subjected to repeated high--velocity stellar passages.  The work begins by considering a single NFW minihalo extending to infinity, and subjecting it to a stellar encounter with a single star. We adopt the K2021 analytical procedure for computing the survival fraction of the minihalo during such a stellar encounter, and compare it to the data from numerical simulations of S2023. We note that the K2021 model predicts the data from S2023 results well, only for concentration $c=100$. To approximate the data for all concentrations in the range of interest, we extend the K2021 model and develop the sequential stripping model in which the bound structure of the minihalo is resolved shell‑by‑shell to more accurately compute the gravitational potential in the minihalo. We find that the sequential stripping model better approximates the data from S2023 for all concentrations in the range of interest.

We further analytically model the process of the gravitational relaxation of an NFW minihalo that takes place following a stellar encounter. We assume the completely relaxed minihalo to have a Hernquist density profile. Given the concentration and virial mass of the unperturbed NFW minihalo, and the normalized energy injection parameter $E_{\rm frac}$ that characterizes the stellar encounter, we computed the scale parameters of the relaxed Hernquist minihalo and showed that the results agreed with the numerical simulations of S2023. We then incorporate the relaxation model to compute the survival fraction of an NFW minihalo when the concentration $c$ of the unperturbed minihalo and the normalized energy injection parameter are known, to show that we can better approximate the numerical data from S2023 at higher values of $E_{\rm frac}$.

Furthermore, we introduce the switching model to capitalize on the finding that when we do not consider the relaxation of the minihalo, our sequential stripping model approximates the numerical data well at low values of $E_{\rm frac}$, whereas when we do consider relaxation of the minihalo to a Hernquist profile, the sequential stripping model approximates the numerical data well at high values of $E_{\rm frac}$. We show that the switching model approximates the numerical data well for the whole range of $E_{\rm frac}$, and is comparable to the predictions of the mathematical fitting functions developed by S2023. However, the switching model is expected to better approximate the data for values of $E_{\rm frac}$ even outside the range considered in Chapter~\ref{ch:journal_paper_1} because it incorporates the underlying physics governing the stellar encounter, while the S2023 fitting functions may not extrapolate well to such out-of-bound $E_{\rm frac}$ values.

Next, we present an analytical procedure to compute the survival fraction of an NFW minihalo undergoing successive stellar encounters, with a long enough time in between encounters allowed for relaxation of the minihalo. Given the concentration of the unperturbed minihalo, we derive a set of $R$-factors to relate the concentration of an $n^{\rm th}$ generation minihalo to that of the $(n+1)^{\rm th}$ generation minihalo. This enables us to compute the survival fraction of the NFW minihalo after each stellar encounter. We show that our results approximate the numerical data from S2023 well.

Finally, we adopt an existing analytical formula for the effective energy injection into a minihalo undergoing successive stellar encounters and curve-fit this formula to the results of our analytical model for multiple stellar encounters, and find that $p\lesssim 1$ gives a good approximation of the effective energy injection.

Thus, we have extended the K2021 model of mass loss, and incorporated the sequential stripping model and the relaxation process to develop a more refined analytical model for mass loss estimation that approximates the S2023 numerical data well.

\section{Minihalo mass retention in the Galactic environment}

Building upon the theoretical framework of minihalo mass loss presented in Chapter~\ref{ch:journal_paper_1}, Chapter~\ref{ch:journal_paper_2} extends this framework to the galactic environment by constructing a Monte‑Carlo framework.

We start by quantifying the pre-infall mass function of axion minihalos and the collapse fraction of the adiabatic halos that host galaxies. We then adopt a formula to compute the final mass function of minihalos that have fallen into adiabatic halos, using the previous two quantities. This formula has an analytical form that uses an integral over the redshift of infall which is a continuous variable. Inspired by this formula, we adopt a Monte-Carlo procedure to compute the final mass function of minihalos.

We first randomly generate minihalo orbits through another Monte Carlo procedure assuming that the minihalos are distributed in the galaxy according to a singular isothermal sphere, for simplicity. In doing so, we assign an energy $E$ and angular momentum $L$ to each orbit such that the minihalo is present at the solar system position today. We then compute the initial velocity vector of the minihalo in each orbit using $E$ and $L$ for that orbit. 

We then use the \texttt{lbparticles} code to evolve each minihalo's trajectory backward in time to find the past position and velocity vectors of the minihalo. We now model the galactic stellar volume density and integrate it over the 1-dimensional trajectory of the minihalo to compute the effective stellar surface density as seen by the minihalo. We use the stellar surface density corresponding to a single galactic disk pass by the minihalo, to compute the normalized energy injection into the minihalo during that disk pass.

We add up the energy injection from consecutive individual disk passes linearly (with $p=2$) if the time between consecutive disk passes is smaller than the minihalo's dynamical time, otherwise we add them non-linearly (with $p=1$). Thus, we compute the effective energy injection for each minihalo during its lifetime in the galaxy, which corresponds to its infall redshift.

We set up a Monte Carlo procedure by constructing a 2-dimensional mass-infall redshift grid $(M_i, z_j)$, and compute the contribution (called weight $w_{ij}$) to the final mass function coming from each incremental infall redshift. We then compute the collapse fraction of minihalos inside the galaxy with and without considering the effect of stellar disruption on the mass retention of these minihalos. We estimate that the ratio of the mass of all axions retained in minihalos with considering the effect of stellar disruption, to that without considering this effect, is around 30\% for axion mass $25\mu \mathrm{eV}$, a significant drop from previous literature estimates of around 60\%. This difference arises primarily when correctly accounting for the relaxation process in between galactic disk passes, which changes the density profile of the minihalo. This result is in qualitative agreement with Chapter~\ref{ch:journal_paper_1}, where we showed that incorporating the relaxation process into our computation decreases the mass retention by the minihalo during multiple stellar encounters.

This result then predicts that the local axion density at the solar system position is higher than what was previously predicted in the literature, thereby increasing the relative chances of direct detection of axions by Earth-based haloscope experiments.

\section{Forward outlook}\label{sec:summary_outlook}

Combined together, Chapters \ref{ch:journal_paper_1} and \ref{ch:journal_paper_2} quantify the mass retention in minihalos in the galactic population and predict a higher local axion density at the Earth's location.

As a next step, we would like to more closely examine the axions that become unbound from the minihalo during a stellar interaction. Ref. \cite{OHare2023, Schneider2010}, for example, state that the mass that becomes unbound will form tidal streams with leading and trailing tails, with the minihalo at the center. These trails will continue to orbit the galactic center with the minihalo. The Galactic potential will further elongate these streams.
Ref.~\cite{OHare2023} did not consider the effects of minihalo relaxation. The inclusion of this relaxation will, in turn, increase the axion density in these streams and eventually increase the local axion density at the Earth's location.

In sum, the integration of a more nuanced disruption law with a Galactic‐scale Monte‐Carlo analysis delivers a substantially revised picture of how axion minihalos survive and ultimately influence observable signatures.  By providing both a sub-parsec and galactic‐scale context, the narrative developed here lays the groundwork for the methodological and observational advances in this field.



\begin{spacing}{1.1}


\bibliographystyle{JHEP}
\cleardoublepage
\bibliography{References/references} 



\end{spacing}


\begin{appendices} 


\chapter{Computing the expression for the normalized relative potential of an untruncated NFW minihalo}\label{app:psi for untruncated NFW}
The Newtonian gravitational potential for an untruncated NFW density profile is Ref.~\cite[equation 2.67]{binneyTremaine}
\begin{equation}\label{eq:Phi for untruncated NFW}
    \Phi(r) = -4\pi G\rho_{\rm s}r_{\rm s}^3 \frac{\ln\left(1+\frac{r}{r_{\rm s}}\right)}{r}
\end{equation}
Making the substitutions $\frac{r}{r_{\rm s}} = \frac{xr_{\rm vir}}{r_{\rm s}} = cx$ and $r_{\rm s} = \frac{r_{vir}}{c}$ in Eq.~(\ref{eq:Phi for untruncated NFW})
\begin{align}\label{eq:Phi(x) intermediate}
    \Phi(x) &= -4\pi G\rho_{\rm s}\frac{r_{\rm vir}^3}{c^3} \frac{\ln\left(1+cx\right)}{xr_{\rm vir}} \nonumber\\
    &= -\Psi_0 \times \frac{4\pi \rho_{\rm s}r_{\rm vir}^3}{M_{\rm vir}c^3}\frac{\ln\left(1+cx\right)}{x} 
\end{align}
where $\Psi_0 \equiv \frac{GM_{\rm vir}}{r_{\rm vir}}$

Substituting for $M_{\rm vir}$ from Eq.~(\ref{eq:M_vir}) in Eq.~(\ref{eq:Phi(x) intermediate})
\begin{equation}\label{eq:Phi intermediate 2}
    \Phi(x) = -\Psi_0  \frac{1}{f_{\rm NFW}(c)}\frac{\ln\left(1+cx\right)}{x} 
\end{equation}
Let the relative potential be $\Psi \equiv -\Phi$ and the normalized relative potential be $\psi \equiv \frac{\Psi}{\Psi_0}$. $\implies \psi = -\frac{\Phi}{\Psi_0}$. From Eq.~(\ref{eq:Phi intermediate 2}), this implies that the normalized relative potential is
\begin{equation}
    \psi(x) =   \frac{1}{f_{\rm NFW}(c)}\frac{\ln\left(1+cx\right)}{x} 
\end{equation}
\chapter{Computing the dimensionless expression for the survival fraction of an NFW minihalo}\label{app:survival fraction expression for NFW minihalo}
We start with Eq.~(\ref{eq:mass loss NFW}). Substituting with $r = xr_{\rm vir}$, $\Psi = \psi\Psi_0$, $\varepsilon = \epsilon\Psi_0$, $\vert\Delta\varepsilon\vert = \vert\Delta\epsilon\vert \Psi_0$ and $f = \frac{\rho_{\rm s}}{\Psi_0^{3/2}} \hat{f}$, we get
\begin{equation}\label{eq:Delta M dimensionless}
    \Delta M = 16\pi^2\rho_{\rm s}r_{\rm vir}^3 \int\limits_{x=0}^1 \mathrm{d}x\ x^2 \int\limits_{\epsilon=0}^{\min\left[\vert\Delta\epsilon(x)\vert, \psi(x)\right]} \mathrm{d}\epsilon
    \sqrt{2(\psi(x)-\epsilon)} \hat{f}(\epsilon)
\end{equation}
From Eq.~(\ref{eq:M_vir}),
\begin{equation}\label{eq:M_vir repeated}
    M_{\rm vir} = 4\pi\rho_{\rm s}r_{\rm vir}^3 \frac{f_{\rm NFW}(c)}{c^3}
\end{equation}
From Eqs.~(\ref{eq:Delta M dimensionless}) and (\ref{eq:M_vir repeated}), it follows that the survival fraction is
\begin{align}\label{eq:surival fraction expression dimensionless double integral REPEATED}
    \text{SF} &\equiv 1 - \frac{\Delta M}{M_{\rm vir}} \nonumber\\
    &= 1 - \frac{4\pi c^3}{f_{\rm NFW}(c)}\int\limits_{x=0}^1\mathrm{d}x\ x^2\int\limits_{\epsilon=0}^{\min[\vert\Delta\epsilon(x)\vert, \psi(x)]} \mathrm{d}\epsilon\ \hat{f}(\epsilon)\nonumber\\
    &\ \ \ \ \ \ \ \ \ \ \ \  \ \ \ \ \ \ \ \ \ \ \ \ \ \ \ \ \ \ \ \ \ \ \ \sqrt{2(\psi(x)-\epsilon)}
\end{align}
\chapter{Computing the mass of the minihalo in a shell of finite thickness}\label{app:Computing mass of minihalo}
K2021 gives the phase space distribution of dark matter particles in a minihalo as:
\rev{
\begin{equation}\label{eq:phase space distribution definition}
    f(\varepsilon) \equiv m_a\frac{\mathrm{d}N}{\mathrm{d}^3\vec{r}\mathrm{d}^3\vec{v}}
\end{equation}
}
\begin{equation}\label{eq:varepsilon definition}
    \varepsilon \equiv \Psi(r) - \frac{v^2}{2}
\end{equation}
where $\varepsilon$ is called the specific relative (total) energy, $\vec{r}$ is the radius vector associated with a dark matter particle, $\vec{v}$ is the corresponding velocity vector, \rev{$\mathrm{d}N$ is the number of axion particles} present in the phase space volume of $\mathrm{d}^3\vec{r}\mathrm{d}^3\vec{v}$. \rev{If $\mathrm{d}M$ is the total mass corresponding to this number of axion particles $\mathrm{d}N$}, and assuming spherical symmetry in physical space and velocity space,
\begin{equation}\label{eq:dM}
    \mathrm{d}M = 16\pi^2 f(\varepsilon) r^2v^2\mathrm{d}v\mathrm{d}r
\end{equation}
To find the mass of the minihalo between $r=r_1$ and $r=r_2$, we integrate Eq.~(\ref{eq:dM})
\begin{equation}\label{eq:DeltaM equals integral of dM}
    \delta M = 16\pi^2 \int\limits_{r=r_1}^{r_2} \int\limits_{v=0}^{v_{\rm max}(r)} f(\varepsilon) r^2v^2\mathrm{d}v\mathrm{d}r
\end{equation}
While performing the $v$-integral, $r$ remains fixed. Thus, differentiating Eq.~(\ref{eq:varepsilon definition})
\begin{equation}\label{eq:dVarepsilon}
    \mathrm{d}\varepsilon = -v\mathrm{d}v
\end{equation}
Moreover, using Eq.~(\ref{eq:varepsilon definition}), solving for $v$
\begin{equation}\label{eq:v solved}
    v = \sqrt{2(\Psi(r) - \varepsilon)}
\end{equation}
Also, we note that when $v = v_{\rm max}(r)$, $\varepsilon=0$ from Eq.~(\ref{eq:varepsilon definition}) since $\varepsilon$ cannot be negative for a dark matter particle bound to the minihalo.

Using Eqs.~(\ref{eq:dVarepsilon}) and (\ref{eq:v solved}), Eq.~(\ref{eq:DeltaM equals integral of dM}) becomes
\begin{equation}\label{eq:Delta M final equation}
    \delta M = 16\pi^2 \int\limits_{r=r_1}^{r_2} \int\limits_{\varepsilon=0}^{\Psi(r)} f(\varepsilon) r^2\sqrt{2(\Psi(r) - \varepsilon)}\mathrm{d}\varepsilon\mathrm{d}r
\end{equation}
Thus, when the upper limit of the $\varepsilon$-integral is $\Psi(r)$, Eq.~(\ref{eq:Delta M final equation}) gives the mass of the minihalo between the radii $r=r_1$ and $r=r_2$.

\chapter{Evaluating the survival fraction using the sequential stripping model}
\label{app:Evaluating the survival fraction using the sequential stripping model}
Eq.~(\ref{eq:cross over radius condition}) can be used to split the triple integral in Eq.~(\ref{eq:surival fraction expression dimensionless triple integral}) into two triple integrals as follows:
\begin{multline}\label{eq:surival fraction expression dimensionless with crossover radius}
    \text{SF} = 1 - 
    \frac{4\pi c^3}{f_{\rm NFW}(c)}\biggr[ 
    \int\limits_{x=0}^{\min[x^*, 1]} \int\limits_{\epsilon=0}^{\vert\Delta\epsilon(x)\vert} \  \int\limits_{\psi^\prime=0}^\epsilon     x^2\frac{1}{\sqrt{8}\pi^2}\ \frac{1}{\sqrt{\epsilon - \psi^\prime}}
    \frac{\mathrm{d}^2\varrho}{\mathrm{d}\psi^{\prime^2}}  \sqrt{2(\psi(x)-\epsilon)} \ \mathrm{d}\psi^\prime\ \mathrm{d}\epsilon\ \mathrm{d}x\\
    +    \int\limits_{x=\min[x^*, 1]}^1 \int\limits_{\epsilon=0}^{\psi(x)} \  \int\limits_{\psi^\prime=0}^\epsilon 
    x^2\frac{1}{\sqrt{8}\pi^2}
    \frac{1}{\sqrt{\epsilon - \psi^\prime}} \frac{\mathrm{d}^2\varrho}{\mathrm{d}\psi^{\prime^2}} \sqrt{2(\psi(x)-\epsilon)} \ \mathrm{d}\psi^\prime\ \mathrm{d}\epsilon\ \mathrm{d}x\biggr]
\end{multline}
Since we are only interested in the mass loss within the virial radius, a $\min[x^*, 1]$ term is introduced in the limits of the $x$-integral in Eq.~(\ref{eq:surival fraction expression dimensionless with crossover radius}) to account for the case when $x^* > 1$, where $x=1$ represents the virial radius.

Let's rewrite Eq.~(\ref{eq:surival fraction expression dimensionless with crossover radius}) as follows:
\begin{equation}\label{eq:survival fraction expression decomposed}
    \text{SF} = 1 - \text{prefactor} \times [I_{\rm A} + I_{\rm B}]
\end{equation}
where
\begin{equation}\label{eq:prefactor}
    \text{prefactor} \equiv \frac{4\pi c^3}{f_{\rm NFW}(c)}
\end{equation}
\begin{multline}\label{eq:I1}
    I_{\rm A} \equiv \int\limits_{x=0}^{\min[x^*, 1]} \int\limits_{\epsilon=0}^{\vert\Delta\epsilon(x)\vert} \  \int\limits_{\psi^\prime=0}^\epsilon x^2\frac{1}{\sqrt{8}\pi^2}\ \frac{1}{\sqrt{\epsilon - \psi^\prime}} \frac{\mathrm{d}^2\varrho}{\mathrm{d}\psi^{\prime^2}} \sqrt{2(\psi(x)-\epsilon)} \ \mathrm{d}\psi^\prime\ \mathrm{d}\epsilon\ \mathrm{d}x\
\end{multline}
\begin{multline}\label{eq:I2}
    I_{\rm B} \equiv \int\limits_{x=\min[x^*, 1]}^1 \int\limits_{\epsilon=0}^{\psi(x)} \  \int\limits_{\psi^\prime=0}^\epsilon x^2\frac{1}{\sqrt{8}\pi^2}\ \frac{1}{\sqrt{\epsilon - \psi^\prime}} \frac{\mathrm{d}^2\varrho}{\mathrm{d}\psi^{\prime^2}} \sqrt{2(\psi(x)-\epsilon)} \ \mathrm{d}\psi^\prime\ \mathrm{d}\epsilon\ \mathrm{d}x
\end{multline}
For illustration purposes, let's assume that $x^*<1$. Then, according to Eq.~(\ref{eq:I2}), $I_{\rm B}$ is non-zero. Here, ``prefactor $\times\ I_{\rm B}$" represents mass loss in the region from $x=x^*$ to $x=1$. It is important to note that when the upper limit of the $\epsilon$-integral is $\psi(x)$, then the triple integral calculates the mass of the minihalo between the lower and upper limits of the $x$-integral (see Appendix \ref{app:Computing mass of minihalo}). This implies a total mass loss in the region $x\in\left[x^*,1\right]$. To be exact, ``prefactor $\times\ I_{\rm B}$" is the mass loss fraction (relative to the virial mass) in the region $x\in\left[x^*,1\right]$. Then, ``1 - prefactor $\times\ I_{\rm B}$" represents the mass fraction of the region $x\in[0, x^*]$, since all mass in the region $x\in[x^*, 1]$ is lost due to stellar interaction. Thus,
\begin{equation}\label{eq:mass fraction x < x^*}
    1 - \text{prefactor} \times I_{\rm B} = \text{mass fraction}_{x < \min[x^*,1]}
\end{equation}
Thus, Eq.~(\ref{eq:survival fraction expression decomposed}) can be written as:
\begin{equation}\label{eq:SF expression in terms of mass fraction x < x^* and I_A}
    \text{SF} = \text{mass fraction}_{x < \min[x^*,1]} - \text{prefactor}\times I_{\rm A}
\end{equation}
According to Fig.~\ref{fig:psi crossing over deltaEpsilon}, for $x<x^*$, $\vert\Delta\epsilon(x)\vert < \psi(x)$. Since the upper limit of the $\epsilon$-integral in Eq.~(\ref{eq:I1}) is $\vert\Delta\epsilon(x)\vert$, ``prefactor $\times\ I_{\rm A}$" represents a partial mass loss occurring in the region $x\in[0, x^*]$.

We now have to mathematically evaluate the two terms of Eq.~(\ref{eq:SF expression in terms of mass fraction x < x^* and I_A}). Since ``mass fraction$_{x < \min[x^*,1]}$" represents the mass of the minihalo in the region $x\in\left[0, \min[x^*,1]\right]$, we can calculate it by simply integrating the NFW density profile from $x=0$ to $x=\min[x^*,1]$, and dividing the result by the virial mass which is a known expression.
\begin{multline}\label{eq:expression for mass fraction x < x^*, in terms of r}
    \text{Mass fraction}_{x < \min[x^*,1]} = \frac{1}{M_{\rm vir}} \int\limits_{r=0}^{\min[r^*, r_{\rm vir}]} \frac{\rho_{\rm s}}{\frac{r}{r_{\rm s}}\left(1 + \frac{r}{r_{\rm s}}\right)^2}
     4\pi r^2\mathrm{d}r
\end{multline}
where $r^*$ is the crossover radius defined as $r^* \equiv x^*r_{\rm vir}$.

It can be shown that Eq.~(\ref{eq:expression for mass fraction x < x^*, in terms of r}) reduces to (see Appendix \ref{app:Computing mass fraction below crossover radius}):
\begin{equation}\label{eq:expression for mass fraction x < x^*, in terms of x}
    \text{Mass fraction}_{x < \min[x^*,1]} = \frac{c^2}{f_{\rm NFW}(c)} \int\limits_{x=0}^{\min[x^*, 1]} \frac{x}{\left(1 + cx\right)^2} \mathrm{d}x
\end{equation}

For $x>x^*$, the resulting relative potential $\psi_{\rm B}(x)$ is given by (see Appendix (\ref{apppsiforxgreaterthanxstarsequentialstrippingmodel})),
\begin{align}\label{eq:psi2(x)}
    \psi_{\rm B}(x) = \frac{1}{f_{\rm NFW}(c)} \left[ \frac{\ln(1+cx)}{x} - \frac{c}{1+cx} \right], && x>x^*
\end{align}
To compute the expression for the relative potential in the region $x<x^*$, while taking the dark matter particle from $x$ to infinity, we assume that all matter is intact from normalized radius $x$ to $x^*$, and all matter is already stripped off in the region $x>x^*$. The resulting relative potential $\psi_{\rm A}(x)$ is given by (see Appendix (\ref{app:psi(x) for x < x^* - sequential stripping model})),
\begin{align}\label{eq:psi1(x)}
    \psi_{\rm A}(x) = \frac{1}{f_{\rm NFW}(c)} \left[ \frac{\ln(1+cx)}{x} - \frac{c}{1+cx^*} \right], && x<x^*
\end{align}
It is important to note that $\psi(x)$ is continuous at $x=x^*$ and hence $x^*$ can be easily evaluated using Eq.~(\ref{eq:cross over radius condition}).

To evaluate the mass loss in the region $x<x^*$, we have to evaluate ``prefactor $\times\ I_{\rm A}$'' in Eq.~(\ref{eq:SF expression in terms of mass fraction x < x^* and I_A}). In the expression for $I_{\rm A}$, as given by Eq.~(\ref{eq:I1}), the $\sqrt{2(\psi(x)-\epsilon)}$ term becomes $\sqrt{2(\psi_{\rm A}(x)-\epsilon)}$ because the $x$-integral in Eq.~(\ref{eq:I1}) ranges from $x=0$ to $x=x^*$. Thus, $\psi_{\rm A}(x)$ from Eq.~(\ref{eq:psi1(x)}) applies here.

In Eq.~(\ref{eq:I1}), there is also a term $\frac{\mathrm{d}^2\varrho}{\mathrm{d}\psi^{\prime^2}}$. Here , we have to ascertain if it is $\psi_{\rm A}(x)$ or $\psi_{\rm B}(x)$ that we differentiate $\varrho$ by. To find out, note that from the $\psi^\prime$-integral in Eq.~(\ref{eq:I1}), we have
\begin{equation}\label{eq:psi_prime condition}
    0 \leq \psi^\prime \leq \epsilon
\end{equation}
From the $\epsilon$-integral in Eq.~(\ref{eq:I1}), we have
\begin{equation}\label{eq:epsilon condition}
    \epsilon \leq \vert\Delta\epsilon(x)\vert\bigg\vert_{x\leq x^*}
\end{equation}
From Eqs.~(\ref{eq:normalized energy injected per unit mass (part 2) as a function of x}) and (\ref{eq:cross over radius condition}) we see that 
\begin{equation}\label{eq:Delat_epsilon(x) condition}
    \vert\Delta\epsilon(x)\vert\bigg\vert_{x\leq x^*} \leq \vert\Delta\epsilon(x^*)\vert = \psi(x^*)
\end{equation}
This can also be seen in Fig.~\ref{fig:psi crossing over deltaEpsilon} for the $c=10$ case.

From Eqs.~(\ref{eq:psi_prime condition}), (\ref{eq:epsilon condition}) and (\ref{eq:Delat_epsilon(x) condition}), it follows that
\begin{equation}\label{eq:psi_prime final condition}
    \psi^\prime \leq \psi(x^*)
\end{equation}
Fig.~\ref{fig:psi crossing over deltaEpsilon} then says that $\psi^\prime$ is in the region $x\geq x^*$. Thus, we must use $\psi_{\rm B}^\prime(x^\prime)$ to differentiate $\varrho$ in $\frac{\mathrm{d}^2\varrho}{\mathrm{d}\psi^{\prime^2}}$. Thus, $I_{\rm A}$ (from Eq.~(\ref{eq:I1})) can be rewritten as
\begin{multline}\label{eq:I1 updated to sequential stripping model}
    I_{\rm A} \equiv \int\limits_{x=0}^{\min[x^*, 1]} \int\limits_{\epsilon=0}^{\vert\Delta\epsilon(x)\vert} \  \int\limits_{\psi_{\rm B}^\prime=0}^\epsilon x^2\frac{1}{\sqrt{8}\pi^2}\ \frac{1}{\sqrt{\epsilon - \psi_{\rm B}^\prime}} \frac{\mathrm{d}^2\varrho}{\mathrm{d}\psi_{\rm B}^{\prime^2}}\left(x^\prime\left(\psi_{\rm B}^\prime\right)\right)\\ \sqrt{2\left(\psi_{\rm A}(x)-\epsilon\right)} \ \mathrm{d}\psi_{\rm B}^\prime\ \mathrm{d}\epsilon\ \mathrm{d}x\
\end{multline}
Using Eqs.~(\ref{eq:expression for mass fraction x < x^*, in terms of x}), (\ref{eq:prefactor}) and (\ref{eq:I1 updated to sequential stripping model}), the survival fraction can be computed using Eq.~(\ref{eq:SF expression in terms of mass fraction x < x^* and I_A}).

\chapter[Mass fraction of NFW minihalo below the normalized crossover radius]{Computing the mass fraction of the NFW minihalo below the normalized crossover radius}\label{app:Computing mass fraction below crossover radius}
In this section, we compute the expression for the mass fraction of the NFW minihalo in the range $x<\min[x^*,1]$. We start with Eq.~(\ref{eq:expression for mass fraction x < x^*, in terms of r}).
\begin{equation}\label{eq:expression for mass fraction x < x^*, in terms of r - repeated}
    \text{Mass fraction}_{x < \min[x^*,1]} = \frac{1}{M_{\rm vir}} \int\limits_{r=0}^{\min[r^*, r_{\rm vir}]} \frac{\rho_{\rm s}}{\frac{r}{r_{\rm s}}\left(1 + \frac{r}{r_{\rm s}}\right)^2} \times 4\pi r^2\mathrm{d}r
\end{equation}
Making the substitution $r=xr_{\rm vir}$ and $c=\frac{r_{\rm vir}}{r_{\rm s}}$ in Eq.~(\ref{eq:expression for mass fraction x < x^*, in terms of r - repeated})
\begin{equation}\label{eq:expression for mass fraction x < x^*, in terms of r - repeated - dimensionless}
    \text{Mass fraction}_{x < \min[x^*,1]} = \frac{4\pi\rho_{\rm s}r_{\rm vir}^3}{M_{\rm vir}} \int\limits_{x=0}^{\min[x^*, 1]} \frac{1}{cx\left(1 + cx\right)^2} x^2\mathrm{d}x
\end{equation}

where $x^* \equiv \frac{r^*}{r_{\rm vir}}$.

The mass enclosed within a sphere of radius $r$ for an NFW density profile is Ref.~\cite[equation 2.66]{binneyTremaine}
\begin{equation}\label{eq:enclosed mass for NFW}
    M_{\rm enc}(r) = 4\pi\rho_{\rm s}r_{\rm s}^3\left[ \ln\left(1+\frac{r}{r_{\rm s}}\right) +\frac{\frac{r}{r_{\rm s}}}{1+\frac{r}{r_{\rm s}}} \right]
\end{equation}
Making the substitutions $\frac{r}{r_{\rm s}} = \frac{xr_{\rm vir}}{r_{\rm s}} = cx$ and $r_{\rm s} = \frac{r_{\rm vir}}{c}$ in Eq.~(\ref{eq:enclosed mass for NFW})
\begin{equation}\label{eq:Menc(x)}
    M_{\rm enc}(x) = \frac{4\pi\rho_{\rm s}r_{\rm vir}^3}{c^3}\left[ \ln(1+cx) - \frac{cx}{1+cx} \right]
\end{equation}
The virial mass $M_{\rm vir}$ is defined as that mass contained within the virial radius ($x=1$). Thus
\begin{align}\label{eq:M_vir}
    M_{\rm vir} &\equiv M_{\rm enc}(x=1)\nonumber\\
    &= \frac{4\pi\rho_{\rm s}r_{\rm vir}^3}{c^3}\left[ \ln(1+c) - \frac{c}{1+c} \right]\nonumber\\
    &= 4\pi\rho_{\rm s}r_{\rm vir}^3 \frac{f_{\rm NFW}(c)}{c^3}
\end{align}
where $f_{\rm NFW}(c) \equiv \ln(1+c) - \frac{c}{1+c}$

Substituting for $M_{\rm vir}$ from Eq.~(\ref{eq:M_vir}) in Eq.~(\ref{eq:expression for mass fraction x < x^*, in terms of r - repeated - dimensionless})
\begin{equation}\label{eq:expression for mass fraction x < x^*, in terms of x - repeated}
    \text{Mass fraction}_{x < \min[x^*,1]} = \frac{c^2}{f_{\rm NFW}(c)} \int\limits_{x=0}^{\min[x^*, 1]} \frac{x}{\left(1 + cx\right)^2} \mathrm{d}x
\end{equation}
\chapter{Computing the expression for the normalized relative potential of an NFW minihalo in the region 
    $x>x^*$ using the sequential stripping model}
\label{apppsiforxgreaterthanxstarsequentialstrippingmodel}
For the sequential stripping model, in the region $x>x^*$ where there is complete mass loss, we start by taking the outermost shell to infinity and then the next outermost shell, and so on. Thus, in the region $x>x^*$, when a dark matter particle is taken to infinity, it doesn't encounter a net force from dark matter particles in outer shells. So, the enclosed mass that the dark matter particle sees remains fixed. The Newtonian gravitational potential is Ref.~\cite[equation (2.67)]{binneyTremaine}
\begin{equation}\label{eq:Phi(r) fundamental equation}
    \Phi(r) = -G \int\limits_{r^\prime=r}^{\infty} \frac{M_{\rm enc}(r^\prime)}{r^{\prime^2}}\mathrm{d}r^\prime
\end{equation}
Making the substitution $r^\prime = x^\prime r_{\rm vir}$, Eq.~(\ref{eq:Phi(r) fundamental equation}) becomes
\begin{equation}\label{eq:Phi(x) intermediate 3}
    \Phi(x) = -\frac{G}{r_{\rm vir}} \int\limits_{x^\prime=x}^{\infty} \frac{M_{\rm enc}(x^\prime)}{x^{\prime^2}}\mathrm{d}x^\prime
\end{equation}
where $x \equiv \frac{r}{r_{\rm vir}}$

But the enclosed mass seen by the dark matter particle is always $M_{\rm enc}(x)$, even as it is taken to infinity. Thus
\begin{align}\label{eq:Phi(x) intermediate 4}
    \Phi(x) &= -\frac{G}{r_{\rm vir}} M_{\rm enc}(x) \int\limits_{x^\prime=x}^{\infty} \frac{1}{x^{\prime^2}}\mathrm{d}x^\prime \nonumber\\
    &= -\frac{G}{r_{\rm vir}} M_{\rm enc}(x) \frac{1}{x}\nonumber\\
    &= -\Psi_0 \frac{M_{\rm enc}(x)}{M_{\rm vir}} \frac{1}{x}
\end{align}
where $\Psi_0 \equiv \frac{GM_{\rm vir}}{r_{\rm vir}}$.

Substituting for $M_{\rm enc}(x)$ from Eq.~(\ref{eq:Menc(x)}) and $M_{\rm vir}$ from Eq.~(\ref{eq:M_vir}), Eq.~(\ref{eq:Phi(x) intermediate 4}) becomes
\begin{align}\label{eq:Phi(x) intermediate 5}
    \Phi(x) &= -\Psi_0 \frac{1}{f_{\rm NFW}(c)} \left[ \ln(1+cx) - \frac{cx}{1+cx} \right] \frac{1}{x} \nonumber\\
    &= -\Psi_0 \frac{1}{f_{\rm NFW}(c)} \left[ \frac{\ln(1+cx)}{x} - \frac{c}{1+cx} \right]
\end{align}
Let $\Psi \equiv -\Phi$ and $\psi \equiv \frac{\Psi}{\Psi_0}$ $\implies \psi = -\frac{\Phi}{\Psi_0}$. From Eq.~(\ref{eq:Phi(x) intermediate 5}), this implies that the normalized relative potential is
\begin{align}
    \psi_{\rm B}(x) = \frac{1}{f_{\rm NFW}(c)} \left[ \frac{\ln(1+cx)}{x} - \frac{c}{1+cx} \right] &&, x>x^*
\end{align}
\chapter{Computing the expression for the normalized relative potential of an NFW minihalo in the region 
    $x<x^*$ using the sequential stripping model}
\label{app:psi(x) for x < x^* - sequential stripping model}
Here, we utilize the sequential stripping model. For the purposes of computing the relative potential in the region $x<x^*$, we assume that there is no mass loss in the region $x<x^*$ and complete mass loss has already occurred in the region $x>x^*$, in keeping with the sequential stripping model. From Eq.~(\ref{eq:Phi(x) intermediate 3}), the Newtonian gravitational potential in the region $x<x^*$ is
\begin{align}\label{eq:Phi(x) intermediate 6}
    \Phi(x) &= -\frac{G}{r_{\rm vir}} \int\limits_{x^\prime=x}^{\infty} \frac{M_{\rm enc}(x^\prime)}{x^{\prime^2}}\mathrm{d}x^\prime \nonumber\\
    &= -\frac{G}{r_{\rm vir}} \left[\ \int\limits_{x^\prime=x}^{x^*} \frac{M_{\rm enc}(x^\prime)}{x^{\prime^2}}\mathrm{d}x^\prime + M_{\rm enc}(x^*)\int\limits_{x^\prime=x^*}^{\infty} \frac{1}{x^{\prime^2}}\mathrm{d}x^\prime \right]
\end{align}
because in the region $x>x^*$, the dark matter particle only sees an enclosed mass of $M_{\rm enc}(x^*)$ since all the mass in the region $x>x^*$ is already stripped off.

Substituting for $M_{\rm enc}(x)$ from Eq.~(\ref{eq:Menc(x)}), Eq.~(\ref{eq:Phi(x) intermediate 6}) becomes
\begin{align*}
    \Phi(x) = -\frac{G}{r_{\rm vir}} \frac{4\pi\rho_{\rm s}r_{\rm vir}^3}{c^3} \biggr[ &\int\limits_{x^\prime=x}^{x^*} \frac{1}{x^{\prime^2}}\left[ \ln(1+cx^\prime) - \frac{cx^\prime}{1+cx^\prime} \right] \mathrm{d}x^\prime \\
    &+ \left[ \ln(1+cx^*) - \frac{cx^*}{1+cx^*} \right] \frac{1}{x^*} \biggr]
\end{align*}
\begin{align*}
    \Phi(x) = -\frac{G}{r_{\rm vir}} \frac{4\pi\rho_{\rm s}r_{\rm vir}^3}{c^3} \biggr[&\left( \frac{\ln(1+cx)}{x} - \frac{\ln(1+cx^*)}{x^*} \right)\\\
    &+ \left( \frac{\ln(1+cx^*)}{x^*} - \frac{c}{1+cx^*} \right) \biggr]
\end{align*}
\begin{equation}\label{eq:Phi(x) intermediate 7}
    \Phi(x) = -\Psi_0 \frac{4\pi\rho_{\rm s}r_{\rm vir}^3}{M_{\rm vir}c^3} \left[ \frac{\ln(1+cx)}{x} - \frac{c}{1+cx^*} \right]
\end{equation}
Substituting for $M_{\rm vir}$ from Eq.~(\ref{eq:M_vir}), Eq.~(\ref{eq:Phi(x) intermediate 7}) becomes
\begin{equation}\label{eq:Phi(x) intermediate 8}
    \Phi(x) = -\Psi_0 \frac{1}{f_{\rm NFW}(c)} \left[ \frac{\ln(1+cx)}{x} - \frac{c}{1+cx^*} \right]
\end{equation}
Let $\Psi \equiv -\Phi$ and $\psi \equiv \frac{\Psi}{\Psi_0}$. $\implies \psi = -\frac{\Phi}{\Psi_0}$. From Eq.~(\ref{eq:Phi(x) intermediate 8}), this implies that the normalized relative potential is
\begin{align}
    \psi_{\rm A}(x) = \frac{1}{f_{\rm NFW}(c)} \left[ \frac{\ln(1+cx)}{x} - \frac{c}{1+cx^*} \right], && x<x^*
\end{align}

\chapter{Computing the expression for the normalized relative potential of the Hernquist density profile using the sequential stripping model}
\label{app:psi for Hernquist minihalo}
For a Hernquist profile, the mass enclosed within radius $r$ is Ref.~\cite[equation 2.66]{binneyTremaine}
\begin{equation}\label{eq:enclosed mass for Hernquist minihalo as a functino of r}
    M_{\rm enc}(r) = 2\pi\rho_{\rm s}r_{\rm s}^3 \frac{\left(\frac{r}{r_{\rm s}}\right)^2}{\left(1+\frac{r}{r_{\rm s}}\right)^2}
\end{equation}
Using Eqs. (\ref{eq:x definition}) and (\ref{eq:concentration defintion}), Eq.~(\ref{eq:enclosed mass for Hernquist minihalo as a functino of r}) can be rewritten as
\begin{equation}\label{eq:enclosed mass for Hernquist minihalo as a functino of x}
    M_{\rm enc}(x) = \frac{2\pi\rho_{\rm s}r_{\rm vir}^3}{c} \frac{x^2}{\left(1+cx\right)^2}
\end{equation}
The virial mass is, by definition
\begin{equation}\label{eq:virial mass of Hernquist minihalo}
    M_{\rm vir} \equiv M_{\rm enc}(x=1) = \frac{2\pi\rho_{\rm s}r_{\rm vir}^3}{c} \frac{1}{(1+c)^2}
\end{equation}
From Eqs.~(\ref{eq:enclosed mass for Hernquist minihalo as a functino of x}) and (\ref{eq:virial mass of Hernquist minihalo}), it follows that
\begin{equation}\label{eq:enclosed mass for Hernquist minihalo as a functino of x, part 2}
    M_{\rm enc}(x) = M_{\rm vir} (1+c)^2 \frac{x^2}{\left(1+cx\right)^2}
\end{equation}
According to Eq.~(\ref{eq:Phi(x) intermediate 3}),
\begin{equation}\label{eq:Phi(x) intermediate 9}
    \Phi(x) = -\frac{G}{r_{\rm vir}} \int\limits_{x^\prime=x}^{\infty} \frac{M_{\rm enc}(x^\prime)}{x^{\prime^2}}\mathrm{d}x^\prime
\end{equation}
The normalized crossover radius $x^*$ is defined by Eq.~(\ref{eq:cross over radius condition}). We now look at two cases.

\noindent\underline{Case $x>x^*$}:

In the sequential stripping model, when computing the normalized relative potential $\psi$ at normalized radius $x$, we assume that all shells outward from $x$ have already been stripped off. Thus, as the dark matter particle is taken from $x$ to infinity, the enclosed mass is always $M_{\rm enc}(x)$. Thus Eq.~(\ref{eq:Phi(x) intermediate 9}) becomes
\begin{equation}\label{eq:Phi(x) intermediate 10}
    \Phi(x) = -\frac{G}{r_{\rm vir}} M_{\rm enc}(x) \frac{1}{x}
\end{equation}
\begin{equation}\label{eq:Phi(x) intermediate 11}
    \Phi(x) = -\Psi_0 (1+c)^2\frac{x}{\left(1+cx\right)^2}
\end{equation}
where $\Psi_0 \equiv \frac{GM_{\rm vir}}{r_{\rm vir}}$.

Defining $\Psi \equiv -\Phi$, and $\psi \equiv \frac{\Psi}{\Psi_0} = -\frac{\Phi}{\Psi_0}$, we compute the normalized relative potential $\psi$ as
\begin{align}\label{eq:psi for x>x^* for a Hernquist profile}
    \psi_{\rm B}(x) = (1+c)^2\frac{x}{\left(1+cx\right)^2}, && x>x^*
\end{align}
\noindent\underline{Case $x<x^*$}:

We assume complete mass loss in the region $x>x^*$. In the sequential stripping model, when computing $\psi(x)$, we assume all shells for which $x>x^*$ have already been stripped off. We also assume no shell stripping has occurred in the range $[x,x^*]$. Thus, Eq.~(\ref{eq:Phi(x) intermediate 9}) becomes
\begin{equation}\label{eq:Phi(x) intermediate 12}
    \Phi(x) = -\frac{G}{r_{\rm vir}} \left[\ \ \int\limits_{x^\prime=x}^{x^*} \frac{M_{\rm enc}(x^\prime)}{x^{\prime^2}}\mathrm{d}x^\prime + \int\limits_{x^\prime=x^*}^{\infty} \frac{M_{\rm enc}(x^*)}{x^{\prime^2}}\mathrm{d}x^\prime \right]
\end{equation}
Substituting for $M_{\rm enc}$ from Eq.~(\ref{eq:enclosed mass for Hernquist minihalo as a functino of x, part 2}), Eq.~(\ref{eq:Phi(x) intermediate 12}) becomes
\begin{equation}\label{eq:Phi(x) intermediate 13}
    \Phi(x) = -\Psi_0 (1+c)^2 \left[ \frac{x^*-x}{(1+cx^*)(1+cx)} + \frac{x^*}{(1+cx^*)^2} \right]
\end{equation}
Again, by definition, $\psi = -\frac{\Phi}{\Psi_0}$. We can then compute the normalized relative potential $\psi$ as
\begin{align}\label{eq:psi for x<x^* for a Hernquist profile}
    \psi_{\rm A}(x) = (1+c)^2 \left[ \frac{x^*-x}{(1+cx^*)(1+cx)} + \frac{x^*}{(1+cx^*)^2} \right], && x<x^*
\end{align}
\chapter{Evaluating the expression for $\alpha^2$ and $\gamma$ for a Hernquist minihalo}
\label{app:alpha squared and beta for Herquist minihalo}
\noindent\underline{$\alpha^2$}:

For a spherically symmetric density profile like the Hernquist profile (S2023)
\begin{equation}\label{eq:alpha squared for Hernquist profile part 1}
    \alpha^2 = \frac{1}{M_{\rm vir}r_{\rm vir}^2}\ \int\limits_{r=0}^{r_{\rm vir}} \mathrm{d}^3\vec{r}\ r^2\rho(r)
\end{equation}
Since the density profile is spherically symmetric, Eq.~(\ref{eq:alpha squared for Hernquist profile part 1}) becomes
\begin{equation}\label{eq:alpha squared for Hernquist profile part 2}
    \alpha^2 = \frac{4\pi}{M_{\rm vir}r_{\rm vir}^2} \int\limits_{r=0}^{r_{\rm vir}} \rho(r)r^4\mathrm{d}r
\end{equation}
Instead of $r$, writing the variable of integration as $x$, Eq.(\ref{eq:alpha squared for Hernquist profile part 2}) can be written as
\begin{equation}\label{eq:alpha squared for Hernquist profile part 3}
    \alpha^2 = \frac{4\pi r_{\rm vir}^3}{M_{\rm vir}}\ \int\limits_{x=0}^1 \rho(x)x^4\mathrm{d}x
\end{equation}
Substituting for $\rho(x)$ from Eq.~(\ref{eq:Hernquist profile in terms of x}), we can then compute the one-dimensional integral in Eq.~(\ref{eq:alpha squared for Hernquist profile part 3}). Thus, Eq.~(\ref{eq:alpha squared for Hernquist profile part 3}) becomes
\begin{equation}\label{eq:alpha squared for Hernquist profile part 4}
    \alpha^2 = \frac{4\pi\rho_{\rm s}r_{\rm vir}^3}{M_{\rm vir}} \left[ \frac{\frac{c(6+9c+2c^2)}{(1+c)^2} - 6\ln(1+c)}{2c^5} \right]
\end{equation}
Substituting for $M_{\rm vir}$ from Eq.~(\ref{eq:virial mass of Hernquist minihalo}), Eq.~(\ref{eq:alpha squared for Hernquist profile part 4}) can be written as
\begin{equation}\label{eq:alpha squared for Hernquist profile part 5}
    \alpha^2 = \frac{c(6+9c+2c^2) - 6(1+c)^2\ln(1+c)}{c^4}
\end{equation}
\noindent\underline{$\gamma$}:

For a spherically symmetric density profile like the Hernquist profile (K2021)
\begin{equation}\label{eq:beta for Hernquist profile part 1}
    \gamma = \frac{4\pi r_{\rm vir}}{M_{\rm vir}^2}\ \int\limits_{r=0}^{r_{\rm vir}} M_{\rm enc}(r)\rho(r)r\mathrm{d}r
\end{equation}
Changing the variable of integration from $r$ to $x$, Eq.~(\ref{eq:beta for Hernquist profile part 1}) becomes
\begin{equation}\label{eq:beta for Hernquist profile part 2}
    \gamma = \frac{4\pi r_{\rm vir}^3}{M_{\rm vir}^2}\ \int\limits_{x=0}^1 M_{\rm enc}(x)\rho(x)x\mathrm{d}x
\end{equation}
Substituting for $M_{\rm enc}(x)$ from Eq.~(\ref{eq:enclosed mass for Hernquist minihalo as a functino of x, part 2}), $\rho(x)$ from Eq.~(\ref{eq:Hernquist profile in terms of x}) and $M_{\rm vir}$ from Eq.~(\ref{eq:virial mass of Hernquist minihalo}), the one dimensional integral in Eq.~(\ref{eq:beta for Hernquist profile part 2}) can be evaluated and Eq.~(\ref{eq:beta for Hernquist profile part 2}) can be succinctly written as
\begin{equation}\label{eq:beta for Hernquist profile part 3}
    \gamma = \frac{4+c}{6}
\end{equation}

\chapter{Computing expressions for the disrupted minihalo's parameters}
\label{app:Computing expressions for the disrupted minihalo's parameters}
We now compute each of the terms in Eq.~(\ref{eq:mass condition for transition from NFW minihalo to first-generation minihalo}). For an NFW profile, the mass enclosed within radius $r$ is (Ref.~\cite[Eq. (2.66)]{binneyTremaine})
\begin{equation}\label{eq:enclosed mass for NFW profile}
    M_{\text {enc,s }}(r)=4 \pi \rho_{\rm s} r_{\rm s}^3\left[\ln \left(1+\frac{r}{r_{\rm s}}\right)-\frac{\frac{r}{r_{\rm s}}}{1+\frac{r}{r_{\rm s}}}\right]
\end{equation}
Now,
\begin{equation}\label{eq:frac{r}{r_s}}
    \frac{r}{r_{\rm s}}=\frac{r}{r_{\rm vir, s}} \frac{r_{\text {vir,s}}}{r_{\rm s}}=x_{\rm s} c_{\rm s}
\end{equation}
where
\begin{equation}\label{eq:c_s defintion}
    c_{\rm s} \equiv \frac{r_{\rm vir,s}}{r_{\rm s}}
\end{equation}
Substituting Eq.~(\ref{eq:frac{r}{r_s}}) in Eq.~(\ref{eq:enclosed mass for NFW profile})
\begin{align}
    M_{\text {enc,s }}\left(x_{\rm s}\right)&=4 \pi \rho_{\rm s} r_{\rm s}^3\left[\ln \left(1+c_{\rm s} x_{\rm s}\right)-\frac{c_{\rm s} x_{\rm s}}{1+c_{\rm s} x_{\rm s}}\right] \nonumber\\
    &=4 \pi \rho_{\rm s} r_{\rm s}^3 f_{\rm NFW}(c_{\rm s} x_{\rm s})
\end{align}
\begin{align}\label{eq:mass enclosed within crossover radius for NFW profile}
    \Rightarrow M_{\text {enc,s }}\left(x_{\rm s}^*\right)&=4 \pi \rho_{\rm s} r_{\rm s}^3f_{\rm NFW}(c_{\rm s} x_{\rm s}^*)
\end{align}
Next, prefactor $\times I_{\rm A}$ from Eqs.~(\ref{eq:prefactor}) and (\ref{eq:I1 updated to sequential stripping model}), computes the mass loss fraction between $x=0$ and $x=\min \left[x^{*}, 1\right]$. We can convert a mass loss fraction term to a mass loss term by multiplying with $M_{\rm vir}$. Thus,
\begin{multline}\label{eq:partial mass loss within crossover radius for NFW minihalo}
    \Delta M_{x_{\rm s}=0 \rightarrow x_{\rm s}^*}=M_{\rm vir} \times \frac{4 \pi c_{\rm s}^3}{f_{\rm NFW}(c_{\rm s})} \int\limits_{x=0}^{x_{\rm s}^*} \int\limits_{\epsilon=0}^{\vert\Delta\epsilon(x)\vert} \  \int\limits_{\psi_{\rm B}^\prime=0}^\epsilon x^2\frac{1}{\sqrt{8}\pi^2}\\ \times\frac{1}{\sqrt{\epsilon - \psi_{\rm B}^\prime}} \frac{\mathrm{d}^2\varrho}{\mathrm{d}\psi_{\rm B}^{\prime^2}}\left(x^\prime\left(\psi_{\rm B}^\prime\right)\right) \sqrt{2\left(\psi_{\rm A}(x)-\epsilon\right)} \ \mathrm{d}\psi_{\rm B}^\prime\ \mathrm{d}\epsilon\ \mathrm{d}x\
\end{multline}
Substituting for $M_{\rm vir}$ from Eq.~(\ref{eq:M_vir}) in Eq.~(\ref{eq:partial mass loss within crossover radius for NFW minihalo})
\begin{align}\label{eq:partial mass loss within crossover radius for NFW minihalo, intermediate 2}
    \Delta M_{x_{\rm s}=0 \rightarrow x_{\rm s}^*} &= 16\pi^2\rho_{\rm s}r_{\rm s}^3c_{\rm s}^3 \int\limits_{x=0}^{x_{\rm s}^*} \int\limits_{\epsilon=0}^{\vert\Delta\epsilon(x)\vert} \  \int\limits_{\psi_{\rm B}^\prime=0}^\epsilon x^2\frac{1}{\sqrt{8}\pi^2} \frac{1}{\sqrt{\epsilon - \psi_{\rm B}^\prime}} \frac{\mathrm{d}^2\varrho}{\mathrm{d}\psi_{\rm B}^{\prime^2}}\left(x^\prime\left(\psi_{\rm B}^\prime\right)\right) 
     \sqrt{2\left(\psi_{\rm A}(x)-\epsilon\right)} \ \mathrm{d}\psi_{\rm B}^\prime\ \mathrm{d}\epsilon\ \mathrm{d}x\nonumber\\
    &= 16\pi^2\rho_{\rm s}r_{\rm s}^3c_{\rm s}^3 \times I_{\rm s}
\end{align}
where
\begin{multline}\label{eq:I_s definition}
    I_{\rm s} = \int\limits_{x=0}^{x_{\rm s}^*} \int\limits_{\epsilon=0}^{\vert\Delta\epsilon(x)\vert} \  \int\limits_{\psi_{\rm B}^\prime=0}^\epsilon x^2\frac{1}{\sqrt{8}\pi^2} \frac{1}{\sqrt{\epsilon - \psi_{\rm B}^\prime}} \frac{\mathrm{d}^2\varrho}{\mathrm{d}\psi_{\rm B}^{\prime^2}}\left(x^\prime\left(\psi_{\rm B}^\prime\right)\right)
    \sqrt{2\left(\psi_{\rm A}(x)-\epsilon\right)} \ \mathrm{d}\psi_{\rm B}^\prime\ \mathrm{d}\epsilon\ \mathrm{d}x
\end{multline}
Next, in general, for a broken power law of the form 
given by Eq. (\ref{eq:broken power law})
the total enclosed mass is
\begin{align}\label{eq:total mass of broken power law profile with k = 2 + Delta}
    \lim _{x_1 \rightarrow \infty} M_{\rm enc,1}\left(x_1\right) &= \lim _{r \rightarrow \infty} M_{\text {enc,1}}(r) \nonumber\\
    & =\int\limits_{r=0}^{\infty} \rho_{k=2+\Delta}(r) \times 4 \pi r^2 d r \nonumber\\
    & =\int\limits_{r=0}^{\infty} \frac{\rho_1}{\frac{r}{r_1}\left(1+\frac{r}{r_1}\right)^{2+\Delta}} 4 \pi r^2 d r \nonumber\\
    & =\frac{4 \pi \rho_1 r_1^3}{\Delta+\Delta^2}
\end{align}
Finally, substituting Eqs.~(\ref{eq:mass enclosed within crossover radius for NFW profile}), (\ref{eq:partial mass loss within crossover radius for NFW minihalo, intermediate 2}) and (\ref{eq:total mass of broken power law profile with k = 2 + Delta}) in Eq.~(\ref{eq:mass condition for transition from NFW minihalo to first-generation minihalo})

\begin{align}\label{eq:mass condition for transition from NFW minihalo to first-generation minihalo substituted}
    4 \pi \rho_{\rm s} r_{\rm s}^3 f_{\rm NFW}(c_{\rm s} x_{\rm s}^*)-16 \pi^2 \rho_{\rm s} r_{\rm s}^3 c_{\rm s}^3 I_{\rm s} =\frac{4 \pi \rho_1 r_1^3}{\Delta+\Delta^2}
\end{align}

Dividing Eq.~(\ref{eq:mass condition for transition from NFW minihalo to first-generation minihalo substituted}) by $4 \pi \rho_{\rm s} r_{\rm s}^3$
\begin{equation}\label{eq:mass condition for transition from NFW minihalo to first-generation minihalo substituted, intermediate 2}
    f_{\rm NFW}(c_{\rm s} x_{\rm s}^*)-4 \pi c_{\rm s}^3 I_{\rm s}=\frac{1}{\Delta + \Delta^2}\left(\frac{\rho_1 r_1}{\rho_{\rm s} r_{\rm s}}\right)\left(\frac{r_1}{r_{\rm s}}\right)^2
\end{equation}

Substituting Eq.~(\ref{eq:small radius condition}) in Eq.~(\ref{eq:mass condition for transition from NFW minihalo to first-generation minihalo substituted, intermediate 2}), we can then compute $r_1$ in terms of $r_{\rm s}$ as follows
\begin{align}\label{eq:r_1 final expression - broken power law k=3.2}
    r_1&=r_{\rm s} \sqrt{(\Delta + \Delta^2)\left[f_{\rm NFW}(c_{\rm s} x_{\rm s}^*)-4 \pi c_{\rm s}^3 I_{\rm s}\right]} \nonumber\\
    &=r_{\rm s} \times R_{\rm s}
\end{align}
where
\begin{equation}\label{eq:r_s defintion - broken power law k=3.2}
    R_{\rm s} \equiv \sqrt{(\Delta + \Delta^2)\left[f_{\rm NFW}(c_{\rm s} x_{\rm s}^*)-4 \pi c_{\rm s}^3 I_{\rm s}\right]}
\end{equation}
Substituting Eq.~(\ref{eq:r_1 final expression - broken power law k=3.2}) in Eq.~(\ref{eq:small radius condition}), we get
\begin{equation}\label{eq:rho_1 final condition}
    \rho_1 = \frac{\rho_{\rm s}}{R_{\rm s}}
\end{equation}

\chapter{Computing the expression for survival fraction, incorporating relaxation}
\label{app:Computing the expression for survival fraction with relaxation}
Here, we compute the expression for the survival fraction, assuming that the remnant minihalo following a stellar encounter (with an NFW minihalo) relaxes to a Hernquist profile. Our region of interest for calculating the survival fraction is the physical virial radius of the unperturbed NFW minihalo. Thus, the survival fraction is the ratio of the mass enclosed by the region of interest for the relaxed Hernquist profile to the mass enclosed by the same region of interest for the unperturbed NFW profile which is just the virial mass $M_{\text{vir,s}}$ of the NFW minihalo.

\begin{equation}\label{eq:survival fraction incorporating relaxation - appendix version}
    \text {SF} \equiv \frac{M_{\text {enc,1}}\left(x_1^{r_{\text{vir,s}}}\right)}{M_{\text{vir,s}}}
\end{equation}
Here, $x_1^{r_{\text{vir,s}}}$ is the physical virial radius of the unperturbed NFW minihalo expressed in the ``local" normalized radial distance variable of the relaxed Hernquist minihalo.

\begin{align}
    x_1^{r_{\text{vir,s}}} &\equiv \frac{r_{\rm vir,s}}{r_{\rm vir,1}} \\
    & =\frac{c_{\rm s} r_{\rm s}}{c_1 r_1} \nonumber\\
    & =\frac{c_{\rm s}}{c_1} \frac{1}{R_{\rm s}} 
\end{align}
where $r_{\rm vir,s}$ and $r_{\rm vir,1}$ are the physical virial radii of the unperturbed NFW minihalo and the relaxed Hernquist minihalo, respectively. $c_{\rm s}$ and $c_1$ are the concentrations of the NFW and Hernquist minihalos respectively.
According to  Ref.~\cite[Eq.(2.66)]{binneyTremaine}
\begin{align}
    M_{\rm enc,1}(r) &= 2 \pi \rho_1  r_1^3 \frac{\left(r / r_1\right)^2}{\left(1+r / r_1\right)^2} \\
    \implies M_{\rm enc,1}\left(x_1\right) &= 2 \pi \rho_1 r_1^3 \frac{\left(c_1 x_1\right)^2}{\left(1+c_1 x_1\right)^2}
\end{align}
where $x_1 \equiv \frac{r}{r_{\rm vir, 1}}$.
Thus,

\begin{align}\label{eq:mass enclosed within physical virial radius (of NFW minihalo) for a Hernquist profile}
    M_{\rm enc,1}\left(x_1^{r_{\rm vir,s}}\right) &= 2 \pi \rho_1 r_1^3 \frac{\left(c_1 x_1^{r_{\rm vir,s}}\right)^2}{\left(1+c_1 x_1^{r_{\rm vir,s}}\right)^2} \nonumber\\
    & =2 \pi \rho_1 r_1^3 f_{\rm Hern}(c_1 x_1^{r_{\rm vir,s}})
\end{align}
Next, $M_{\rm vir,s}$ is given by Eq.~(\ref{eq:M_vir}) but it can be rewritten as

\begin{equation}\label{eq:virial mass of NFW profile using r_s}
    M_{\rm vir, s} = 4 \pi \rho_{\rm s} r_{\rm s}^3 f_{\rm NFW}(c_{\rm s})
\end{equation}
Substituting Eqs.~(\ref{eq:mass enclosed within physical virial radius (of NFW minihalo) for a Hernquist profile}) and (\ref{eq:virial mass of NFW profile using r_s}) in Eq.~(\ref{eq:survival fraction incorporating relaxation - appendix version})
\begin{align}\label{eq:survival fraction incorporating relaxation; pref-final expression}
    \text{SF} &=\frac{2 \pi \rho_1 r_1^3 f_{\rm Hern}(c_1 x_1^{r_{\rm vir,s}})}{4 \pi \rho_{\rm s} r_{\rm s}^3 f_{\rm NFW}(c_{\rm s})} \nonumber\\
    & =\frac{1}{2}\left(\frac{\rho_1 r_1}{\rho_{\rm s} r_{\rm s}}\right)\left(\frac{r_1}{r_{\rm s}}\right)^2 \frac{f_{\rm Hern}(c_1 x_1^{r_{\rm vir,s}})}{f_{\rm NFW}(c_{\rm s})}
\end{align}
Substituting Eqs.~(\ref{eq:small radius condition}) and (\ref{eq:r_1 final expression - broken power law k=3.2}) in Eq.~(\ref{eq:survival fraction incorporating relaxation; pref-final expression})

\begin{equation}
    \text{SF} =\frac{1}{2} R_{\rm s}^2 \frac{f_{\rm Hern}(c_1 x_1^{r_{\rm vir,s}})}{f_{\rm NFW}(c_{\rm s})}
\end{equation}
\chapter{Evaluating mass loss under multiple stellar encounters of an NFW minihalo}
\label{app:multiple stellar encounters}
\section{Computing the concentration of a Hernquist minihalo given its scale density}\label{app:computing the concentration of Hernquist minihalo, given its scale density}
We start with the definition of the virial radius of the Hernquist minihalo, analogous to Eq.~(\ref{eq:Virial radius condition}). This leads us to an equation similar to Eq.~(\ref{eq:Virial radius condition, intermediate 2}) but for the first-generation Hernquist minihalo. Thus, we have

\begin{equation}\label{eq:virial radius condition intermediate for Hernquist profile}
    \int\limits_{x_1=0}^1 \rho_{\text {Hern }}\left(x_1\right) x_1^2 d x_1=\frac{200}{3} \rho_{\text {crit }}
\end{equation}
where
\begin{equation}
    x_1 \equiv \frac{r}{r_{\rm vir,1}}
\end{equation}
and $r_{\rm vir,1}$ is the virial radius of the first-generation Hernquist minihalo. Substituting Eq.~(\ref{eq:Hernquist profile in terms of x}) in Eq.(\ref{eq:virial radius condition intermediate for Hernquist profile}) and performing the integral with respect to $x_1$ in the L.H.S. of Eq.(\ref{eq:virial radius condition intermediate for Hernquist profile}), we get

\begin{equation}\label{eq:relating concentration and scale radius for Hernquist profile, appendix version}
    \frac{1}{2 c_1\left(1+c_1\right)^2}=\frac{200}{3}\frac{\rho_{\text {crit}}}{\rho_1}
\end{equation}
\section{Evaluating the normalized virial radius of the unperturbed NFW minihalo, expressed in the local variable of the first-generation Hernquist minihalo}\label{app:normalized radius of NFW minihalo, expresssed in terms of local variable of first-generation minihalo}
Here, we compute $x_1^{r_{\rm vir,s}}$, the normalized virial radius of the unperturbed NFW minihalo, expressed in the local variable of the first-generation Hernquist minihalo, as follows

\begin{align}
    x_1^{r_{\text {vir,s}}} & \equiv \frac{r_{\text {vir,s}}}{r_{\text {vir,1}}} \nonumber\\
    & =\frac{c_{\rm s} r_{\rm s}}{c_1 r_1} \nonumber\\
    & =\frac{c_{\rm s}}{c_1} \frac{1}{R_{\rm s}}
\end{align}
\section{Computing the ratio of scale radii of the $(n+1)^{\text{th}}$ and $n^{\text{th}}$ generation minihalos. Also, computing the scale density of the $(n+1)^{\text{th}}$ generation minihalo}\label{app:rn+1_by_rn and rho_n+1 derivation}
We start with Eq.~(\ref{eq:mass condition for transition from n^th to (n+1)^th generation Hernquist minihalo}) and evaluate each of the three terms in this equation. Both the $n^{\rm th}$ and $(n+1)^{\rm th}$ generation minihalos have a Hernquist density profile. Firstly, adapting Eq.~(\ref{eq:enclosed mass for Hernquist minihalo as a functino of r}), the mass enclosed by the $n^{\rm th}$ generation Hernquist minihalo is

\begin{equation}\label{eq:enclosed mass for (n-1)th generation Hernquist minihalo as a function of r}
    M_{\text {enc,} n}(r)=2 \pi \rho_n r_n^3 \frac{\left(\frac{r}{r_n}\right)^2}{\left(1+\frac{r}{r_n}\right)^2}
\end{equation}
But

\begin{equation}\label{eq:r by r_(n-1) expression}
    \frac{r}{r_n}=\frac{r}{r_{\text{vir},n}} \frac{r_{\text{vir},n}}{r_n}=x_n c_n
\end{equation}
where $r_{\text{vir},n}$ is the viral radius of the $n^{\text{th}}$ generation minihalo, and

\begin{equation}
    x_n \equiv \frac{r}{r_{\text{vir},n}}
\end{equation}
\begin{equation}
    c_n \equiv \frac{r_{\text{vir},n}}{r_n}
\end{equation}
Substituting Eq.~(\ref{eq:r by r_(n-1) expression}) in Eq.~(\ref{eq:enclosed mass for (n-1)th generation Hernquist minihalo as a function of r}),
\begin{align}\label{eq:mass enclosed by the n^th generation Hernquist minihalo as a function of x}
    M_{\text {enc,}n}(x_n)&=2 \pi \rho_n r_n^3 \frac{\left(c_n x_n\right)^2}{\left(1+c_n x_n\right)^2} \nonumber\\
    &= 2 \pi \rho_n r_n^3 f_{\rm Hern}(c_n x_n)
\end{align}
When $x_n=x_n^*$, the normalized crossover radius of the $n^{\text{th}}$ generation Hernquist minihalo,
\begin{equation}\label{eq:mass enclosed within normalized crossover radius for (n-1)th generation minihalo}
    M_{\text {enc,}n}(x_n^*)=2 \pi \rho_n r_n^3 f_{\rm Hern}(c_n x_n^*)
\end{equation}

Secondly, Eq.~(\ref{eq:Delta M dimensionless}) gives the mass loss between $x=0$ and $x=1$ for a Hernquist (as well as NFW) profile. To evaluate $\Delta M_{x_n=0 \rightarrow x_n^*}$, we need to evaluate the mass loss between $x=0$ and $x=x^*$. Between these limits, Fig.~\ref{fig:psi crossing over deltaEpsilon} tells us that $\min[\vert\Delta\epsilon(x)\vert, \psi(x)] = \vert\Delta\epsilon(x)\vert$. Thus, Eq.~(\ref{eq:Delta M dimensionless}) turns into
\begin{equation}\label{eq:mass loss between x=0 and x=min[x^*,1]}
    \Delta M_{x_n=0 \rightarrow x_n^*} = 16\pi^2\rho_n r_{\text{vir},n}^3 \int\limits_{x_n=0}^{x_n^*} \mathrm{d}x_n\ x_n^2
    \int\limits_{\epsilon=0}^{\vert\Delta\epsilon(x_n)\vert} \mathrm{d}\epsilon\sqrt{2(\psi_{\rm A}(x_n)-\epsilon)}\hat{f}(\epsilon)
\end{equation}
Substituting $\hat{f}(\epsilon)$ from Eq.~(\ref{eq:f_hat}) in Eq.~(\ref{eq:mass loss between x=0 and x=min[x^*,1]}),
\begin{align}\label{eq:partial mass lost inside normalized crossover radius for (n-1)th generation minihalo}
    \Delta M_{x_n=0 \rightarrow x_n^*} &= 16\pi^2\rho_nr_{\text{vir},n}^3  \int\limits_{x_n=0}^{x_n^*}\int\limits_{\epsilon=0}^{\vert\Delta\epsilon(x_n)\vert} \int\limits_{\psi_{\rm B}^\prime=0}^\epsilon \frac{1}{\sqrt{8}\pi^2} x_n^2 \sqrt{2(\psi_{\rm A}(x_n)-\epsilon)} \nonumber\\
    &\ \ \ \ \ \ \ \ \ \ \ \times\frac{1}{\sqrt{\epsilon - \psi_{\rm B}^\prime}} \frac{\mathrm{d}^2\varrho}{\mathrm{d}\psi_{\rm B}^{\prime^2}} \left(x_n^\prime(\psi_{\rm B}^\prime)\right) \mathrm{d}\mathrm{\psi_{\rm B}^\prime} \mathrm{d}\epsilon \mathrm{d}x_n \nonumber\\
    &= 16\pi^2\rho_nc_n^3r_n^3 \times I_n
\end{align}
where
\begin{multline}
    I_n \equiv \int\limits_{x_n=0}^{x_n^*}\int\limits_{\epsilon=0}^{\vert\Delta\epsilon(x_n)\vert} \int\limits_{\psi_{\rm B}^\prime=0}^\epsilon \frac{1}{\sqrt{8}\pi^2} x_n^2 \sqrt{2(\psi_{\rm A}(x_n)-\epsilon)}
    \frac{1}{\sqrt{\epsilon - \psi_{\rm B}^\prime}} \frac{\mathrm{d}^2\varrho}{\mathrm{d}\psi_{\rm B}^{\prime^2}} \left(x_n^\prime(\psi_{\rm B}^\prime)\right) \mathrm{d}\mathrm{\psi_{\rm B}^\prime}\mathrm{d}\epsilon \mathrm{d}x_n
\end{multline}
Thirdly,

\begin{align}\label{eq:total mass of (n)th generation minihalo}
    \lim _{x_{n+1} \rightarrow \infty} M_{\text{enc},n+1}\left(x_{n+1}\right) & =\lim _{r \rightarrow \infty} M_{\text {enc,}n+1}(r) \nonumber\\
    & =\lim _{r \rightarrow \infty} 2 \pi \rho_{n+1} r_{n+1}^3 \frac{\left(\frac{r}{r_{n+1}}\right)^2}{\left(1+\frac{r}{r_{n+1}}\right)^2} \nonumber\\
    & =2 \pi \rho_{n+1} r_{n+1}^3
\end{align}
Substituting Eqs.~(\ref{eq:mass enclosed within normalized crossover radius for (n-1)th generation minihalo}), (\ref{eq:partial mass lost inside normalized crossover radius for (n-1)th generation minihalo}) and (\ref{eq:total mass of (n)th generation minihalo}) in Eq.~(\ref{eq:mass condition for transition from n^th to (n+1)^th generation Hernquist minihalo})
\begin{equation}\label{eq:mass condition for transition from (n-1)th gneeration Hernquist minihalo to (n)th generation Hernquist minihalo; expressions substituted}
    2 \pi \rho_n r_n^3 f_{\rm Hern}(c_n x_n^*)-16 \pi^2 \rho_n c_n^3 r_n^3 I_n =2 \pi \rho_{n+1} r_{n+1}^3
\end{equation}
Dividing Eq.~(\ref{eq:mass condition for transition from (n-1)th gneeration Hernquist minihalo to (n)th generation Hernquist minihalo; expressions substituted}) by $2 \pi \rho_n r_n^3$

\begin{equation}\label{eq:mass condition for transition from (n-1)th gneeration Hernquist minihalo to (n)th generation Hernquist minihalo; expressions substituted and divided throughout by apprpriate quantity}
    f_{\rm Hern}(c_n x_n^*) - 8 \pi c_n^3 I_n = \frac{\rho_{n+1} r_{n+1}}{\rho_n r_n} \left( \frac{r_{n+1}}{r_n} \right)^2
\end{equation}
Here too, we assume that at small radii, the $n^{\text{th}}$ and $(n+1)^{\text{th}}$ generation minihalos are indistinguishable from each other. Thus, we arrive at a similar ``small radius condition" as Eq.~(\ref{eq:small radius condition}):
\begin{equation}\label{eq:small radius condition for (n-1)th and (n)th generation minihalos}
    \rho_n r_n = \rho_{n+1} r_{n+1}
\end{equation}
Substituting Eq.~(\ref{eq:small radius condition for (n-1)th and (n)th generation minihalos}) in Eq.~(\ref{eq:mass condition for transition from (n-1)th gneeration Hernquist minihalo to (n)th generation Hernquist minihalo; expressions substituted and divided throughout by apprpriate quantity}), we get the ratio
\begin{align}\label{eq:r_n by r_(n-1); appendix version}
    \frac{r_{n+1}}{r_n} & =\sqrt{f_{\rm Hern}(c_n x_n^*)-8 \pi c_n^3 I_n} \nonumber\\
    & =R_n
\end{align}
where
\begin{equation}\label{eq:R_(n-1) definition; appendix version}
    R_n \equiv \sqrt{f_{\rm Hern}(c_n x_n^*)-8 \pi c_n^3 I_n}
\end{equation}
Substituting Eq.~(\ref{eq:r_n by r_(n-1); appendix version}) in Eq.~(\ref{eq:small radius condition for (n-1)th and (n)th generation minihalos})

\begin{equation}\label{eq:rho_n; appendix version}
    \rho_{n+1} = \frac{\rho_n}{R_n}
\end{equation}
\section{Computing the survival fraction of the $n^{\text{th}}$ generation minihalo}\label{app:SF_n derivation}
Here, we compute the survival fraction of the $n^{\text{th}}$ generation Hernquist minihalo, assuming it relaxes to the $(n+1)^{\text{th}}$ generation Hernquist minihalo. Our region of interest for calculating the mass loss is the physical virial radius of the unperturbed NFW minihalo.

We subject the $n^{\text{th}}$ generation Hernquist minihalo to a stellar encounter and let the remnant minihalo relax to an $(n+1)^{\text{th}}$ generation Hernquist minihalo. The survival fraction of the $n^{\text{th}}$ generation minihalo is then given by the ratio of the mass enclosed by the relaxed $(n+1)^{\text{th}}$ generation Hernquist minihalo within the region of interest to the mass enclosed by the unperturbed NFW minihalo within the same region of interest. Thus,

\begin{equation}\label{eq:survival fraction of n^th generation minihalo; definition}
    \text{SF}_n \equiv \frac{M_{\text{enc},n+1}\left(x_{n+1}^{r_{\text{vir,s}}}\right)}{M_{\text{vir,s}}}
\end{equation}
where

\begin{equation}
    x_{n+1}^{r_{\text{vir,s}}}=\frac{r_{\text{vir,s}}}{r_{\text{vir,}n+1}}
\end{equation}
is the physical virial radius of the unperturbed NFW minihalo expressed in the normalized ``local" radial distance variable of the $(n+1)^{\text{th}}$ generation minihalo. Thus,

\begin{align}
    x_{n+1}^{r_{\rm vir,s}} & =\frac{c_{\rm s} r_{\rm s}}{c_{n+1} r_{n+1}} \nonumber\\
    & =\frac{c_{\rm s}}{c_{n+1}} \frac{1}{\frac{r_{n+1}}{r_n} \frac{r_n}{r_{n-1}} \cdots \frac{r_2}{r_1} \frac{r_1}{r_{\rm s}}} \nonumber\\
    & =\frac{c_{\rm s}}{c_{n+1}} \frac{1}{R_n R_{n-1} \cdots R_1 R_{\rm s}}
\end{align}
where
\begin{equation}
    R_i=\frac{r_{i+1}}{r_i}
\end{equation}
Next, adapting Eq.~(\ref{eq:mass enclosed by the n^th generation Hernquist minihalo as a function of x}), the mass enclosed by the $(n+1)^{\text{th}}$ generation Hernquist profile is

\begin{align}\label{eq:mass enclosed by (n+1)^th generation Hernquist minihalo inside the physical virial radius of the NFW minihalo}
    M_{\text {enc,}n+1}\left(x_{n+1}\right) & =2 \pi \rho_{n+1} r_{n+1}^3 f_{\rm Hern}(c_{n+1} x_{n+1}) \nonumber\\
    \implies M_{\text {enc,}n+1}\left(x_{n+1}^{r_{\rm vir,s}}\right) & =2 \pi \rho_{n+1} r_{n+1}^3 f_{\rm Hern}(c_{n+1} x_{n+1}^{r_{\rm vir,s}})
\end{align}
Substituting Eqs.~(\ref{eq:mass enclosed by (n+1)^th generation Hernquist minihalo inside the physical virial radius of the NFW minihalo}) and (\ref{eq:virial mass of NFW profile using r_s}) in Eq.~(\ref{eq:survival fraction of n^th generation minihalo; definition})

\begin{align}\label{eq:survival fraction of n^th generation minihalo; intermediate 1}
    \text{SF}_n &= \frac{2 \pi \rho_{n+1} r_{n+1}^3 f_{\rm Hern}(c_{n+1} x_{n+1}^{r_{\rm vir,s}})}{4 \pi \rho_{\rm s} r_{\rm s}^3 f_{\rm NFW}(c_{\rm s})} \nonumber\\
    & =\frac{1}{2} \frac{\rho_{n+1} r_{n+1}}{\rho_{\rm s} r_{\rm s}}\left(\frac{r_{n+1}}{r_{\rm s}}\right)^2 \frac{f_{\rm Hern}(c_{n+1} x_{n+1}^{r_{\rm vir,s}})}{f_{\rm NFW}(c_{\rm s})} 
\end{align}
Mandating that all generations of minihalos are indistinguishable at small radii, we have

\begin{equation}\label{eq:small radius condition for NFW and (n+1)^th generation minihalos}
    \rho_{\rm s} r_{\rm s} = \rho_{n+1} r_{n+1}
\end{equation}
Moreover,

\begin{align}\label{eq:ratio of scale radii of (n+1)^th generation minihalo to NFW minihalo}
    \frac{r_{n+1}}{r_{\rm s}} & =\frac{r_{n+1}}{r_n} \frac{r_n}{r_{n-1}} \cdots \frac{r_2}{r_1} \frac{r_1}{r_{\rm s}} \nonumber\\
    & =R_n R_{n-1} \cdots R_1 R_{\rm s}
\end{align}
Substituting Eqs.~(\ref{eq:small radius condition for NFW and (n+1)^th generation minihalos}) and (\ref{eq:ratio of scale radii of (n+1)^th generation minihalo to NFW minihalo}) in Eq.~(\ref{eq:survival fraction of n^th generation minihalo; intermediate 1})

\begin{equation}
    \text{SF}_n = \frac{1}{2}\left(R_n R_{n-1} \cdots R_1 R_{\rm s}\right)^2 \frac{f_{\rm Hern}(c_{n+1} x_{n+1}^{r_{\rm vir,s}})}{f_{\rm NFW}(c_{\rm s})}
\end{equation}

\graphicspath{{Appendices/Figs/}}

\chapter{Rescaling the \texttt{CAMB} isocurvature growth function}\label{app: renormalizing the CAMB growth function}
\citelink{X2021}{X2021} gave the following approximate formula for the isocurvature growth function: 
\begin{equation}\label{eq: linear growth function}
	D(a) = \frac{2}{3} + \frac{a}{a_{\rm eq}}\ ,
\end{equation}
where $a$ is the scale factor at which the growth function is evaluated, and $a_{\rm eq}$ is the scale factor at matter-radiation equality. This growth function varies linearly with the scale factor $a$. It does not take into account the contribution from dark energy and hence is not accurate during the dark energy-dominated epoch (and consequently today). Dark energy causes the growth function to become sub-linear at late times. The isocurvature growth function that we calculate using \texttt{CAMB} is accurate at all redshifts in consideration. 

\begin{figure}
	\includegraphics[width=\columnwidth]{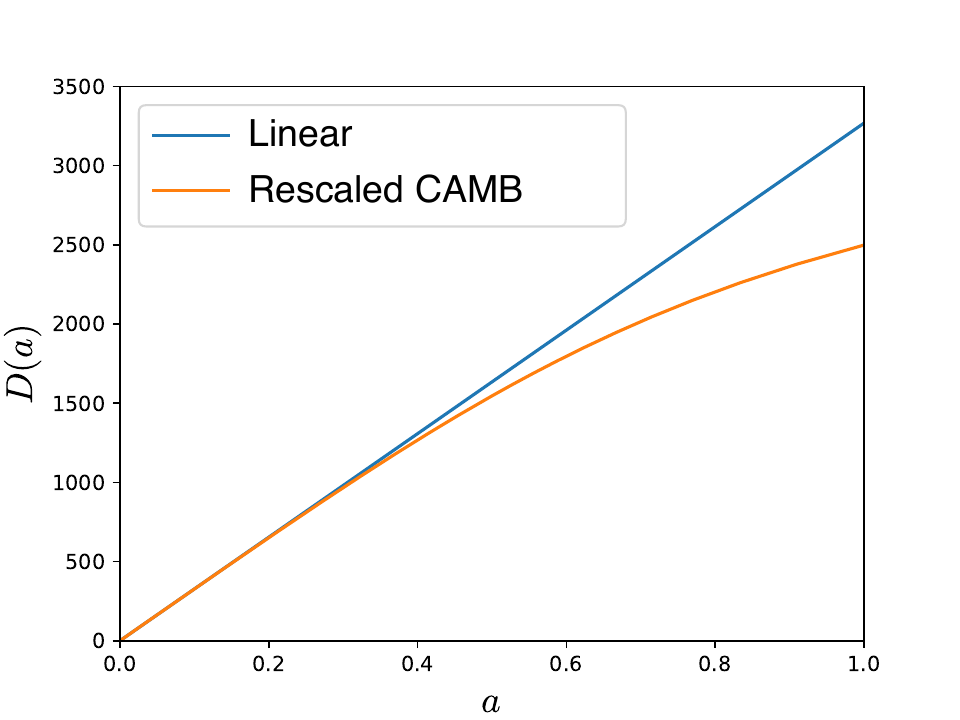}
	\caption[Isocurvature growth function $D$ of axions plotted against the scale factor $a$]{The isocurvature growth function $D$ is shown as a function of the scale factor $a$. The linear growth function is a solution to the M\'esz\'aros equations and doesn't take into account the dark energy contribution. The rescaled \texttt{CAMB} growth function is obtained by using the \texttt{CAMB} package. It takes into account the dark energy contribution. The two growth functions match during the matter- and radiation-dominated epochs but differ in the dark energy-dominated epoch.}
	\label{fig: growth_factor}
\end{figure}

The \texttt{CAMB} growth function,  $D_{\rm CAMB}$ is calculated using the power spectrum at some reference redshift (in our case, $z=100$) in the matter-dominated epoch.
For $z\gg 1$,
\begin{equation}
    D_{\rm CAMB}(z)=AD(z)
    \label{eq:DCAMB}
\end{equation}
where $A$ is some normalization constant to be determined and $D(a)$ is given in Eq.~(\ref{eq: linear growth function})).
We calculate $A$ using two reference redshifts:  $z=10$ and $z_{\rm eq} = 3266$ (the redshift of matter-radiation equality). Thus, using Eqs.~\ref{eq: linear growth function} and \ref{eq:DCAMB} we get
\begin{equation}
	 A=a_{\rm eq}\frac{D_{\rm CAMB}(a(z=10)) - D_{\rm CAMB}(a_{\rm eq})}{a(z=10) - a_{\rm eq}} \ ,
 \label{eq:evaluating A}    
\end{equation}
where $a(z) = 1 / (1+z)$, and $a_{\rm eq} = a(z_{\rm eq})$.  

Let $\widetilde{D}_{\rm CAMB}$ be the rescaled CAMB growth function:
\begin{equation}\label{eq: preliminary expression for renormalized growth function}
	\widetilde{D}_{\rm CAMB}(a) = \frac{{D}_{\rm CAMB}(a)}{ A}\ ,
\end{equation}
where $A$ is obtained from Eq.~\ref{eq:evaluating A}.

Fig.~\ref{fig: growth_factor} shows the growth function $D$ as a function of the scale factor $a$. It can be seen that the rescaled \texttt{CAMB} growth function is linear during the matter and radiation-dominated epochs but becomes suppressed to sub-linear during the dark energy-dominated epoch.

\chapter{Generating the pre-infall mass function}\label{app: pre-infall mass function}
Here, we generate the pre-infall mass function of axion minihalos using a modified Sheth-Tormen formalism \citelink{X2021}{X2021}. For minihalos that do not get captured by the halo of a galaxy, the  mass function $\mathrm{d}n_0/\mathrm{d}M$ is governed by the following equation: 
\begin{equation}\label{eq: mass_function_governing_equation}
	\frac{M^2 (\mathrm{d} n_0 / \mathrm{d} M)}{\bar{\rho}_{\rm c}} \frac{\mathrm{d} M}{M}=\nu f(\nu) \frac{\mathrm{d} \nu}{\nu} \, ,
\end{equation}
where $M$ represents the mass of a minihalo and $\bar{\rho}_{\rm c}$ is the comoving density of cold dark matter.
The preinfall comoving number density of minhalos between masses $M_{\rm min}$ and $M_{\rm max}$ is given by
\begin{equation}
     \int_{M_{\rm min}}^{M_{\rm max}} {\rm d}M\, \frac{\mathrm{d} n_0}{\mathrm{d}M} \,.
\end{equation}
 The parameter $\nu$ is defined by: 
\begin{equation}\label{eq: nu definition}
	\nu(M, z) \equiv \frac{\delta_{\rm c}^2}{\sigma^2(M, z)} \ ,
\end{equation}
where $\delta_{\rm c} = 1.686$ is the critical overdensity for spherical collapse of axion density perturbations and  $\sigma^2(M, z)$ is the variance of the initial density perturbations when smoothed with a spherical top-hat filter of length scale $R=(3 M / 4 \pi \bar{\rho}_{\rm c})^{1 / 3}$. It can be shown that the variance of the primordial white-noise matter power spectrum corresponding to the axion is given by \citelink{X2021}{X2021} and  \citelink{S2023}{S2023} as 
\begin{equation}\label{eq: sigma definition}
	\sigma(M, z)=D(z) \sqrt{\frac{3 A_{\mathrm{osc}}}{2 \pi^2} \frac{M_0}{M}} \ ,
\end{equation}
where $D$ is the isocurvature growth function of axion density perturbations. The amplitude of the white-noise matter power spectrum that arises from the axion isocurvature perturbations is $A_{\mathrm{osc}} = 0.1$.  The characteristic mass corresponding to the comoving Hubble length scale when the axion potential starts to oscillate is  \cite{dai2020gravitational}
\begin{equation}\label{eq: characteristic mass of minihalo as a function of axion mass}
	M_0=2.3 \times 10^{-10}\left(\frac{50\, \mu \mathrm{eV}}{m_{\rm a}}\right)^{0.51} M_{\odot} \ ,
\end{equation}
where $m_{\rm a}$ is the axion mass in $\mu \mathrm{eV}$. The function $f(\nu)$ is defined by: 
\begin{equation}\label{eq: nu times f(nu) definition}
	\nu f(\nu)=A\left(1+(q \nu)^{-p }\right)\left(\frac{q \nu}{2 \pi}\right)^{1 / 2} \exp (-q \nu / 2) \ .
\end{equation} 
\citelink{X2021}{X2021} performed numerical simulations and fitted Eq.~(\ref{eq: nu times f(nu) definition}) to their resulting mass function. They found the best-fit parameters to be $A = 0.374$, $p = 0.19$, and $q = 1.2$. Note that these values do not correspond to the standard Sheth-Tormen mass function. Hence, we call this the modified Sheth-Tormen formalism.

To find the expression for the mass function of axion minihalos, we can rearrange Eq.~(\ref{eq: mass_function_governing_equation}) as follows: 
\begin{equation}\label{eq: mass function final expression}
	\frac{\mathrm{d}n_0}{\mathrm{d}M}(M, z) = \frac{\nu f(\nu)}{\nu} \frac{\bar{\rho}_{\rm c}}{M} \frac{\mathrm{d}\nu}{\mathrm{d}M} \ .
\end{equation}

We now need to evaluate $\mathrm{d}\nu / \mathrm{d}M$. We do this by differentiating Eq.~(\ref{eq: nu definition}) with respect to $M$ at a fixed redshift: 
\begin{equation}\label{eq: dNu / dM - first expression}
	\frac{\mathrm{d}\nu}{\mathrm{d}M} = \delta_{\rm c}^2 \left(-\frac{2}{\sigma^3}\right) \frac{\mathrm{d}\sigma}{\mathrm{d}M} \ .
\end{equation}

Next, we need to evaluate $\mathrm{d}\sigma / \mathrm{d}M$. We do this by differentiating Eq.~(\ref{eq: sigma definition}) with respect to $M$ at a fixed redshift: 
\begin{equation}\label{eq: dSigma / dM}
	\frac{\mathrm{d}\sigma}{\mathrm{d}M} = D(z) \sqrt{\frac{3 A_{\mathrm{osc}}M_0}{2 \pi^2}} \left(-\frac{1}{2}\frac{1}{M^{3/2}}\right) \ .
\end{equation}

Substituting Eqs.~(\ref{eq: sigma definition}) and (\ref{eq: dSigma / dM}) in Eq.~(\ref{eq: dNu / dM - first expression}), we get: 
\begin{equation}
	\frac{\mathrm{d}\nu}{\mathrm{d}M} = \frac{\delta_{\rm c}^2}{D^2(z)}\left(\frac{3 A_{\mathrm{osc}}M_0}{2 \pi^2}\right)^{-1} \ .
\end{equation}

Thus, we can find the value of the mass function $\mathrm{d}n_0 / \mathrm{d}M$ of axion minihalos at a given minihalo mass and redshift using Eq.~(\ref{eq: mass function final expression}).

\chapter{Code to evolve an orbit}
\label{app: Code to evolve an orbit}
The position and velocity of each particle at arbitrary times are evaluated using the \texttt{LBparticles} code\footnote{\url{https: //github.com/lbparticles/lbparticles}}. The code is a \texttt{Python} implementation of the high-order epicyclic approximation developed by Ref.~\cite{lyndenbell2015} with several practical improvements. Given an initial 3D position and 3D velocity and a static potential, the code computes two series of coefficients for series in $\cos(n\eta)$ and $\cos(n\chi)$, where $\eta$ and $\chi$ are fictional angles related to the particle's  angular coordinate in the potential and the time elapsed respectively, and $n$ denotes the element of the series. The orbits are not fully analytic because their properties depend on the peri- and apocenter, which must be found numerically, and the relationship between $\chi$ and $t$ must be computed numerically for each orbit. This latter relationship can be quickly constructed for arbitrary orbits by precomputing a series of integrals on a grid of $\chi$, $e$, and $k_{\rm LB}$. Here $e$ is the eccentricity of the orbit and $k_{\rm LB}$ is a closely-related quantity chosen in Ref.~\cite{lyndenbell2015} to make the 0th order version of this approximation as accurate as possible. Given an arbitrary orbit's value of $k_{\rm LB}$ and $e$, we find the closest values of $k_{\rm LB}$ and $e$ from the precomputed grid, and perform a 2D Taylor series to evaluate $t(\chi)$.

The improvements relative to Ref.~\cite{lyndenbell2015} include the following. First, several algebraic errors in the expressions for the $\cos$ series are corrected. Second, the numerical prescription for evaluating $t(\chi)$ replaced a prescription which was highly-accurate at evaluating the period of the radial oscillations of the particle, but not incredibly accurate within a single oscillation, both of which are necessary for evaluating the particle's position and velocity at arbitrary times. Third, prescriptions for the vertical oscillation of particles embedded in thin disks were added, though we do not use them in the present work.

While it is straightforward to integrate a particle's motion in a smooth central potential, \texttt{LBparticles} allows us to simply evaluate the position and velocity of the particle at any time with a low cost and a high accuracy. In contrast, numerical integration increases in expense as $t$ advances away from the time of initialization, and would require interpolation of the solution or pre-ordained evaluation points to evaluate the position at arbitrary times.

\chapter[Local minimum of $\rho_* v$ vs.\ time]{The local minimum of the $\rho_* v$ against time curve does not coincide with $\protect\dot{Z}=0$}\label{app: local minimum of rho_*v does not coincide with z_dot=0}

The term $\rho_* v$ can be plotted against time $t$. The local minima of the $\rho_* v$ curve occur when $\mathrm{d} (\rho_* v)/ \mathrm{d}t  = 0$. Thus,
\begin{equation}\label{eq: d/dt of rho_* v - part 1}
	\frac{\mathrm{d}}{\mathrm{d} t}\left(\rho_*(R, Z) v\right)=\left[\frac{\partial \rho_*}{\partial R} \dot{R}+\frac{\partial \rho_*}{\partial Z} \dot{Z}\right] v+\rho_* \dot{v}\ ,
\end{equation}
where the superscript ``.'' indicates derivative with respect to time. Differentiating Eq.~(\ref{eq: rho_* expression}) with respect to $R$,
\begin{align}\label{eq: partial rho_* / partial R}
	\frac{\partial \rho_*}{\partial R} & =\sum_{d=t,T} \frac{\Sigma_{\mathrm{d}, 0}}{2 Z_{\mathrm{d}}}\left(\frac{-1}{R_\mathrm{d}}\right) \exp \left(-\frac{|Z|}{Z_{\mathrm{d}}}\right) \exp \left(-\frac{R}{R_{\mathrm{d}}}\right) \nonumber\\
	& =-\frac{\rho_*}{R_\mathrm{d}}\ .
\end{align}

The second equality in Eq.~(\ref{eq: partial rho_* / partial R}) comes from making use of the definition of $\rho_*$ in Eq.~(\ref{eq: rho_* expression}). Similarly,
\begin{align}\label{eq: partial rho_* / partial z}
	\frac{\partial \rho_*}{\partial Z} & =\sum_{d=t, T} \frac{\Sigma_{\mathrm{d}, 0}}{2 Z_{\mathrm{d}}}\left( \pm \frac{1}{Z_{\mathrm{d}}}\right) \exp \left(-\frac{|Z|}{Z_{\mathrm{d}}}\right) \exp \left(-\frac{R}{R_{\mathrm{d}}}\right) \nonumber\\
	& = \pm \frac{\rho_*}{Z_{\mathrm{d}}} \ .
\end{align}

Note that we use ``$\pm$'' in Eq.~(\ref{eq: partial rho_* / partial z}). The ``$+$'' sign is applicable when $Z<0$ since $|Z|=-Z$ in this regime. On the other hand, the ``$-$'' sign is applicable when $Z>0$ since $|Z|=Z$ in this regime.

Substituting Eqs.~(\ref{eq: partial rho_* / partial R}) and (\ref{eq: partial rho_* / partial z}) in eqn (\ref{eq: d/dt of rho_* v - part 1}),
\begin{align}\label{eq: d/dt of rho_* v - part 2}
	\frac{\mathrm{d}}{\mathrm{d} t}\left(\rho_* v\right) & =\left[-\rho_* \frac{\dot{R}}{R_{\mathrm{d}}} \pm \rho_* \frac{\dot{Z}}{Z_{\mathrm{d}}}\right] v+\rho_* \dot{v} \nonumber\\
	& =\rho_* v\left[-\frac{\dot{R}}{R_{\mathrm{d}}} \pm \frac{\dot{Z}}{Z_{\mathrm{d}}}+\frac{\dot{v}}{v}\right] \ .
\end{align}

In the R.H.S. of Eq.~(\ref{eq: d/dt of rho_* v - part 2}), $\rho_*$ and $v$ are never zero in practice. Thus, in the L.H.S., the $\rho_* v$ curve has a local minimum when the term in the square parentheses is zero. For this, it is not sufficient that $\dot{Z}=0$. There are contributions from $\dot{R}$ and $\dot{v} / v$ as well. Thus, the local minimum of the $\rho_* v$ curve doesn't exactly coincide with the instant that $\dot{Z}=0$.

\chapter{Converting infall redshift to look back time}\label{app: converting infall redshift to look back time}

We start by rewriting eqn.~(\ref{eq:derived from Friedmann's first equation}) for the Hubble parameter $H$: 
\begin{align}\label{eq: Hubble parameter expression in terms of Hubble constant}
	H^2(z)&=H_0^2\left[\Omega_{\rm m}(1+z)^3+\Omega_{\rm r}(1+z)^4+\Omega_{\Lambda}\right] \nonumber\\
	\implies H(z)&=H_0\left[\Omega_{\rm m}(1+z)^3+\Omega_{\rm r}(1+z)^4+\Omega_{\Lambda}\right]^{1 / 2} \ .
\end{align}

By definition,
\begin{equation}\label{eq: hubble paramter definition}
	H=\frac{\dot{a}}{a}=\frac{1}{a} \frac{\mathrm{d} a}{\mathrm{d} t} \ .
\end{equation}

But
\begin{equation}\label{eq: scale factor formula}
	a=\frac{1}{1+z}
\end{equation}
and
\begin{equation}\label{eq: time derivative of scale factor formula}
	\frac{\mathrm{d} a}{\mathrm{d} t}=\frac{\mathrm{d} a}{\mathrm{d} z} \frac{\mathrm{d} z}{\mathrm{d} t}=-\frac{1}{(1+z)^2} \frac{\mathrm{d} z}{\mathrm{d} t} \ .
\end{equation}

Substituting Eqs.~(\ref{eq: scale factor formula}) and (\ref{eq: time derivative of scale factor formula}) in Eq.~(\ref{eq: hubble paramter definition}),
\begin{equation}
	H(z) = -\frac{1}{1+z} \frac{\mathrm{d} z}{\mathrm{d} t}
\end{equation}
\begin{equation}\label{eq: pre-integrated lookback time}
	\implies \mathrm{d}t =-\frac{1}{H(z)} \frac{1}{1+z} \mathrm{d}z \ .
\end{equation}

Integrating Eq.~(\ref{eq: pre-integrated lookback time}) from infall redshift until today
\begin{equation}\label{eq: integrated lookback time}
	\int_t^{t_0} \mathrm{d} t^\prime= - \int_{z}^0 \frac{1}{H(z^\prime)} \frac{1}{1 + z^\prime} \mathrm{d} z^\prime \ ,
\end{equation}
where $t_0$ is the time today (as measured since the big bang) and $t$ is the time corresponding to the infall redshift. But the L.H.S. of Eq.~(\ref{eq: integrated lookback time}) is just the lookback time $T$ corresponding to the infall redshift. Substituting Eq.~(\ref{eq: Hubble parameter expression in terms of Hubble constant}) into the R.H.S. of Eq.~(\ref{eq: integrated lookback time}) gives the required expression:
\begin{equation}
	T(z) = \frac{1}{H_0} \int_0^{z} \frac{1}{1+z^\prime} 
	\left[\Omega_{\rm m}(1+z^\prime)^3 + \Omega_{\rm r}(1+z^\prime)^4 + \Omega_\Lambda\right]^{-1/2} \mathrm{d}z^\prime\, .
\end{equation}

\chapter{Estimating the effect of updating the dynamical time after each disk pass}\label{app:updating dynamical time}

\begin{color}{black}

For simplicity, we do not update the dynamical time of the minihalo after each disk pass. To update $t_{\rm dyn}$, we would need to compute the virial radius and virial mass after each disk pass, which would be computationally intensive. In our current approach, $t_{\rm dyn}$ is set to the dynamical time at infall before any stellar disruption begins, using Eqs.~(\ref{eq: dynamical time}) and (\ref{eq:virial density}). But we now present a back-of-the-envelope calculation to estimate the effect of updating $t_{\rm dyn}$.
	
	Consider an isolated NFW profile minihalo (as considered in our previous article \cite{dsouza2024}) undergoing a stellar encounter, and then gravitationally relaxing to a Hernquist profile minihalo. For the purposes of determining the mass loss incurred by the minihalo, let our physical region of interest be the volume $V_{\mathrm{vir}, i}$ inside the virial radius of the NFW minihalo. Then, the average density $\bar{\rho}_i$ of the NFW minihalo inside volume $V_{\mathrm{vir}, i}$ is given by
	\begin{equation}\label{E1}
		\bar{\rho}_i = \frac{M_i}{V_{\mathrm{vir}, i}} ,
	\end{equation}
	where $M_i$ is effectively the virial mass of the unperturbed NFW minihalo. Next, after relaxation to the Hernquist profile, let $M_f$ be the mass of the Hernquist minihalo enclosed by the same physical volume $V_{\mathrm{vir}, i}$. 
    Then the average density $\bar{\rho}_f$ of the Hernquist minihalo inside $V_{\mathrm{vir}, i}$ is given by
	\begin{equation}\label{E2}
		\bar{\rho}_f = \frac{M_f}{V_{\mathrm{vir}, i}} .
	\end{equation}
    \rev{Note that we have assumed the same volume for both the unperturbed and perturbed minihalos. This is because we are fixing the physical region of interest within which to evaluate the mass loss. This is in keeping with our developed analytical method in Chapter~\ref{ch:journal_paper_1}, where we used the physical virial radius of the unperturbed NFW minihalo to define our region of interest.}	Dividing Eq.~(\ref{E2}) by Eq.~(\ref{E1}), we have
	\begin{equation}\label{E3}
		\frac{\bar{\rho}_f}{\bar{\rho}_i} = \frac{M_f}{M_i} .
	\end{equation}
	
	$M_f / M_i$ represents the ratio of the mass that has survived after the stellar encounter to the mass that was initially present, both evaluated inside $V_{\mathrm{vir}, i}$.
	
	Now, in Table~\ref{table:mass ratios} of our article, for the case of $p=$ hybrid, $m_{\rm a} = 25 \mu$eV, $M_{\rm min} = 10^{-2}M_\odot$, we have the ratio of the mass of the Milky Way (MW) population of minihalos that survives after stellar disruption, to the mass of the minihalo population without any stellar disruption, i.e., $M_{\rm surv}/M_{\rm ori} = 37.6\%$. In the spirit of a back-of-the-envelope calculation, we relate the isolated minihalo case to the MW minihalo population case by following the approximate equality:
	\begin{equation}\label{E4}
		\frac{M_f}{M_i} \approx \frac{M_{\rm surv}}{M_{\rm ori}}.
	\end{equation}

	Comparing Eqs.~(\ref{E3}) and (\ref{E4}), we have
	\begin{equation}\label{E5}
		\frac{\bar{\rho}_f}{\bar{\rho}_i} \approx 0.376 .
	\end{equation}
	For the MW minihalo population case, we should now think of $\bar{\rho}_i$ as an approximate measure of the dynamical time $t_{\mathrm{dyn},i}$ before any stellar disruption begins. We should also think of $\bar{\rho}_f$ as an approximate measure of the dynamical time $t_{\mathrm{dyn},f}$ of the population after all stellar encounters, i.e., today ($z=0$). Now, looking at Eq.~(\ref{eq: dynamical time}), we can state that
	\begin{equation}
		\frac{t_{\mathrm{dyn},f}}{t_{\mathrm{dyn},i}} = \sqrt{\frac{\bar{\rho}_i}{\bar{\rho}_f}} \approx \sqrt{\frac{1}{0.376 }} = 1.63 
	\end{equation}
	\begin{equation}
		\implies t_{\mathrm{dyn},f} \approx 1.63 \times t_{\mathrm{dyn},i}
	\end{equation}

	Now, coming back to our Monte-Carlo method, for each grid point $(M, z_{\rm i})$, we have assigned an orbit. We can say that, to a good approximation, the dynamical time at the start of stellar interactions (i.e., at $z = z_{\rm i}$) is $t_{\rm dyn}(z_{\rm i})$. On average, as time passes by, the dynamical time should increase gradually up to $1.63 \times t_{\rm dyn}(z_{\rm i})$ at the end of stellar disruptions, i.e., today $(z=0)$. Instead, if we set $t_{\rm dyn} = t_{\rm dyn}(z_{\rm i})$ and not update it in the Monte-Carlo procedure, we would be underestimating the true value of our final result: $M_{\rm surv}/M_{\rm ori}$. This is because by having a smaller $t_{\rm dyn}$, the effect of gravitational relaxation is more emphasized, leading to more mass loss, resulting in a smaller value of $M_{\rm surv}/M_{\rm ori}$ (=37.6\%). On the other hand, if we set $t_{\rm dyn} = 1.63 \times t_{\rm dyn}(z_{\rm i})$ and not update the dynamical time, we would be overestimating the value of $M_{\rm surv}/M_{\rm ori}$. We implemented this case and found out that $M_{\rm surv}/M_{\rm ori} = 38.9\%$. Thus, we conclude that by updating the dynamical time with each disk pass, it would not change the value of $M_{\rm surv}/M_{\rm ori}$ from the value presented in Table~\ref{table:mass ratios} by more than $38.9 - 37.6 = 1.3\%$.
	
	We performed a similar analysis for another case in Table~\ref{table:mass ratios}: $p=$hybrid, $m_{\rm a} = 25 \mu$eV, $M_{\rm min} = 10^2 M_\odot$. Here, we found that the value of $M_{\rm surv}/M_{\rm ori}$ wouldn't change by more than 3\% from the value presented in Table~\ref{table:mass ratios}.

\end{color}

\end{appendices}

\printthesisindex 

\end{document}